\documentclass[12pt,a4paper]{article}
\usepackage{jheppub}
\usepackage{amsmath}
\usepackage{amsfonts}
\usepackage{mathtools}
\usepackage{amssymb}
\usepackage{caption}
\usepackage{subcaption}
\allowdisplaybreaks[2]

\title{Universal corrections to entanglement entropy of local quantum  quenches}
\author{Justin R. David ${}^{a}$, Surbhi Khetrapal ${}^{a}$, S. Prem Kumar ${}^{b}$ }
\affiliation{${}^{a}$ Centre for High Energy Physics, Indian Institute of Science,\\
C. V. Raman Avenue, Bangalore 560012, India. \\
${}^{b}$ Department of Physics, Swansea University, \\
Singleton Park, Swansea SA2 8PP, UK. }
\emailAdd{justin, surbhi@cts.iisc.ernet.in, s.p.kumar@swansea.ac.uk}
\abstract{ 
We study the time evolution  of single interval R\'{e}nyi and entanglement entropies following 
local quantum quenches in two dimensional conformal field theories at finite temperature  
for which the locally excited states have a finite temporal width $\epsilon$. 
We show that, for local quenches produced by the action of a conformal primary field, the time dependence of R\'enyi and entanglement entropies at order $\epsilon^2$ is universal. 
It is determined by the expectation value of the stress tensor in the replica geometry and proportional to  the conformal dimension of the primary field generating the local excitation.  
We  also show that in CFTs with a gravity dual,
the $\epsilon^2$ correction to the holographic entanglement entropy following a local quench precisely agrees with the CFT prediction. 
We then consider CFTs admitting a higher spin symmetry and turn on a higher spin chemical potential  $\mu$. We calculate the 
time dependence of the order $\epsilon^2$ correction to the entanglement entropy for small $\mu$, and show that the contribution at order $\mu^2$ is universal. 
We verify our arguments against exact  results for minimal models and the  free fermion theory.  
}

\begin{document}
\def\be{\begin{equation}}
\def\ee{\end{equation}}
\def\bea{\begin{eqnarray}} 
\def\eea{\end{eqnarray}}

\maketitle
\flushbottom

\section{Introduction}

The study of out-of-equilibrium dynamics in strongly coupled quantum systems is of fundamental importance for unravelling the physical mechanisms leading to eventual thermalisation or equilibration in such systems. Remarkably, these questions are now accessible experimentally, particularly in lower dimensions e.g. \cite{kww}. Furthermore, nonequilibrium dynamics of systems with hidden symmetries, such as integrable models, which are generally expected not to thermalise, are being extensively studied  \cite{gge1, gge2,gge3, gge4, gge}  and their steady state behaviour is believed to be described by Generalized Gibbs Ensembles (GGE) rather than a thermal state. The simplest  way to drive a system away from equilibrium, and follow the ensuing dynamics, is  via a so-called quantum quench in which the  Hamiltonian of the theory is changed abruptly at some time and the system allowed to evolve unitarily  subsequently  \cite{calcardy1, calcardy2}. 
Such quenches can be both ``global'' or ``local'', the former corresponding to a rapidly changing coupling in the Hamiltonian whilst the latter is generated by a localized change in the initial quantum state or density matrix so that it departs from the ground state only locally. The time dependence of  observables following a local quench 
in quantum field theories 
has been the focus of several recent studies \cite{Calcard,Eispes,Stepdub,Asplund:2013zba}, particularly in conformal field theories (CFTs) in 1+1 dimensions \cite{calcardy3}.

A local quench in a two dimensional CFT can be achieved using different initial  conditions or ``protocols'' \cite{calcardy3}. One natural approach is via the intsertion  of a local operator (typically a conformal primary) at some point, creating a local change in the initial density matrix\footnote{A different, so-called ``cut and glue'' local quench protocol has been the subject of study in \cite{Calcard, calcardy3}.}. Such local quenches have also been keenly pursued in recent holographic studies of CFTs that admit gravity duals
\cite{Asplund:2011cq,Nozaki:2013wia,Nozaki:2014hna,Caputa:2014vaa} (see \cite{Nozaki:2014uaa} for  a brief review). In this paper, we will focus attention on this type of quenches, and point out new universal features of these.

The physical observables whose time evolution we want to track are the R\'enyi and entanglement entropies  (RE/EE) of a single interval $A$, following the local quench characterised by a density matrix of the form
\begin{eqnarray}\label{initstate}
&&\hat{\rho}_\epsilon\,=\, \mathcal{N}\,   e^{-iHt}\,  \left( e^{-\epsilon H} \mathcal{O}(0) 
e^{  \epsilon H } \right)\,  \rho_\beta \,\left(  e^{ \epsilon H}    
\mathcal{O}^\dagger(0)e^{-\epsilon H} \right) \,e^{iHt}\,,\\\nonumber
&&\rho_\beta\,=\,e^{-\beta H}\,.
\end{eqnarray}
Here $\rho_\beta$ is the thermal density matrix and ${\cal O}$ a conformal primary field generating an excitation at the origin at time $t=0$. The parameter $\epsilon$, which plays an important role in our work, represents the { temporal width} of the localized excitation in a sense that can be made precise. It can be viewed as a (small) translation in imaginary time of the operator ${\cal O}$ and its Hermitian conjugate, so that the operator product of ${\cal O}$ with ${\cal O}^\dagger$ is regulated in the ultraviolet (UV) and well defined. The density matrix can then be interpreted as the analytic continuation of a correlation function in Euclidean time \cite{Nozaki:2014hna}.

In this paper, we will show that the single interval RE possesses universal features at finite $\epsilon$. In particular, for  CFTs with  fixed central charge $c$, we will find that 
at finite $\epsilon$ there exists a correction to the R\'{e}nyi and entanglement entropies at order $\epsilon^2$ whose  time 
dependence is universal.

\begin{figure}
\centering
\includegraphics[scale=0.45]{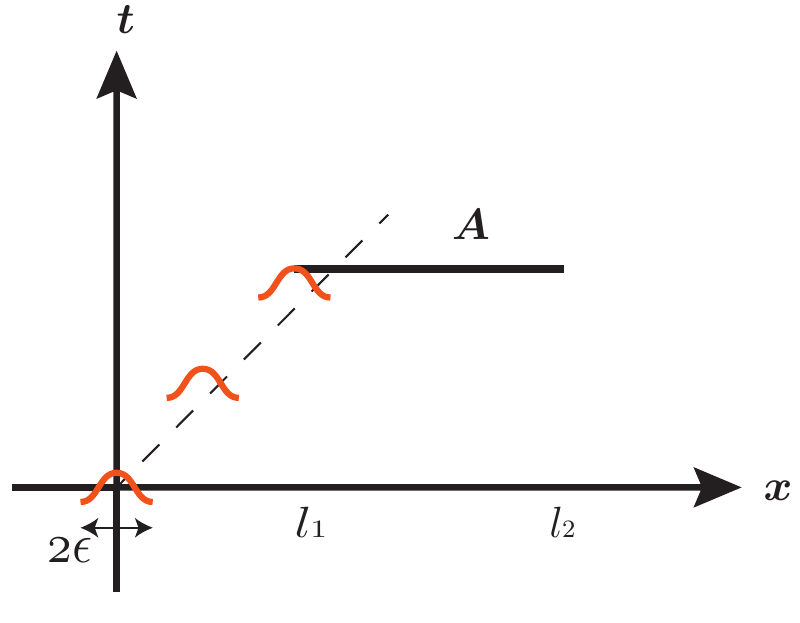}\hspace{0.4in}
\includegraphics[scale=0.75]{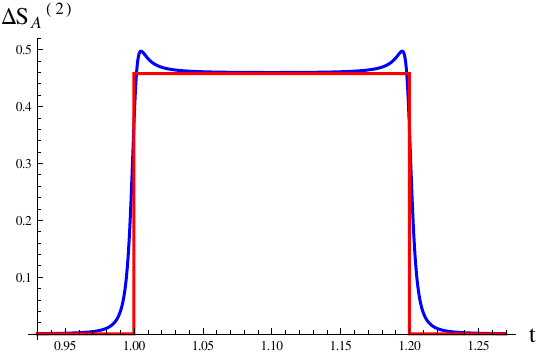}
\caption{\small {{\bf Left}}: A finite width pulse generated by a local quench propagates along the lightcone and reaches the interval $A$ after time $t=l_1$. {{\bf Right}:} Typical profile for the time evolution of the change in the (second) R\'enyi entropy after the quench, for a minimal model CFT. Following an initial spike or overshoot after (before) the pulse enters (exits) the interval, $\Delta S_A^{(n)}$ settles toward a constant value and subsequently vanishes when the pulse exits the interval.}
\label{introfig}
\end{figure}

Taking an interval $A$  on the real line  with end points $(l_1, l_2)$ such that $l_2>l_1>0$, the finite width excitation created by the conformal primary ${\cal O }$ moves outward on the light cone as in figure \ref{introfig}. When the pulse encounters the  interval after time $t \simeq l_1$, we expect 
the R\'enyi/entanglement entropy of the interval $A$ to then evolve with time. We will assume that this change in the R\'enyi entropy of entanglement admits an expansion in powers of $\epsilon$:
\bea
\Delta S_A^{(n)}\,=\,\Delta S_A^{(n;\,0)}\,+\,\epsilon\,\Delta S_A^{(n;\,1)}\,+\,\epsilon^2\,\Delta S_A^{(n;\,2)}\,+\ldots\,.
\eea
In \cite{He:2014mwa} it was shown that for any rational CFT, the change  in the entanglement entropy $\Delta S_A^{(1)}$ of the interval $A$ as a function of  time, in  the limit $\epsilon\rightarrow 0$,  is a step function with a step size determined by the so-called quantum dimension of the operator ${\cal O}$:
\begin{eqnarray}\label{jump}
 \Delta S _{A}^{(1;\,0)}
 \,  &=& \, 0  \qquad 
 \quad   t <l_1 \quad{\mbox{or}}\quad  t > l_2\,, \\\nonumber \\\nonumber
 &=&\, \ln d_{\cal{ O} }\qquad   
 \quad l_1\leq t\leq l_2\,.
  \end{eqnarray}
Here $d_{\cal{O}}$ denotes the quantum dimension (not to be confused with the scaling dimension) of the  primary field ${\cal O}$.  In fact, this result has also been shown to hold for quenching operators which are descendants within a conformal family and where $d_{\cal O}$ is then the quantum dimension of the  associated primary \cite{Caputa:2015tua}.
For finite width quenches, i.e. $\epsilon\neq 0$, one naturally expects the step function to be smoothed out and, as we will show below, we generically encounter smooth profiles of the type displayed in figure \ref{introfig}.  
In general, the smoothening of the step function depends on the details of the CFT. 
However, the growth regime is driven by universal effects at order $\epsilon^2$ which we establish using general arguments and specific examples: 
\begin{itemize}
\item{We show that time dependence of the first non-trivial correction to the $n$-th R\'enyi entropy  $\Delta S_A^{(n)}$ at order $\epsilon^2$ is determined  by the expectation value of the stress tensor in the replica geometry, i.e. the $n$-sheeted Riemann surface  branched along the interval $A$. The overall multiplicative coefficient is such that this correction is independent of the central charge $c$, but depends on the conformal dimension of the quenching operator ${\cal O}$.  Furthermore, the correction can be trusted, i.e. subleading terms in the $\epsilon$-expansion are small,  only when the centre of the pulse is outside a certain neighbourhood of the end-points of the interval $A$. The universal correction takes the general form,
\be\label{introgeneral}
\Delta S_A^{(n;\,2)}\,=\,4\,\frac{\Delta_{\cal O}}{c}\,\frac{n}{n-1}\left(\langle T(0)\rangle_n\,-\,\langle T\rangle_\beta\right)\,+\,{(t\to -t)}\,,
\ee
}
with the putative correction at the first order in $\epsilon$  vanishing identically, i.e. $\Delta S_A^{(n;\,1)}=0$. The result is determined by  the expectation values $\langle T \rangle_n$ and $\langle \overline T \rangle_n$ of the holomorphic and antiholomorphic components of the stress tensor, at the point of insertion of the local operator ${\cal O}$, in the $n$-sheeted replica geometry.  The locations of the branch-points of the $n$-sheeted Riemann surface are time dependent and given by the lightcone coordinates of the end-points of the interval  $A$, namely $(x_2,\,\bar x_2)\,=\,(l_1-t,\,l_1+t)$ and $(x_3,\,\bar x_3)\,=\,(l_2-t,\,l_2+t)$.
The resulting functional form is therefore time dependent, and  is  obtained via a uniformization map from the replica geometry to the complex plane. This is independent of the details of the CFT (when formulated on the infinite spatial line). We will show that the universal contribution at order $\epsilon^2$ is the leading correction to the zero-width result provided the dimension of the lowest lying primary is greater than one.

\item{ The holographic dual to the excited (quenched) state is given by  the backreacted geometry produced by a  massive infalling particle  in AdS$_3$ spacetime, or in the BTZ black hole geometry when the CFT is at a finite temperature \cite{Caputa:2014vaa,Asplund:2014coa, Caputa:2014eta}. The temporal width $\epsilon$ of the excitation is  naturally incorporated in the infalling matter geometry. Expanding the  result for the holographic entanglement entropy we are able to show that the $O(\epsilon^2)$ term is in  precise agreement with the CFT prediction summarised above.
}
\item{While our arguments  are general,  we  work out in detail, the finite $\epsilon$ contributions to the second R\'{e}nyi entropy for two specific examples. These are the minimal model CFTs and the free fermion theory. In both these theories, the second R\'enyi entropy can be computed exactly as a function of the width $\epsilon$, for appropriately chosen quenching operators ${\cal O}$.
The minimal models exhibit precise agreement of the order $\epsilon^2$ correction to the second R\'enyi entropy with the universal prediction summarised above.
The jump at order $\epsilon^0$ is smoothed out in the exact result as depicted in figure \ref{introfig}. The behaviour of the R\'enyi entropy on either side of the overshoot in figure \ref{introfig} is well approximated by the universal correction at finite width.
The free fermion CFT, which is the second example where the exact result for the second R\'enyi entropy can be obtained (following an appropriate local quench), illustrates an exception to the universal result due to the presence of a dimension one current. In this case we will be able to track precisely how the universal contribution described above, is modified by the $U(1)$ current which appears in the OPE of the quenching operators at order $\epsilon^2$.
This theory also possesses  the ${\cal W}_{1+ \infty}$  symmetry 
and is therefore a suitable example to study the effects of  higher spin chemical potential as summarised below. 
}
\item{One of our aims in this paper is to  initiate a study of the effects of finite higher spin chemical potential $\mu$  on 
the time evolution of entanglement entropy.  CFTs with higher spin symmetry are known to be dual to higher spin theories of gravity \cite{Henneaux:2010xg, Prokushkin:1998bq} and the CFT state with a higher spin chemical potential is mapped to a black hole carrying higher spin charge \cite{Gutperle:2011kf}.
We will show that for any such CFT with fixed central charge $c$,  the 
step-function change (\ref{jump}) in the $\epsilon\rightarrow 0$ limit of the quench, is unchanged at  $O(\mu^2)$ in the chemical potential\footnote{Closely related observations have been made in \cite{Chen:2015usa}.}.  
However there exists a nontrivial correction at $O(\mu^2 \epsilon^2)$ whose time dependence  is again universal.  This time dependence is determined by the three point function involving the stress tensor $T$, and the higher spin current $W$,  
$\langle T(x)  W (y_1) W(y_2)  \rangle_n $ evaluated in the replica geometry and hence the result is universal.  Again this correction does not depend on the  central charge and is sensitive to the conformal dimension of the  operator ${\cal O}$ which appears as an overall multiplicative factor.  We perform checks of these results against explicit computations in the free fermion theory.
}

\item{It is important to note that the temporal growth of R\'enyi/entanglement entropy discussed in this paper is distinct from a different growth region known and well understood \cite{Caputa:2014vaa,Asplund:2014coa} for CFTs with large central charge $c \gg 1$. In large-$c$ CFTs,  using properties of Virasoro conformal blocks it can be shown that the  entanglement entropy is parametrically (vanishingly) small for $ t<l_1$ and $t>l_2$. Close to the end-points of the interval $A$, however, it grows logarithmically. In particular, for local quenches generated by 
operators with large conformal dimensions $\Delta_{\cal O }\gg 1$, such that  
$c\gg \Delta_{\cal O}$ \cite{Caputa:2014vaa} in the large-$c$ limit, the entanglement entropy grows as,
 \begin{equation}\label{regime1}
 \Delta S_{A}\, =\, 2 \Delta_{\cal O } \ln \frac{(t-l_1)}{\epsilon}  \qquad\qquad l_1 <  t  <   l_ 1 \,+\,  
 \epsilon\, d_{\cal O}^{1/2 \Delta_{{\cal O }}}\,.
\end{equation}
On the other hand, when the  operator dimension scales with  the central charge $\Delta_{\cal O } /{c}  \sim O(1)$, one obtains a different logarithimic growth 
in the  same window, given by  \cite{Asplund:2014coa}:
\begin{equation}\label{regime2}
 \Delta S_{A} \,=\,\frac{c}{6}  \ln \frac{(t-l_1)}{\epsilon}  \qquad  \qquad l_1 <  t <  l_ 1 \,+\,  
 \epsilon\, d_{\cal O}^{1/2 \Delta_{{\cal O }}}\,.
\end{equation}
Outside this small window\footnote{There is a similar window for  $l _2  \,-\, \epsilon \,d_{\cal O}^{1/2 \Delta_{{\cal O }}}< t< l_2$.}, representing a growth phase,
the entanglement entropy is expected to behave as given in (\ref{jump}) in the $\epsilon \rightarrow 0$ limit. In practice, the finite step jump is not visible in large-$c$ or holographic limits because the quantum dimension diverges in the limit. Nevertheless, we will show that when $t < l_1$ ( and $t> l_2$), there is another growth (and relaxation) phase for a finite width quench, when the time dependence is determined by the universal 
function following from eq.\eqref{introgeneral}. 
Generalizations of the logarithmic growth in the regimes (\ref{regime1}) and (\ref{regime2}) 
to  CFTs held at finite temperature have been discussed  in \cite{Caputa:2014eta,Caputa:2015waa} and verified in the holographic dual setup \cite{Caputa:2014vaa,Asplund:2014coa}  of a massive infalling particle  falling in the BTZ black hole background.  For operators with $\Delta_{\cal O}\sim c \gg 1$ leading to the regime (\ref{regime2}) it becomes necessary  to incorporate the backreaction 
on the geometry due to the massive particle \cite{Asplund:2014coa}. Obtaining generalisations of these holographic, large-$c$ limits for CFTs with non-zero higher spin chemical potential (going beyond the small $\mu$ limits)  is one of various follow-up directions we hope to address in future work.
}
\end{itemize}

The organization of this paper is as follows: 
In the following section, we will review in some detail the setup for the local quench and the correlators  involved in 
evaluating the R\'{e}nyi /entanglement entropy of excited states using the the replica trick. 
In section 3 we present the derivation  of the $\epsilon^2$ contribution to the 
R\'{e}nyi/entanglement entropy in general, and subsequently compare it with the exact calculation of the  
second R\'{e}nyi entropy of the excited  state in minimal models and the free fermion CFT. 
We also obtain the $\epsilon^2$ correction for  the entanglement entropy using the 
holographic description for the quenched state as  a massive infalling particle in the BTZ black hole geometry. 
In section 4 we turn on higher spin chemical potentials $\mu$ in CFTs with ${\cal W}$-symmetry and obtain  general results on the entanglement entropy following local quenches.  
We explicitly evalute the correction at $O(\mu^2 \epsilon^2) $ for the spin-three case and show that the time dependence
is universal. The general arguments are  supported by explicit calculations in the 
free fermion CFT. 
Section 5 contains our conclusions and discussions. In the Appendix, we present detailed derivations of higher spin current correlators in the replica geometry by applying the uniformization map. We also make explicit, the results of all correlator computations at order $\mu^2$, and subsequent tedious integrations that yield the exact time dependence of the leading finite width corrections in the presence of spin-three chemical potential.

\section{Finite width local quenches: general features} \label{setup}

\begin{figure}
\center
\includegraphics[width=2.7in]{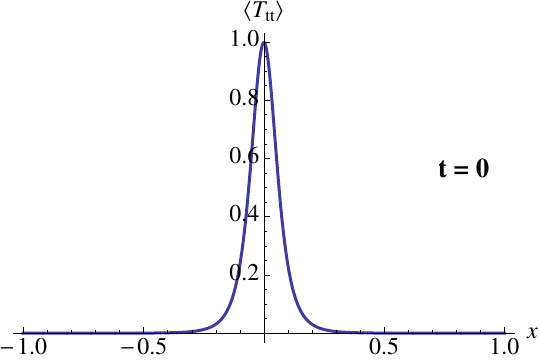}\hspace{0.2in}
\includegraphics[width=2.7in]{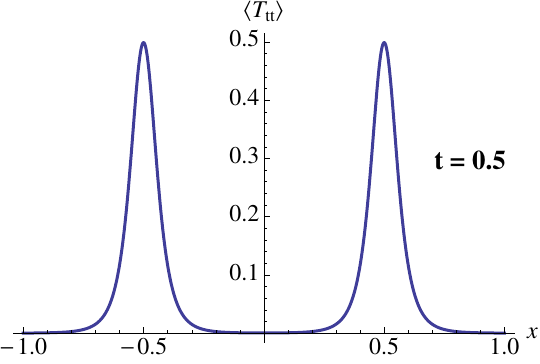}
\caption{\small{A lump of energy density of width $\sim \epsilon =0.1$ and height normalized to unity at $t=0$ splits into left- and right-moving pulses moving along the light-cone, shown centred at $x=\pm0.5$ at $t=0.5$.}}
\label{fig:pulses}
\end{figure}

In this section we review the setup in 2d CFT for the study of local quenches, and also summarise  general arguments leading to the universal features that we elaborate on in this paper. It will also serve to introduce our notations and conventions. 
Consider a 2d CFT at finite temperature $\beta^{-1}$, which is excited by the action of a local operator  at the (spatial) origin at time $t=0$. 
The resulting perturbation will  propagate along the light cone  under time evolution. 
We are interested in the effect of this perturbation on the entanglement entropy of the spatial interval 
$A$  between the points $(l_1,l_2)$.  In thermal equilibrium, the entanglement entropy of the interval  is \cite{Calabrese:2004eu}, 
\begin{equation}
  S_A\, =\, \frac{c}{3}\, \ln
 \left[ \frac{\beta}{\pi\, \Lambda_{\rm UV} }\,  \sinh\tfrac{\pi}{\beta} ( l_2 - l_1)\right] \,.
\end{equation}
By causality we expect this to change when the excitation reaches the edge of the interval, i.e. when $t\sim l_1$. 
The state with a perfectly localized excitation at the origin is described by the (time evolved) density matrix
\begin{equation}
\hat{\rho} \, =\, \mathcal{N}   e^{-iHt}\,\mathcal{O}(0) \,\rho_\beta\,\mathcal{O}^\dagger(0)\, e^{iHt}\,,
\end{equation}
where $\rho_\beta = e^{-\beta H}$ is the thermal density matrix, and the normalisation ${\cal N}$ chosen so that ${\rm Tr\, \hat \rho}=1$. Computation of correlators or expectation values in this state requires  regularisation of the operator product of ${\cal O}$ and ${\cal O}^\dagger$ at the same point. This can be achieved by separating the two operators infinitesimally in {\it imaginary time} as in eq.\eqref{initstate}, so that
\begin{eqnarray}\label{densmat}
\hat{\rho}_\epsilon \,& =&\, 
\mathcal{N} e^{-iHt}\, \mathcal{O}(x_1,\, \bar x_1 )\, \rho_\beta 
\,\mathcal{O}^\dagger(x_4,\,\bar x_4 )\,e^{iHt}\,,
\end{eqnarray}
where the position coordinates of the operators are given in terms of $\epsilon$\,: 
\begin{equation}
x_1 = -i \epsilon, \qquad \bar x_1 = +i \epsilon, \qquad x_4 = +i \epsilon, \qquad \bar x_4 = -i\epsilon\,.
\end{equation}
We label the coordinates of the operators in Lorentzian signature as
\begin{equation}
(z,\,\bar z)\,\equiv\,\left(x -t, \,x +t\right)\,,
\end{equation}
where $x$ and $t$ denote the spatial position and time, respectively.  These are naturally continued to holomorphic and anti-holomorphic coordinates in Euclidean signature. 
Taking the operator ${\cal O}$ to be a conformal primary of weight $(h,\bar h)$ with $h=\bar h = \frac{\Delta_{\cal O}}{2}$, we can see that $\epsilon$ parametrises the width of the excitation by 
evaluating the expectation value of the  energy density in the excited state:
\begin{eqnarray}
\langle T_{tt}\rangle_\epsilon\,\equiv\, {\rm Tr}\, \left( \hat \rho_\epsilon\, T_{tt}\right)\,     =\,  
 \frac{ \langle {\cal O}^\dagger( x_4, \bar x_4 )\, T_{tt} ( x - t, x+ t) \,{\cal O} ( x_1, \bar x_1 )  \rangle_{\beta} }
 {  \langle {\cal O }^\dagger ( x_4, \bar x_4)\, {\cal O} ( x_1, \bar x_1)  \rangle_{\beta}  } \,.
\end{eqnarray}
Since the  expectation values on the right hand side are evaluated in the thermal state, we can compute the relevant correlator in Euclidean signature by utilising the conformal (exponential) map from the plane to the cylinder. Taking into account the fact that the stress tensor acquires a one-point function in the thermal state, and that ${\cal O}$ is primary, we obtain 
\begin{eqnarray}
 && \langle T_{tt}\rangle_\epsilon\,  =\, \frac{\pi^2 c}{ 3\beta^2}\,+\,\frac{4\pi^2 \Delta_{\cal O} }{\beta^2}
  \sin^2 \left( \frac{2\pi \epsilon}{\beta} \right) 
  \times   \\ \nonumber
 &&\qquad  \times\left[ \left( \cosh \frac{2\pi }{\beta} ( x -t)  - \cos \frac{2\pi \epsilon}{\beta} \right)^{-2}\,  +\, 
  \left( \cosh \frac{2\pi }{\beta} ( x +t)  - \cos \frac{2\pi \epsilon}{\beta} \right)^{-2} \right]\,.
\end{eqnarray}
The time dependence of this profile represents two lumps of energy density (left- and right-moving) each of width $\sim \epsilon$ travelling on the light cone (see figure \ref{fig:pulses}).  

\subsection{Universal features}
For a finite width excitation, the time evolution of physical quantities such as the R\'enyi and entanglement entropies in the quenched state will, in general, depend on the details of the CFT in question. Nevertheless, it is possible to pin-point certain universal contributions to these. We already know that the contributions at order $\epsilon^0$ are universal, given by eq.\eqref{jump}. To understand what happens at the next nontrivial order in $\epsilon$, we note that the OPE of any primary field ${\cal O}$ with itself, contains  a term proportional to the stress tensor that is generic and universal for all CFTs:
\bea
\mathcal{O}^\dagger(x_4,\bar{x}_4)\mathcal{O}(x_1,\bar{x}_1) \,\sim\,&&
 |x_4-x_1|^{-2\Delta_{\cal O}}\times\label{ope1p}\\\nonumber 
&&\left[1\,+  \frac{\Delta_{\cal O}}{c} \left(( x_4 - x_ 1)^2\, T(x_1)\,+
\,( \bar x_4 - \bar x_1)^2\, \overline T(\bar{x}_1)\right) +  \cdots \right]\,. \nonumber
\eea
 Given that $x_4-x_1 = 2i\epsilon$, in the limit of small width, the leading terms in this OPE will determine the resulting time evolution of physical observables in the quenched state. The normalisation and the appearance of the stress tensor in the OPE can both be verified by computing the three-point function (in vacuum, say) $\langle T(w) {\cal O}^\dagger(x_4){\cal O}(x_1) \rangle$ using the standard stress tensor Ward identity and taking its limit when the points $x_1$ and $x_4$ approach each other. 
 
 An important consequence of this is that at the order $\epsilon^2$, which controls the first non-trivial correction in the width of the local quench, expectation values of physical observables in the quenched state will be determined purely by corresponding thermal expectation values with one insertion of the stress tensor at the origin. In particular, for the R\'enyi entropies of entanglement of a single interval (with the CFT on the infinite spatial line), which are computed by the partition function on an  $n$-sheeted  Riemann surface, the order $\epsilon^2$ correction will be determined by the expectation value of the stress tensor on this Riemann surface and therefore can be shown to be universal, as we explain below.
 
\subsection{R\'enyi entropies}
\label{sec:basics}
 To compute the  R\'enyi and entanglement entropies, we make use of the replica trick \cite{Calabrese:2004eu, Holzhey:1994we}, which involves obtaining the reduced density matrix $\hat \rho_{A,\epsilon}$\,, by performing a partial trace over the complement of 
the  spatial interval $A\,\equiv\, (l_1, l_2)$. In the Euclidean path integral formulation,  we need to perform the 
path integral over the thermal cylinder  with a branch cut along the interval $A$ accompanied by  
two insertions of the operator ${\cal O}$ at $(x_1,\bar x_1)$ and $(x_4,\bar x_4)$. 
The replica trick on the quenched state is implemented by performing the trace over $n$ powers of the reduced density matrix which, in  path integral language, yields a  $2n$-point correlation function of the operator ${\cal O}$   on an $n$-fold 
cover of the cylinder branched over the 
interval $A$. 
We write this as 
\begin{eqnarray} \label{n_densitymatrix}
\mathrm{Tr} \, \left(\hat \rho_{A,\epsilon}\right)^n \,=\, 
\mathcal{N} \,\,\langle\,\prod_{j=1}^n \,\mathcal{O}^\dagger(x_4^{(j)},\,\bar{x}_4^{(j)})\,\, \mathcal{O}(x_1^{(j)},\,\bar{x}_1^{(j)})
\,\rangle _{\beta, n}\,.
\end{eqnarray}
Each copy of the thermal cylinder is accompanied by two insertions of the operator ${\cal O}$. The points of insertion of the pair of operators ${\cal O}$ and ${\cal O}^\dagger$ in each copy are given as
\begin{eqnarray}\label{cylpts}
&& x_1^{(j)}  \,=\, -i\epsilon + i (j-1) \beta\,, \qquad x_4^{(j)}\, =\,  i\epsilon + i (j-1)\beta\,,\\\nonumber\\\nonumber
&& \bar x_1^{(j)}\,  =\, i\epsilon - i ( j-1) \beta\,, \qquad\,\,\,\, \bar x_4^{(j)}\,=\,  -i\epsilon- i (j-1)\beta\,, 
\end{eqnarray}
where $j=1,2,\ldots n$. In addition to this, the (Lorentzian signature) lightcone coordinates of the locations of the end-points of the interval $A$ (corresponding to the Euclidean branch points) are,
\be\label{lightcone}
 (x_2, \,\bar x_2)\,\equiv\,( l_1 -t,\, l_1+t)\,,\qquad 
  (x_3,\,\bar x_3)\, =\, (l_2-t,\,l_2+t)\,.
 \ee
Now, the change in the $n$-th R\'enyi entropy of the interval, due to the local quench, is given in terms of the $2n$-point correlator (\ref{n_densitymatrix}):
\begin{align} \label{RE_n}
\Delta S_A^{(n)} \,=\, \frac{1}{1-n} \,\ln\left[ \frac{\mathrm{Tr}\, \left(\hat\rho_{A,\epsilon}\right)^{\,n}}{(\mathrm{Tr} \,\hat\rho_{\epsilon})^n} \right]\,.
\end{align}
\begin{figure}
\centering
\includegraphics[width=3.5in]{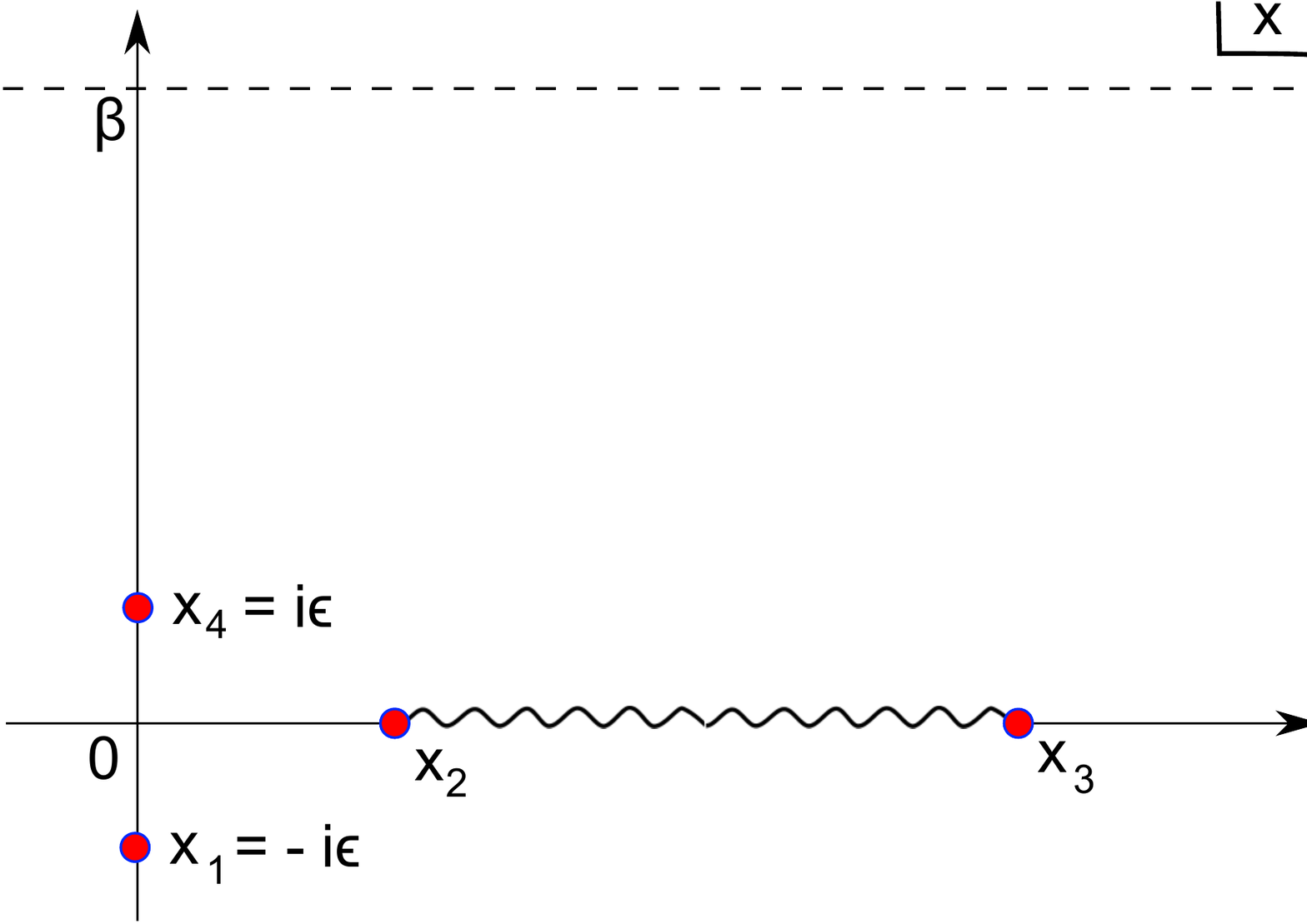}
\caption{\small{A strip of width $\beta$ on the $x$-plane, representing the thermal cylinder ${\mathbb R}\times S^{1}_\beta$. The operators ${ {\cal O}^\dagger}$ and ${\cal O}$ are inserted at $x_4$ and $x_1$ respectively. The end-points of the interval $A$ are at $x_2$ and $x_3$ whose values are given by (Lorentzian) lightcone coordinates at any given real time $t$.
}}
\label{cutpoints}
\end{figure}
Before discussing the detailed evaluation of this correlator, let us note that for small $\epsilon$, the pair of insertions of ${\cal O}$ and ${\cal O}^\dagger$ in a given sheet approach each other (figure \ref{cutpoints}), whilst operator insertions in distinct sheets remain well separated, as long as the excitations created by ${\cal O}$ remain outside the interval $A$ i.e. for $t < l_1$ and $t> l_2$.  In this regime, whilst the excitation remains outside the interval $A$, it is clear that the $2n$-point correlator is determined by the OPE (having taken pairs of operators on a given sheet close to each other)\footnote{Similar OPEs in the context of entanglement entropies for CFTs in excited states have been encountered and exploited in \cite{Giusto:2014aba} where the OPEs in question arise in the short interval limit.}
\bea
\prod_{j=1}^n\,\mathcal{O}^\dagger(x_4^{(j)},\,\bar{x}_4^{(j)})\,\, \mathcal{O}(x_1^{(j)},\,\bar{x}_1^{(j)})&&\,\sim\,\label{2nptearly}\\\nonumber
\frac{1}{(2\epsilon)^{2n\Delta_{\cal O}}}&&\left[1\,-\,\frac{4\epsilon^2\,\Delta_{\cal O}}{c}\sum_{j=1}^n\left\{T\left(x_1^{(j)}\right)\,+\,\bar{T}\left(\bar x_1^{(j)}\right)\right\}\right]\,.
\eea
In order to compute the correlator we only need the one-point function of the stress tensor in each sheet. This is most easily evaluated using the uniformization map from the branched Riemann surface to the complex $w$-plane wherein the stress tensor has vanishing expectation value. Then the one-point function is given by the Schwarzian derivative for this map.
When the excitations enter the interval $l_1 < t < l_2$, the uniformization map shows that the pair of operator insertions on a given sheet lie on either side of the branch cut along the entangling interval. In this situation the pair of insertions in a given sheet remain well separated. Instead, each of them approaches an image insertion from an adjacent Riemann sheet (a neighbouring wedge in the $w$-plane),  thus changing the leading contribution to the OPE. We will examine this carefully below, since it is precisely this phenomenon which is responsible for the jump in $\Delta S_A^{(n)}$ in the zero width quench.

\subsection{The uniformisation map}\label{sec:uniformmap}
\begin{figure}
\centering
\includegraphics[width=3.5in]{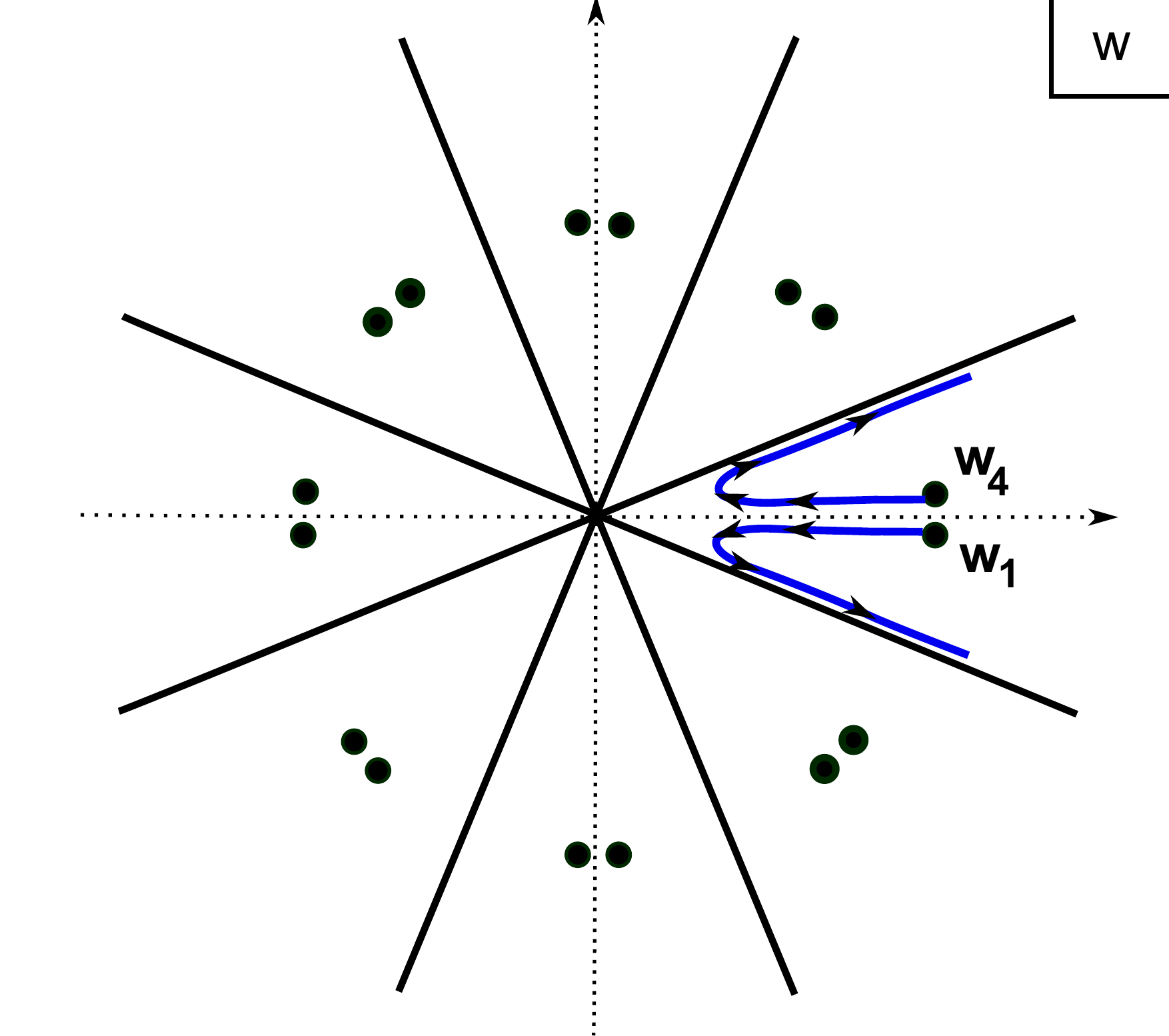}
\caption{\small{The uniformization map takes the branched $n$-sheeted cover of the cylinder to the $w$-plane. Each sheet maps to a wedge with opening angle $2\pi/n$. Shown are the locations of pairs of operator insertions in the fundamental wedge, $w_1\equiv w(x_1^{(1)})$ and $w_4\equiv w(x_4^{(1)})$, and their images. The relative locations of the operator insertions/excitations as a function of time are indicated in blue, in the fundamental domain.}}
\label{wofx}
\end{figure}
As noted above, the  $2n$-point function is readily evaluated by using
the uniformisation  map from the $n$-sheeted branched cover of the thermal cylinder to the complex $w$-plane. 
Under this map, each sheet of the branched Riemann surface  gets mapped to a wedge in the complex $w$-plane with opening angle ${2\pi}/{n}\,$ (figure \ref{wofx}): 
 \bea \label{uniform_map}
&&w(x)\, =\, e^{\frac{\pi(x_2-x_3)}{n \beta}} \left[\frac{\sinh\frac{\pi}{\beta}(x-x_2)}
{\sinh\frac{\pi}{\beta}(x-x_3)}\right]^{\frac{1}{n}}\,, \\\nonumber\\\nonumber
&& \bar w(\bar{x})\,=\, e^{\frac{\pi(\bar x_2- \bar x_3)}{n \beta}} \left[\frac{\sinh\frac{\pi}{\beta}(\bar x-\bar{x}_2)}{\sinh\frac{\pi}{\beta}(\bar x- \bar{x}_3)}\right]^{\frac{1}{n}}.
\eea
The map has branch-points at $(x_2, \,\bar {x}_2)$ and $(x_3, \,\bar x_3)$ which are given by the lightcone coordinates \eqref{lightcone}.
These  branch points, which are the end-points of the interval $A$,  are mapped to $w=0$ and $w=\infty$, respectively. For a natural choice of phases, points immediately above (below) the branch cut along the interval $A$ correspond to the  semi-infinite segment ${\rm arg}(w)= -(+){\pi}/{n}$ joining $w=0$ and $w=\infty$. The $2n$-point function on the branched Riemann surface can be transformed to the plane using the above coordinate transformation so that:
\begin{align}\label{2npt}
 &\left\langle \prod_{j=1}^n\mathcal{O}^\dagger(x_4^{(j)},\bar x_4^{(j)}) \,\mathcal{O}(x_1^{(j)},\bar x_1^{(j)})\right\rangle_n\,=\,  \\ \nonumber
&
\hspace{1.8in}\left\langle \prod_{j=1}^{n} \left|w^{(j)\prime}(x_1)\,w^{(j)\prime}(x_4)\right| \mathcal{O}^\dagger(w_4^{(j)},\bar w_4^{(j)}) \,\mathcal{O}(w_1^{(j)},\bar w_1^{(j)})\right\rangle_w\,,
\end{align}
where $w^{(j)}_p \equiv w(x^{(j)}_p)$ and  a primed coordinate denotes a  derivative with respect to the argument. 
The $j$-th sheet of the branched cylinder is mapped to the $j$-th wedge of  the uniformised plane, and  the locations of the images of the operator insertions on different sheets are related by rotations 
on the complex $w$-plane:
\begin{align}\label{relslice}
w_p^{(j)}\, =\, e^{i\frac{2\pi }{n}(j-1)}\, w_p^{(1)}\,, \qquad 
\bar w_p^{(j)}\,  =\, e^{-i\frac{2\pi }{n}(j-1)}\, \bar w_p^{(1)}\,, 
\qquad 
p\,=\,1,4\,,
\end{align}
with $j=1,2,\ldots n$. The difference between the two temporal regimes $t < l_{1}$ (or $t> l_{2}$) and $l_1 < t < l_2$ is clear upon examining the phases of the locations of the operator insertions in the ``fundamental'' or $j=1$ wedge, in the limit of vanishing $\epsilon$. For a natural choice of phases in the complex $x$-plane, it follows immediately that 
\bea
&&\lim_{\epsilon\to 0}\,\,{\rm arg}\left(w_1^{(1)}\right)\,=\,0\,,\qquad\qquad \lim_{\epsilon\to 0}\,\,{\rm arg}\left(w_4^{(1)}\right)\,=\,0\,,\qquad\qquad \quad{t<l_{1,2}}\,,
\\\nonumber\\\nonumber
&&\lim_{\epsilon\to 0}\,\,{\rm arg}\left(w_1^{(1)}\right)\,=\,\frac{\pi}{n}\,,\qquad
\qquad \lim_{\epsilon\to 0}\,\,{\rm arg}\left(w_4^{(1)}\right)\,=\,-\frac{\pi}{n}\,,\qquad\qquad l_1 < t < l_2\,.
\eea
This change in the relative locations of the operator insertions is depicted in figure \ref{wofx} for finite width $\epsilon$. Importantly, it shows that at early and late times ($t< l_{1,2}$ and $t > l_{1,2}$), the pair of points 
$(w_1^{(j)}, w_4^{(j)})$ in a given wedge remain close to each other. For intermediate times ($l_1<t<l_2$) they separate and approach image  insertions in neighbouring wedges (or adjacent sheets on the Riemann surface). The same analysis for the anti-holomorphic coordinates shows that there is no change in their relative positions over time.

The phase changes above are encapsulated in corresponding phase changes in conformal cross-ratios on which correlation functions  depend nontrivially. We define the cross-ratio,
\be\label{defcrossratio}
z\,\equiv\,\frac{ \sinh\frac{\pi}{\beta} ( x_1  -x_2)\, \sinh\frac{\pi}{\beta} ( x_4  -x_3) }
{  \sinh\frac{\pi}{\beta} ( x_1  -x_3)\, \sinh\frac{\pi}{\beta} ( x_4  -x_2) }\,,
\ee
and similarly $\bar z$, which is obtained from the same expression with the replacement $\{x_p\}\to \{\bar x_p\}$. These are related to positions of the operator insertions in the $w$-plane as:
\be
\frac{w_1}{w_4}\, =\, z^{\frac{1}{n}}\,,\qquad \qquad \qquad
\frac{\bar w_1}{\bar w_4}\, =\, {\bar z}^{\frac{1}{n}}\,.
\ee
In the limit $\epsilon\to 0$, we obtain,
\bea\label{holbran}
&&\lim_{\epsilon \to 0}\,
z^{\frac{1}{n}}\,=\,1\,,\qquad\qquad
\lim_{\epsilon \to 0}\,
\bar z^{\frac{1}{n}}\,=\,1\,,\qquad t<{l_{1,2}}\\\nonumber\\\nonumber
&&\lim_{\epsilon \to 0}\,
z^{\frac{1}{n}}\,=\,e^{2\pi i/n}\,,\qquad
\lim_{\epsilon \to 0}\,
\bar z^{\frac{1}{n}}\,=\,1\,,\qquad l_1<t<{l_{2}}\,.
\eea
This means that the holomorphic coordinate of the operator insertion in the $j$-th sheet is at a location which is adjacent to the $(j+1)$-th sheet while the anti-holomorphic coordinate remains unchanged for all times. 
The positions (\ref{relslice}) in the $w$-plane of the operator insertions on different sheets, 
and the phase change argued in (\ref{holbran}), 
together imply that we need to examine the $2n$-point function in the short distance regime when the following points come together pairwise, depending on the two distinct temporal regimes in question:
\bea
&&w_1^{(j)}\,\to\,w_4^{(j)}\,,\qquad\qquad t < l_1\,,\quad t> l_2\,,\label{approach}\\\nonumber\\\nonumber
&&w_1^{(j)}\,\to\,w_4^{(j+1)}\,,\qquad\qquad l_1< t< l_2\,.
\eea
The index $j$ is defined modulo $n$ and $j=1,2,\ldots n$. For anti-holomorphic coordinates the OPEs always involve short distance expansions of operator insertions within a given wedge: $\bar w_1^{(j)}\to \bar w_4^{(j)}$ for all times. 

The evaluation of the OPEs requires the expansion of  the cross-ratios in powers of $\epsilon$. Only the first non-trivial term in this expansion will be of relevance to us and its time dependence will make repeated appearances in our discussion below. Therefore, for convenience, we  denote it as ${\cal S}_{l_1 l_2}(t)$, where  
\begin{eqnarray} \label{limitscrossratio}
&&z  \,=\, 1\,+\,2i \epsilon\, {\cal S}_{l_1 l_2}(t)\,+\,O(\epsilon^2)\,,\qquad \bar z \,=\, 1\,-\, 2i\epsilon\, {\cal S}_{l_1 l_2}(-t)\,+\,O(\epsilon^2)\nonumber \\\nonumber\\
&& {\cal S}_{l_1 l_2}(t)\,=\,\frac{ \pi \sinh \frac{\pi}{\beta}(l_2-l_1)}{\beta\, \sinh \frac{\pi}{\beta}(l_1-t) \,\sinh \frac{\pi}{\beta}(l_2-t) }\,.
\end{eqnarray}
For the holomorphic ratio $z$, the sign of the  term linear in $\epsilon$ 
changes when moving from the regime $t< l_1$ to $l_1< t < l_2$, whilst there is no such sign change for  the corresponding term in $\bar z$. The function ${\cal S}_{l_1 l_2}$ is also closely  related to the Schwarzian of the map $w(x)$ from the branched cover of the cylinder to the complex $w$-plane.

\section{The universal correction at finite width}

We now turn to an explicit evaluation of  the ${O}(\epsilon^2)$ correction to the entanglement entropy, making use of the uniformization map and the short distance OPEs relevant for analysing the two distinct regimes $t< l_{1}$ (or $t> l_2$) and $l_1< t< l_2$. Crucially, for the latter regime we will make use of fusion rules. We establish the CFT result using two different routes. First, we will present  a general conformal block argument for  the {\em second} R\'{e}nyi entropy of the perturbed state which 
is determined by the four-point function of the operator ${\cal O}$, and obtain the universal, leading time  dependence at finite width. Then we will obtain the same correction at order $\epsilon^2$ for all R\'{e}nyi entropies by computing the expectation value of the stress tensor on the branched cover of the thermal cylinder. 
We  will also verify the result for the examples of  minimal model CFTs and the free fermion theory where the exact time dependence of the R\'enyi entropies is known.  Finally, we will also see that the holographic description of the CFT excited state given in terms of an infalling massive particle in the BTZ black hole geometry yields the  same universal time dependence for the leading finite width correction to the entanglement entropy.

\subsection{The conformal block argument}

Let us consider the  second R\'{e}nyi entropy for a quenched CFT. From eq.\eqref{RE_n} we see that  it is determined by the 
normalized four-point function of the operator ${\cal O}$.  We continue to treat ${\cal O}$ as an  operator with $ h = \bar h = \frac{\Delta_{\cal O}}{2}$ \footnote{The analysis can be 
easily generalized to operators with $h\neq \bar h$.}.  We also normalise its  two-point function to unity. 
 Following \cite{He:2014mwa}  we parametrise the four-point function relevant for the second R\'enyi entropy in terms of the variables $(u,\bar u)$ related to the cross-ratios $(z,\bar z)$, defined in eq. \eqref{defcrossratio},
\begin{eqnarray}\label{RE2confbG}
&&  \frac{\left\langle\prod_{j=1}^2 \mathcal{O}(x_1^{(j)},\,\bar{x}_1^{(j)})\,\mathcal{O}^\dagger(x_4^{(j)},\,\bar{x}_4^{(j)}) 
\right\rangle_2 }
{\left\langle  \mathcal{O}(x_1,\,\bar{x}_1)\,\mathcal{O}^\dagger(x_4,\,\bar{x}_4) \right\rangle_1^2} 
\,= \,|u |^{ 4 h} | 1 - u |^{4 h} G_{\cal O } ( u , \bar u ) \,, 
\end{eqnarray}
where 
\begin{eqnarray}\label{uzrel}
 u \,=\, -  \frac{( \sqrt{z} - 1) ^2}{ 4 \sqrt{z} }, \qquad \qquad\bar 
 u\, =\, - \frac{ ( \sqrt{\bar z} - 1) ^2 }{ 4 \sqrt{ \bar z}} \,, 
\end{eqnarray}
and the function $G_{\cal O}$ possesses an expansion in conformal blocks.
To obtain the relation between the variable $u$ as defined in  \cite{He:2014mwa} and the cross-ratio $z$ as defined in eq.\eqref{defcrossratio}, we have used the fact that $w_1^{(2)} \,= \,- w_1^{(1)}$ and $ w_4^{(2)} \,=\, - w_4^{(1)}$. 
For an arbitrary CFT with central charge $c$ the function
$G_{\cal O}(u, \bar u) $ has a decomposition in terms of conformal blocks, schematically, 
\begin{equation} \label{G_confb}
 G_{\cal O } ( u, \bar u ) \,=\, \sum_{b} ( C_{\cal O}^{{\cal O}_b} )^2\,   F_{\cal O } ( {\cal O}_b |u )\, 
 \bar F_{\cal O } ( {\cal O}_b |\bar u )\,, 
\end{equation}
where the sum runs over all the primaries $\{{\cal O}_b\}$ of the theory with weights $(h_b,\,\bar h_b)$. As argued in the previous section, the two temporal regimes of interest to us, $z^{1/2}\,=\,1$ and $ z^{1/2}\,=\, -1$, correspond to $u\,\to\,0$ and $u\to1$ respectively, with $\bar u $ vanishingly small at all times. We therefore need to evaluate the OPEs in these two limits.

Before proceeding we recall a few properties of Virasoro conformal blocks. First, the Virasoro block  can be written as
\begin{eqnarray} \label{confb}
 F_{\cal O }(  {\cal O}_b |u) \,=\, u ^{ h_b - 2 h}  \,\tilde F_{\cal O }(  {\cal O}_b |u)\,,
\end{eqnarray}
with $\tilde F_{\cal O}$ regular near $u=0$.
We will be interested in the vacuum Virasoro block for which 
 ${\cal O}_b$  is the identity $(b=0)$.  For factorization onto the identity, normalisation of the operators can be used to fix $C_{\cal O}^{{\cal O}_0 } = 1$. 
 The expansion for the vacuum block is  determined in 
 terms of hypergeometric functions  and is given by  \cite{Perlmutter:2015iya}
\begin{equation} \label{vacexp}
 \tilde F_{\cal O; \rm{ vac }  }\,=\,   \sum_{q = 0}^\infty \chi_{\rm{vac}; q } ( c, h)\,\, u^q\,\, 
 {}_2F_1( q, q, 2 q; u )\,, 
\end{equation}
where $\chi_{{\rm vac};0}=1$ and $\chi_{{\rm vac};1}=0$.
The first few non-zero coefficients $ \chi_{\rm{vac}; q }$ are listed in \cite{Perlmutter:2015iya}, for example,
\begin{eqnarray}\label{vacexp1}
 \chi_{\rm{vac}; 2} \,=\,  2 \frac{h^2}{c}\,,  \qquad\qquad
 \chi_{\rm{vac}; 4 } \, =\,  \frac{10 h^2 }{c} \frac{ (   h + \frac{1}{5} )^2  }{  ( 5 c + 22) } \,.
\end{eqnarray}
Crucially, for 
generic central charge $c$, the first nontrivial term in the expansion begins at order $u^2$ for small $u$:
\be
\tilde F_{\cal O;{\rm vac}}\,=\,1\,+\,\frac{2h^2}{c}\,u^2\,+\ldots
\ee
Note also that for  generic central charge and fixed conformal dimension $h$ the coefficients vanish when $c\rightarrow \infty$. 
One further important property of conformal blocks that we will need before we proceed 
with our analysis, follows from  the fusion transformation rule:
\begin{equation}\label{fusi}
  F_{\cal O }(  {\cal O}_b |u) \, =\, \sum_c F_{bc} [ {\cal O} ] \,F_{\cal O} ( {\cal O}_b|1-u)\,.
\end{equation}
The constants $ F_{bc} [ \cal O]$ form the fusion matrix. 
This transformation allows to expand the conformal blocks around $u=1$ 
using the expansion around $u=0$. 

We also assume that 
 the lowest lying primary has dimension greater than one\footnote{The example of 
the free fermion theory violates this condition since this theory contains a $U(1)$ current 
of dimension $1$.}. If this is not the case, the leading finite width corrections at order $\epsilon^2$ can receive (non-universal) contributions from such states.
Let us now examine the regime $t<l_{1}$. From the discussion around 
eq.\eqref{holbran}, we learnt that we need to expand  $(z, \bar z)$ around $(1, 1)$ to describe the early time behaviour. This in turn implies (using eq.\eqref{uzrel}) that we need the expansion of the conformal blocks around 
$(u, \bar u) = (0, 0)$.  To this end we examine the conformal cross-ratios $(z,\bar z)$ for small $\epsilon$ and obtain
\begin{eqnarray} \label{uexp}
 u\, =\,  \frac{\epsilon^2}{4}\,{\cal S}_{l_1l_2}(t)^2 \,+\, O(\epsilon^4) 
\,, \qquad\qquad
  \bar u\, =\, \frac{\epsilon^2}{4}\,{\cal S}_{l_1l_2}(-t)^2 \,+\, O(\epsilon^4)\,.  
\end{eqnarray}
The time dependent function ${\cal S}_{l_1l_2}$ was defined in eq.\eqref{defcrossratio}.
Substituting this in  (\ref{confb}) 
and using  the expansion for the vacuum block (\ref{vacexp})  we obtain
the following expression for the normalized  four-point function in the early time regime $t < l_{1,2}$ (and also the late time regime $t>l_{1,2}$):
\begin{eqnarray} 
  \frac{\left\langle\prod_{j=1}^2 \mathcal{O}(x_1^{(j)},\,\bar{x}_1^{(j)})\,\mathcal{O}^\dagger(x_4^{(j)},\,\bar{x}_4^{(j)})
\right\rangle_2 }
{\left\langle  \mathcal{O}(x_1,\,\bar{x}_1)\,\mathcal{O}^\dagger(x_4,\,\bar{x}_4) \right\rangle_1^2}\,=\,1\,-\,\Delta_{\cal O}\,(u\,+\,\bar u)\,+\,O( \epsilon^4)\,.
  \end{eqnarray}
  Here, both $u$ and $\bar u$ are given by eq.\eqref{uexp} at order $\epsilon^2$. It is important that the first sub-leading term in the expansion of the vacuum block given in 
  (\ref{vacexp1}) starts at order $u^2 \sim \epsilon^4$. 
Therefore, we unambiguously conclude that  the correction to the second R\'{e}nyi entropy for $t<l_{1,2}$  is given by the universal time dependence:
\begin{eqnarray} \label{RE_confb0}
  \Delta S^{(2)}_A \,=\,\Delta_{\cal O}\frac{\epsilon^2}{ 4}\left[{\cal S}_{l_1l_2}(t)^2\,+\,{\cal S}_{l_1l_2}(-t)^2  \right]\,+\,{O}(\epsilon^4).
 \end{eqnarray}
 where 
 \begin{eqnarray}  
  {\cal S}_{l_1l_2}(t)^2\,=\,&&\frac{ \pi^2}{\beta^2}
  \frac{\sinh^2 \frac{\pi}{\beta} (l_2 - l_1)  }{
  \sinh^2  \frac{\pi}{\beta} (l_1 - t )\,\, 
  \sinh^2  \frac{\pi}{\beta} (l_2 - t ) }\,. 
  \end{eqnarray}
  An important caveat to this conclusion is that the  expansion in the width of the perturbation is only valid for times
  $|t - l_1| \gg \epsilon$.  
  For these times the growth in the second R\'{e}nyi entropy is universal 
  and independent of the CFT. Only the prefactor $\Delta_{\cal O}$  depends on the perturbing operator ${\cal O}$. 
  For times  $t$ approaching $l_1$, higher order terms in powers of $\epsilon$ will need to be included and these will, in general, depend on the details of the theory.  All conclusions above  also apply for late times 
  $t>l_{1,2}$. 
  
  Finally, let us turn to the behaviour for intermediate times $l_1 <t <l_2$. Now, the cross-ratios admit an expansion 
  around $(\sqrt{z}, \sqrt{\bar z} ) \,=\, ( -1, 1) $ or $(u,\bar u)\,=\,(1,0)$:
  \begin{eqnarray} \label{ulimits}
  u  \,=\, 1\, -\, \frac{\epsilon^2}{4}\, {\cal S}_{l_1l_2}(t)^2\,+\, O(\epsilon^4)\,,\qquad
  \bar u \,=\, \frac{\epsilon^2 }{4 }\, {\cal S}_{l_1l_2}(-t)^2\,+\, O(\epsilon^4)\,.
  \end{eqnarray}
  Substituting this expansion and using the 
  fusion relation (\ref{fusi}) for the conformal blocks we find that
the change 
in the second R\'{e}nyi entropy  in the intermediate time regime $l_1<t<l_2$ is given by 
\begin{eqnarray} \label{fin2re}
 && \Delta S^{(2)}_A \,=\, -\, \ln F_{00}[{\cal O}] \,+\, \Delta_{\cal O}(1\,-\,u\,+\bar u)\,+\, O( \epsilon^4)\,,
    \end{eqnarray}
yielding exactly the same (universal) time dependence for the term at $O(\epsilon^2)$.
    Here again we have assumed that the lowest lying primary in the theory has
conformal dimension greater than one. 
Therefore the leading correction at order $\epsilon^2$, due to the width of the perturbation, is 
universal. In fact, the time dependence of this correction is the same for all temporal regimes, except for $|t-l_{1,2}|\sim O(\epsilon)$ when higher order corrections in the width must necessarily be included and the resulting behaviour will depend on detailed properties of the CFT in question. 

Another interesting point is that the height of the jump,  $-\ln F_{00}$, originally seen at order $\epsilon^0$, is affected by the finite width effect at ${O}(\epsilon^2)$. This is most clearly visible in the limit of large interval and late times: $l_2\to \infty$, followed by $t\to\infty$. In this limit, we find,
\be
\Delta S_A^{(2)}\left.\right|_{l_2\to\infty,\,t\to\infty}\,\to\,-\ln F_{00}[{\cal O}]\,+\,\epsilon^2\,\Delta_{\cal O}\frac{\pi^2}{\beta^2}\,.
\ee
The additional jump is a purely thermal effect at this order in the width.  A general physical interpretation for it will be given below.

\subsection{The OPE  argument} \label{sec-OPE}

\paragraph{Early times: $t<l_{1}$:} In section \ref{sec:basics}, we argued that the leading finite width correction is controlled by the expectation value of the stress tensor on the branched Riemann surface $\Sigma_n$. 
Specifically, we have established that for times $t<l_{1}$ (and $t> l_2$), the two insertion points on each sheet of $\Sigma_n$ (or each wedge in the uniformized $w$-plane) remain close, so that the relevant $2n$-point function is determined by expectation value of the stress tensor in each sheet (see eq.\eqref{2nptearly}):
\bea
\left\langle\prod_{j=1}^n\,\mathcal{O}^\dagger(x_4^{(j)},\,\bar{x}_4^{(j)})\,\, \mathcal{O}(x_1^{(j)},\,\bar{x}_1^{(j)})\right\rangle_n&&\,\sim\,\label{2nptearly2}\\\nonumber
\frac{1}{(2\epsilon)^{2n\Delta_{\cal O}}}&&\left[1\,-\,\frac{4\epsilon^2\,\Delta_{\cal O}}{c}\sum_{j=1}^n\left\{\left\langle T(x_1^{(j)})\right\rangle_n\,+\,\left\langle\bar{T}(\bar x_1^{(j)})\right\rangle_n\right\}\right]\,.
\eea
Since the one-point function of the stress tensor must vanish on the uniformized $w$-plane, the expectation values required above are determined by the Schwarzian in the transformation law for the stress tensor:
\be
T(x)\,=\,w'(x)^2\,T(w)\,+\,\frac{c}{12}\,\{w,\,x\}\,,
\ee
with $w(x)$ given by \eqref{uniform_map}. The Schwarzian derivative  $\{w,\,x\}\,=\,
(w'''\,w'\,-\,\frac{3}{2}w^{''2})/w^{\prime \,2}$ is easily evaluated and we find that\footnote{This expression can also be deduced via  application of the conformal Ward identity on the cylinder for the three-point correlator of the stress tensor with the branch-point twist fields ${\cal T}_n$ with  $(h,\bar h)\,=\,\left({\Delta_n}/{2},\, {\Delta_n}/{2}\right)$ where $\Delta_n\,=\,
\frac{c}{12n}(n^2-1)$. Then the result \eqref{schw} is reproduced (after dividing by $n$ for each sheet) by
\be
\langle T(x)\rangle_n\,=\,\left(\sum_{a=2,3}\frac{\pi^2}{\beta^2}\frac{\Delta_n/2}{\sinh^2\frac{\pi}{\beta}(x-x_a)}\,+\,\sum_{a=2,3}\frac{\pi}{\beta}\coth\tfrac{\pi}{\beta}(x-x_a)\frac{\partial}{\partial x_a}\ln\langle {\cal T}_n(x_2){\cal T}_{-n}(x_3)\rangle\right)\,+\,\langle T\rangle_{\beta}\,.\nonumber
\ee
}
\be
\frac{c}{12}\,\{w,\,x\}\,=\,c\frac{(n^2-1)}{24 \,n^2}\,\,\left[{\cal S}_{l_1l_2}(t+x)\right]^2
\,\,-\,\frac{c\,\pi^2}{6\beta^2}\,.\label{schw}
\ee
The antiholomorphic case is given by the same expression, with the replacements $x\to\bar x$ and $t\to -t$ (equivalent to ${x}_{2,3}\to \bar x_{2,3}$). The first term on the right hand side of eq.\eqref{schw} can be viewed as the contribution originating from branch points on the Riemann surface, whilst 
the second term is the one-point function of the stress tensor on the thermal cylinder  (the case $n=1$),
\be
\langle T\rangle_\beta\,=\,-\frac{c\,\pi^2}{6\beta^2}\,.\label{tbeta}
\ee
This gets subtracted out when the  change in the R\'enyi entropy $\Delta S_A^{(n)}$ is calculated by normalising the $2n$-point function with respect to the two-point function on the unbranched cylinder: 
\begin{eqnarray}\label{2pt_cyl}
\langle \mathcal{O}^\dagger(x_4^{(1)},\,\bar x_4^{(1)})\,\mathcal{O}(x_1^{(1)},\,\bar x_1^{(1)}) \rangle _1 
&&\,=\,  \left| \frac{\beta}{\pi}\,\sinh \tfrac{\pi}{\beta}(x_1-x_4)\right|^{-2\Delta_{\cal O}} , \\ \nonumber\\\nonumber
&& \xrightarrow{\epsilon \to 0} \frac{1}{(2\epsilon)^{2\Delta_{\cal O}}} 
\left(1\,+\,\epsilon^2\Delta_{\cal O}\,\frac{4\pi^2}{3\beta^2} \, +\, O(\epsilon^4) \right)\,,
\end{eqnarray} 
where $x_{1,4}\,=\,\mp i\epsilon$. The order $\epsilon^2$ term within parentheses  on the right hand side can be traced to the one-point function of the stress tensor on the thermal cylinder, $\sim -8\epsilon^2\Delta_{\cal O}\langle T\rangle_\beta/c$.
On the branched cover of the cylinder, we need the expectation value of the stress tensor at each of the $n$ points: $x_{1}^{(j)}\,=\,\mp i\epsilon\,+\, i(j-1)\beta$, in the limit of vanishing $\epsilon$, which yields an identical value for each such point:
\be
\lim_{\epsilon\to 0}\,
\left\langle T(x_1^{(j)})\right\rangle_n\,=\,\frac{c}{12}\,\{w, x\}\left.\right|_{x=0}\,.
\ee
Note we are always assuming that $|t-l_{1,2}|\gg \epsilon$.
Taking the ratio of the $2n$-point function and the normalization factor,  we obtain the R\'{e}nyi entropy \eqref{RE_n} of the excited state for $ t<l_1<l_2$, 
\begin{eqnarray}
&& \Delta S^{(n)}_A\,=\,\epsilon^2\Delta_{\cal O}\, \frac{1+n }{6 n } \left[{\cal S}_{l_1l_2}(t)^2\,+\,{\cal S}_{l_1l_2}(-t)^2\right]\,+\,{O}(\epsilon^4)\,.
\end{eqnarray}
As a simple  check, note that for $n=2$ it reduces to the expression \eqref{RE_confb0}  for second R\'enyi entropy 
evaluated using the conformal block argument. 
The same result also holds in the  regime $t>l_{1,2}$.
\paragraph{Intermediate times $l_1<t<l_2$:} Now we move to the  domain $l_1 <t<l_2$  wherein the holomorphic coordinates of the quenching operators  in adjacent sheets 
approach each other as indicated in eq.\eqref{approach}, so that $ w_1^{(j)} \to w_4^{(j+1)}$ whilst the anti-holomorphic variables behave exactly as before, $\bar w_1^{(j)}  \to \bar w_4^{(j)} $. 
Thus there is no change in the OPE in the anti-holomorphic variables. 
However, new OPEs in the holomorphic coordinates must  be obtained by performing $n-1$ fusion transformations
as  in the previous section for $n=2$ \cite{He:2014mwa}. We are thus led to the following result for the operator product in the uniformized $w$-plane:
\begin{eqnarray}\label{crosschan}
&&\prod_{j=1}^n\mathcal{O}^\dagger(w_4^{(j)},\,\bar w_4^{(j)})\,  \mathcal{O}(w_1^{(j)},\,\bar w_1^{(j)}) 
\sim   F_{00}^{n-1} 
\prod_{j= 1}^{n} ( w _1^{(j)} \,-\, w_4^{(j+1)} )^{- \Delta_{\cal O} }  \,\prod_{j=1}^n ( \bar w_1^{(j)}\,-\,\bar w_4^{(j)})^{-\Delta_{\cal O} }\,\nonumber\\\nonumber\\
&&\times \left[1\, -\, 4 \epsilon^2 \frac{\Delta_{\cal O}}{c}\,
\sum_{j=1}^{n}  \left\{(w'_1{}^{(j)})^2\,T(w_1^{(j)}) \,+\, (\bar w'_1{}^{(j)})^2\,\overline T(\bar w_1^{(j)})\right\} \right]
\end{eqnarray} 
Utilizing this factorization, we evaluate the $2n$-point correlator  on the branched cylinder by performing a  coordinate transformation from the $w$-plane.  The stress tensor has vanishing expectation value on the $w$-plane. Therefore the leading correction at order $\epsilon^2$ arises from carefully evaluating  the transformed correlator (having set $\langle T(w)\rangle\,=\,\langle\overline T(\bar w)\rangle\,=\,0$): 
\begin{eqnarray}
&&\left\langle\prod_{j=1}^n \mathcal{O}^\dagger(x_4^{(j)},\,\bar x_4^{(j)})\,\mathcal{O}(x_1^{(j)},\bar x_1^{(j)})\right\rangle_n\,\sim  
\\
& &  F_{00}^{(n-1)}\, \prod_{j=1}^{n}
 \left[ w_1^{(j)\,\prime}  \,w_4^{(j)\, \prime}   ( w_1^{(j)}\,-\,w_4^{( j+1) } )^{-2} \right]^{\Delta_{\cal O}/2} \,
 \left[\bar w^{(j)\,\prime}_1\,\bar w^{(j)\,\prime}_4\, (\bar w_4^{(j)}\,-\,\bar w_1^{(j)})^{-2}\right]^{ \Delta_{\cal O}/2}\,.   \nonumber
\end{eqnarray}
Recalling that $w_p^{(j)}\,=\,w(x_p^{(j)})$ and that the coordinates $x_p^{(j)}$ are as specified in eq.\eqref{cylpts}, we expand the correlator in powers of $\epsilon$ and find, 
\begin{eqnarray}
& & \langle\prod_{j=1}^n \mathcal{O}^\dagger(x_4^{(j)},\,\bar x_4^{(j)})\,\mathcal{O}(x_1^{(j)},\bar x_1^{(j)})\rangle_n \\
& &\approx F_{00}^{(n-1)} 
 \left(\frac{1}{2 \epsilon} \right)^{2n \Delta_{\cal O}} \left[1\,-\, \frac{1}{3} \epsilon^2 n \,
\Delta_{\cal O}\, \left(\{w,x\}\left.\right|_{x=0} \,+\{\overline{w},\bar{x}\}\left.\right|_{\bar x =0}\right) 
\,+\,O(\epsilon^4) \right]\,.
\nonumber
\end{eqnarray}
The Schwarzian $\{w,\,x\}$ given by \eqref{schw}, and evaluated at $x=0$, yields exactly the same time dependence at order $\epsilon^2$ seen for early times $t< l_{1,2}$. Normalizing the correlator 
with $n$ powers of the two-point function on the unbranched cylinder,
we obtain the change in the R\'{e}nyi entropy \eqref{RE2confbG}  in the time 
interval $l_1<t<l_2$:
\begin{eqnarray}
&&\Delta S^{(n)}_A  \,=\,  - \ln F_{00} \,+\, 
\epsilon^2{\Delta_{\cal O}} \frac{1+n}{6 n }\left[{\cal S}_{l_1l_2}(t)^2\,+\,{\cal S}_{l_1l_2}(-t)^2\right]\,+\,{O}(\epsilon^4)\,.
\end{eqnarray}
The time dependence at order $\epsilon^2$ is identical to that obtained in the regime $t<l_{1,2}$ and matches the  expression in (\ref{fin2re}) for $n=2$.
 \paragraph{The general result:} Assuming a parametric expansion for the non-thermal change in the  single interval R\'enyi entropy of the general form
\be
\Delta S^{(n)}_A\,=\, \sum_{m=0}^\infty\,\epsilon^m\,\Delta S^{(n;\,m)}_A\,,
\ee 
we have seen that while the leading term at order $\epsilon^0$ is a step function in time and determined by the quantum dimension of the operator ${\cal O}$, 
\begin{eqnarray} 
\Delta S_A^{(n;\,0)}\, =\, \begin{cases}
0, \qquad & t<l_1 \qquad\text{or}\qquad \, t>l_2\\
- \ln \left( F_{00}  \right)\,,\qquad & l_1<t<l_2\,,
\end{cases}
\end{eqnarray}
the first nontrivial correction (at second order in $\epsilon$) happens to be universal, 
\begin{eqnarray}\label{finrene}
\Delta S_A^{(n;\,2)}\,&&=
{\Delta_{\cal O}} \frac{1+n}{6 n }\left[{\cal S}_{l_1l_2}(t)^2\,+\,{\cal S}_{l_1l_2}(-t)^2\right]\,.
\end{eqnarray}
The only dependence on the exciting field ${\cal O}$ is through its conformal dimension $\Delta_{\cal O}$ which appears as the overall coefficient of the correction. We can state the result for the first finite width correction at order $\epsilon^2$ in more general terms as:
\be
\boxed{\Delta S_A^{(n;\,2)}\,=\,4\,\frac{\Delta_{\cal O}}{c}\,n\,\frac{\langle T(0)\rangle_n\,-\,\langle T\rangle_\beta}{n-1}\,+\,{(t\to -t)}}\,,\label{general}
\ee
where the antiholomorphic contribution can be obtained from the holomorphic one by the operation $t\to -t$ \footnote{When ${\cal O}$ is  a $(\Delta_{\cal O}/2,\,\bar {\Delta}_{\cal O}/2)$ operator, so that the holomorphic and antiholomorphic weights of  are unequal, we must also make the replacement ${\Delta}_{\cal O}\to\bar\Delta_{\cal O}$}. 
For small widths, we expect this general form  to apply even in the presence of CFT deformations which are within the reach of conformal perturbation theory.  The time dependence of this expression arises from the fact that the lightcone coordinates (holomorphic and anti-holomorphic) of the branch-points are time dependent as in eq.\eqref{lightcone}. 
An interesting consequence of this general formula is that we can use it to predict certain aspects of the change in R\'enyi entropies in the limit of large entangling intervals  and late times. This limit, which should be thought of as taking $l_2 \to \infty$ first, followed by  $t \gg l_1$, is particularly interesting as the function $w(x)$ in eq.\eqref{uniform_map} effectively becomes a map from a cylinder of radius $n\beta$ to the $w$-plane, with $w\,\sim\,\exp(2\pi\,x/n\beta)$. In the same limit, the map $\bar w (\bar x)$ for antiholomorphic coordinates becomes trivial i.e. the identity map. Therefore, for large entangling intervals $A$ and  at late times, we must have,
\bea
\Delta S_A^{(n;\,2)}\left.\right|_{l_2\to\infty,\,t\gg l_1}\,=\,
4\frac{\Delta_{\cal O}}{c}\,n\,\frac{\langle T\rangle_{n\beta}\,-\,\langle T\rangle_\beta}{n-1}\,=\,
{\Delta_{\cal O}}\frac{2(n+1)}{3n}\frac{\pi^2}{\beta^2}\,,
\eea
where we have used the thermal expectation value of the stress tensor \eqref{tbeta}.
The result is reproduced by the corresponding limit of the real time expression \eqref{finrene}. This can also be viewed as a late time modification of  the zeroth order jump in the R\'enyi entropy $\Delta S_A^{(n;\,0)}$ by a purely thermal component at order $\epsilon^2$. Its physical significance is slightly clearer in the $n\to 1$ limit which yields the finite width contribution to the entanglement entropy of the interval $A$:
\be
\Delta S_A^{(n;\,2)}\left.\right|_{n\to 1}\,=\,4\frac{\Delta_{\cal O}}{c}\,\partial_n\langle T(0)\rangle_n \,\left.\right|_{n=1}\,+\,(t\to -t)\,.
\ee
In the limit of large interval and late time, the antiholomorphic contribution is vanishing, and the result can be viewed as the effect of an infinitesimal rescaling of the temperature:
\be
\Delta S_A^{(1;\,2)}\left.\right|_{l_2\to\infty,\, t\gg l_1}\,=\,4\frac{\Delta_{\cal O}}{c}\,\partial_n\langle T\rangle_{n\beta} \,\left.\right|_{n=1}\,=\,4\frac{\Delta_{\cal O}}{c}\,\beta\,\frac{\partial}{\partial\beta}\langle T\rangle_{\beta} \,.
\ee
This leads to another way of viewing the entanglement entropy change for large intervals. The thermal expectation value \eqref{tbeta} of the holomorphic stress tensor is proportional to the energy density as $\langle T\rangle_\beta \,=\,-\pi\, {\varepsilon}$. Using standard thermodynamic identites we conclude that 
\be
\Delta S_A^{(1;\,2)}\left.\right|_{l_2\to\infty,\, t\gg l_1}\,=\,-4\pi\frac{\Delta_{\cal O}}{c}\,\frac{\partial\,s}{\partial\beta} \,,
\ee 
where $s$ is the thermal entropy density of the CFT. Therefore in this limit, the change in the entanglement entropy at order $\epsilon^2$ can be viewed as a variation in the thermal entropy due to an infinitesimal change in the temperature of the interval $A$.
The  complete time dependence of the entanglement entropy, obtained in the  
limit $n\rightarrow 1$ of (\ref{finrene}), will be shown in section \ref{holography-check}  to match
the ${O}(\epsilon^2)$ term in the holographic entanglement entropy following a finite width local quench.
Before this, we will study the second R\'{e}nyi entropy in
two exactly solvable models for which the relevant four-point function can be explicitly evaluated, and then compare our correction at $O(\epsilon^2)$ with the exact result at 
 finite $\epsilon$ for these models.

\subsection{Comparision with examples and exact results}

\subsection*{Example I: Minimal model}
Let us turn our attention to a specific example, namely a  $(p,p')$ minimal model CFT with $p>p'$.  Four-point functions in minimal models are known in terms of hypergeometric functions. Therefore, following a local quench by a minimal model primary field, for example the operator $\phi_{(2,1)}$, we can obtain the second R\'enyi entropy from the corresponding four-point function.
We follow the 
arguments of \cite{Chen:2015usa} where the R\'enyi entropy for this minimal model at $\epsilon = 0$ was computed. The central charge of the $(p,p')$ minimal model is,
\begin{align}
c \,=\, 1-6 \frac{(p-p')^2}{pp'}\,.
\end{align}
The second R\'enyi entropy is evaluated from the normalised four point function on the branched cylinder, given in eq.\eqref{RE2confbG}. The conformal dimension $h$ of the operator $\phi_{2,1}$ generating the local quench is
\begin{align}
h \,=\, \frac{3}{4}\frac{p}{p'}\,-\,\frac{1}{2}\,.
\end{align}
The change in the single interval R\'enyi entropy is given by the (logarithm of) normalized four-point function \eqref{RE2confbG} which when expressed in terms of the cross-ratios $(u,\bar u)$, takes the form
\be
\Delta S_A^{(2)}\,=\,-\ln\left[|u|^{4h}|1-u|^{4h}\,G(u,\bar u)\right]\,.
\ee 
The function $G(u,\bar{u})$ is known \cite{Dotsenko:1984nm} \cite{Dotsenko:1984ad},
\bea\label{min_model_G}
&&G(u,\bar{u})  \,=\,\, \mathcal{N}^{-2}\,\, |u|^{p/p'}\, |1-u|^{p/p'} \,\left(\frac{\sin\frac{\pi p}{p'}\, \sin\frac{3\pi p}{p'}}{\sin\frac{2\pi p}{p'}}\,\, |{\cal G}_1(u)|^2 \,\,+\,\, \frac{\sin^2\frac{\pi p}{p'} }{\sin\frac{2\pi p}{p'}}\,\, |{\cal G}_2(u)|^2 \right)\nonumber\\\\\nonumber
&&{\cal G}_1(u)  \,=\, \frac{\Gamma(\frac{3p}{p'}-1)\,\Gamma(1-\frac{p}{p'})}{\Gamma(\frac{2p}{p'})}\,\, \ _2F_1\left(\tfrac{p}{p'},\,-1\,+\,\tfrac{3p}{p'},\,\tfrac{2p}{p'},\,u\right)\\\nonumber\\
&&{\cal G}_2(u)\,=\, u^{1-{2p}/{p'}}\,\,\frac{
\Gamma(1-\frac{p}{p'})^2}{\Gamma(2-\frac{2p}{p'})} \,\,\ _2F_1\left(\tfrac{p}{p'},\,1-\tfrac{p}{p'},\,2-\tfrac{2p}{p'},\,u\right).\nonumber
\eea
In order to evaluate the R\'enyi entropies for $t<l_{1,2}$ (and $t>l_{1,2}$), we make use of the limits of the cross-ratios in eq.\eqref{uexp}, so they are both of order $\epsilon^2$. In the limit $(u,\bar{u}) \to (0,0)$, the leading contribution to $G(u,\bar{u})$ comes from the term proportional to $|{\cal G}_2(u)|^2$ so that,
\be
G(u,\bar{u}) \approx |u|^{-4h}  \left(1\,+\,{O}(\epsilon^4) \right)\,,
\ee
where we fix the  normalisation ${\cal N}$ such that the coefficient of the leading term $\sim|u|^{-4h}$ is set to unity.
It then follows  that the order $\epsilon^2$ correction to the second R\'enyi entropy is
\be
\Delta S^{(2)}_A\,=\,2h(u+\bar u)\,+\,O(\epsilon^4)\,,
\ee
matching the general result \eqref{RE_confb0} from the conformal block argument.

The R\'enyi entropy in the time interval $l_1<t<l_2$ is obtained by using the cross-ratios in eq.\eqref{ulimits} which are expanded about the limits $(u,\bar{u}) \,=\, (1,0)$. In order to obtain the expansions about these new limits, the transformation properties of hypergeometric functions can be used to rewrite ${\cal G}_1$ and ${\cal G}_2$ as,
\begin{eqnarray}
{\cal G}_1(u)  &&=\,  \frac{\Gamma(\frac{3p}{p'}-1 )\,\Gamma(1-\frac{2p}{p'})}{\Gamma (\frac{p}{p'})}\,\, \ _2F_1\left(\tfrac{p}{p'},\,-1+\tfrac{3p}{p'},\,\tfrac{2p}{p'},\,1-u\right)\,+\\\nonumber
&& \,+\, (1-u)^{1-{2p}/{p'}}\,\, \frac{\Gamma(1-\frac{p}{p'})\,\Gamma (\frac{2p}{p'}-1)}{\Gamma(\frac{p}{p'})}\,\, \ _2F_1\left(\tfrac{p}{p'},\,1-\tfrac{p}{p'},\,2-\tfrac{2p}{p'},\,1-u\right)\,,
\eea
and 
\bea\label{newI2}
&&{\cal G}_2(u)\,= \, u^{1-{2p}/{p'}}\,\, \frac{\Gamma(1-\frac{p}{p'})\,\Gamma (1-\frac{2p}{p'})}{\Gamma(2-\frac{3p}{p'})}\,\, \ _2F_1\left(\tfrac{p}{p'},\,-1+\tfrac{3p}{p'},\,\tfrac{2p}{p'},\,1-u\right) \\
&& +\,\, (1-u)^{1-{2p}/{p'}}\,\frac{\Gamma(1-\frac{p}{p'})\,\Gamma (\frac{2p}{p'}-1)}{\Gamma(\frac{p}{p'})}\,\, \ _2F_1\left(\tfrac{p}{p'},\,1-\tfrac{p}{p'},\,2-\tfrac{2p}{p'},\,1-u\right)\,. \nonumber
\end{eqnarray}
Now, the leading behaviour of $G(u,\bar{u})$ is governed by the expansion of  ${\cal G}_2(\bar{u})$ near $\bar u=0$ using \eqref{min_model_G}, and by the expansion of  the second term in the above equation \eqref{newI2} for ${\cal G}_2(u)$, around $u=1$,
\begin{align}
G(u,\bar{u})\, \approx\, (1-u)^{-2h} \,\bar{u}^{-2h}\, \left[-\,\left(2 \cos \tfrac{2\pi p}{p'}\right)^{-1}\,+\,{O}(\epsilon^4) \right]\,.
\end{align}
The expressions for $(u,\bar u)$ at this order are given by eq.\eqref{ulimits} 
which yield the order $\epsilon^2$ contribution to R\'enyi entropy for intermediate times $l_1<t<l_2$. Writing out the change in the 
R\'enyi entropy as 
\be
\Delta S^{(2)}_A\,=\,\Delta S_A^{(2;\,0)}\,+\,\epsilon^2\,\Delta S_A^{2,\,2}+\ldots\,,
\ee
we confirm our general arguments for the case of the minimal model quench
\begin{eqnarray}\label{minimal_entropy} 
&&\Delta S_A^{(2;\,0)}\, =\, \begin{cases}
0, \qquad & t<l_1 \qquad\text{or}\qquad \, t>l_2\\
\ln \left( -2 \cos \frac{\pi p}{p'}  \right)\,,\qquad & l_1<t<l_2\,,
\end{cases}\\\nonumber\\\nonumber
&&\Delta S_A^{(2;\,2)}\,=\,\frac{h}{2 }\left[{\cal S}_{l_1l_2}(t)^2\,+\,{\cal S}_{l_1l_2}(-t)^2\right]\,.
\end{eqnarray}
We note that in order for the leading term at $O(\epsilon^0)$ to be real, we must require the argument of the logarithm to be positive. This happens when the argument of the cosine lies in the second and third quadrants. Thus, the allowed ranges of values of $p/p'$ are (taking into account $p>p'$):
\begin{align}
1 < \tfrac{p}{p'} < \tfrac{3}{2}\,, \qquad \text{and} \qquad 2m\,+\,\tfrac{1}{2}<
\tfrac{p}{p'}<2m\,+\,\tfrac{3}{2}\,, \qquad m=1,2, \dots.
\end{align}
Interestingly, the leading order change in R\'enyi entropy for $l_1<t<l_2$ is {\em negative} for a subset of the allowed range above:
\bea
\tfrac{4}{3} < \tfrac{p}{p'} < \tfrac{3}{2}\,, \qquad  2m+\tfrac{1}{2}<\tfrac{p}{p'}<2m+\tfrac{2}{3}\,,\qquad 2m+\tfrac{4}{3}<\tfrac{p}{p'}<2m+\tfrac{3}{2}, \quad m\,=\,1,2,\ldots,\nonumber
\eea
while it remains {\em positive} for all remaining values, specified by  the range
\be
1 < \tfrac{p}{p'} < \tfrac{4}{3}\,, \qquad 2m+\tfrac{2}{3}<\tfrac{p}{p'}<2m+\tfrac{4}{3}, \qquad m\,=\,1,2, \dots\,.
\ee

\begin{figure}
\begin{subfigure}{0.5\textwidth}
\centering
\includegraphics[scale=0.8]{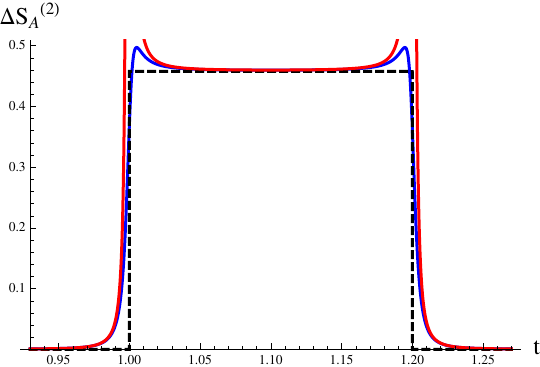}
\caption{•}
\end{subfigure}
\begin{subfigure}{0.5\textwidth}
\centering
\includegraphics[scale=0.8]{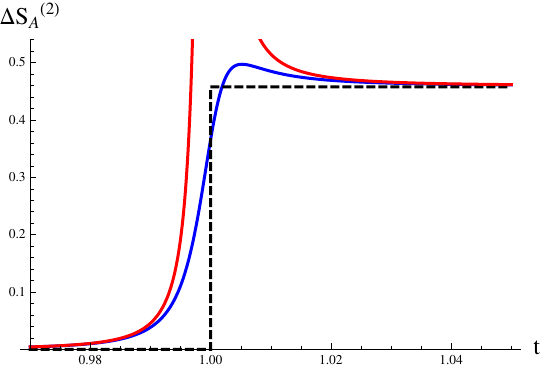}
\caption{•}
\end{subfigure}
\caption{\small{Time dependence of the second R\'enyi entropy for the minimal model quench, with ${p}/{p'} = 1.21$, $\beta=1, l_1=1, l_2=1.2$ and $\epsilon=0.005$. Shown above are the exact result (blue curve), the step jump at order $\epsilon^0$ (dashed black), and the approximation at  order $\epsilon^2$ (in red). 
}}
\label{min_pos}
\end{figure}

\begin{figure}
\begin{subfigure}{0.5\textwidth}
\centering
\includegraphics[scale=0.8]{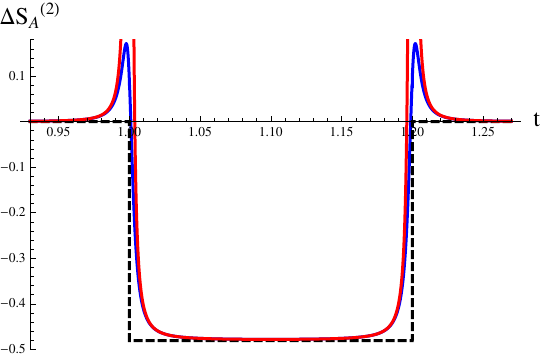}
\caption{•}
\end{subfigure}
\begin{subfigure}{0.5\textwidth}
\centering
\includegraphics[scale=0.8]{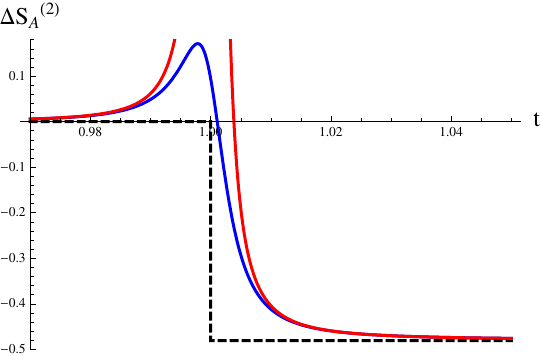}
\caption{•}
\end{subfigure}
\caption{\small{Time dependence of the second R\'enyi entropy for the minimal model quench, with ${p}/{p'} = 1.4$, $\beta=1$, $l_1=1$, $l_2=1.2$ and $\epsilon=0.005$. For this value of $p/p'$ the jump in the R\'enyi entropy at order $\epsilon^0$ is negative. Displayed in this plot are the exact result (blue curve), the step jump at order $\epsilon^0$ (green), and the approximation at  order $\epsilon^2$ (in red).}}
\label{min_neg}
\end{figure}

\paragraph{Comparison with exact formulae:} In Figures \ref{min_pos} and \ref{min_neg}, we have plotted the time dependence of the second R\'enyi entropy for the minimal model quench, for the two cases $p/p'=1.21$ and $p/p'=1.4$, respectively. The dashed lines show the  $\epsilon=0$ step change, and the red curves include the correction at order $\epsilon^2$ given in eq.\eqref{minimal_entropy} for $\epsilon=0.005$. Importantly, the curve in blue is the exact result for the R\'enyi entropy obtained from eq.\eqref{min_model_G} for $\epsilon=0.005$.

\begin{figure}
\begin{subfigure}{0.5\textwidth}
\centering
\includegraphics[scale=0.8]{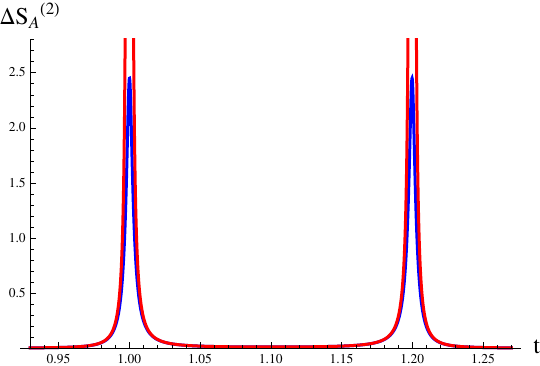}
\caption{•}
\end{subfigure}
\begin{subfigure}{0.5\textwidth}
\centering
\includegraphics[scale=0.8]{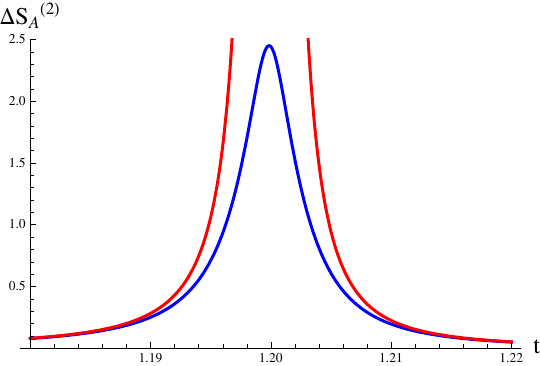}
\caption{•}
\end{subfigure}
\caption{
\small{The evolution of the second R\'enyi entropy for the minimal model quench, with ${p}/{p'} = 10/3$, $\beta=1$, $l_1=1$, $l_2=1.2$ and $\epsilon=0.005$. For this value of $p/p'$, the jump $\sim \ln(- 2\cos(\pi p/p'))$ at order $\epsilon^0$ is vanishing. Displayed in this plot are the exact result (blue curve) and the leading approximation at  order $\epsilon^2$ (in red).}
}
\label{min_small_entropy}
\end{figure}

The figures clearly show that the exact result for R\'enyi entropy at finite $\epsilon$ (blue curve) becomes {\em larger} than the zero-width step function  ($\epsilon=0$ dashed lines). It is accurately tracked by  the ${O}(\epsilon^2)$ correction to R\'enyi entropy (red curve), which is valid for ${|t-l_{1,2}|}\gg \epsilon$, and appears to be a very good approximation to the exact result (blue curve), even for values of $t$ close to $l_{1,2}$.

A striking feature of the exact result for the R\'enyi entropy is the {\em overshoot} or {\em spike} which accompanies the entry and  exit of the excitation into and out of the interval $A$. The universal correction at order $\epsilon^2$ correctly describes the approach towards this spike and the fact that $\Delta S^{(n;\,2)}_A$ is strictly positive is also consistent with the behaviour of the exact R\'enyi entropy.

A further point worth noting is that it is possible to consider  values of the ratio $p/p'$, for which the step function jump at order $\epsilon^0$ is arbitrarily small, so that it is the correction at order $\epsilon^2$ which dominates the time dependence of the R\'enyi entropy.
As an example, figure \ref{min_small_entropy} shows the relevant plots for $\frac{p}{p'} = \frac{10}{3}$. The order $\epsilon^2$  correction is the dominant contribution  even for a small value of $\epsilon =0.005$. The approximate result (red curve) tracks the exact result (blue curve) for all times $t$ such that $|t-l_{1,2}|\gg \epsilon$. Thus, for such values of $\frac{p}{p'}$, the universal correction at $O(\epsilon^2)$ is essential to capture the correct behaviour for the R\'enyi entropy at finite width.

\subsection*{Example II: The free fermion theory}
 
The free fermion theory provides us with the simplest setting in which to examine and compare the time evolution of R\'enyi entropies following the finite width quench, with the universal result at order $\epsilon^2$. Crucially, the free fermion theory has a spin one current and therefore violates the general assumption made in our previous argument based on conformal blocks i.e. that the lowest lying primary should have dimension greater than one.  This feature makes the free fermion theory an instructive example to study.

We consider a  single complex free fermion which, in the bosonized picture, corresponds to a single 
free scalar with holomorphic and antiholomorphic parts $\varphi$ and $\bar\varphi$. We would like to examine the R\'enyi entropies following a local quench by  the operator,
\bea\label{def_expop}
&&\mathcal{O}(x,\,\bar x)  \,=\, e^{i \alpha(\varphi(x)+\bar\varphi(\bar x))} \,+\, \kappa \, e^{-i\alpha(\varphi(x)+\bar\varphi(\bar x))}\,,\\\nonumber\\
&&\mathcal{O}^\dagger(x,\,\bar x) \,=\, \kappa^*\, e^{i\alpha(\varphi(x)+\bar\varphi(\bar x))} \,+\, e^{-i\alpha(\varphi(x)+\bar\varphi(\bar x))}\,, \nonumber
\eea
where $\kappa$ is a complex number.
The operator has conformal weights $(h,\bar h)\,=\, \left(\frac{\alpha^2}{2},\,\frac{\alpha^2}{2}\right)$, and we will assume that $h>\frac{1}{4}$. In order to compute the second R\'enyi entropy, we employ the uniformization map $w(x)$ \eqref{uniform_map} which maps the branched double cover of the cylinder to the complex $w$-plane, and perform the necessary Wick contractions to obtain the required four-point function on the branched cylinder,
\bea \label{exp4point}
&& \left\langle \prod_{j=1}^2\mathcal{O}(x_1^{j},\,\bar x_1^{(j)})\,\mathcal{O}^\dagger(x_4^{(j)},\bar x_4^{(j)}) \right\rangle_2\,=\, (|w'_1w'_4||2w_1||2w_4||w_1^2-w_4^2|^{-2})^{2\alpha^2}\\\nonumber\\\nonumber
&&\qquad\qquad\qquad\qquad\qquad\times\left[ 1+|\kappa|^4 +2|\kappa|^2 \frac{|w_1-w_4|^{8\alpha^2} +|w_1+w_4|^{8\alpha^2}}{(|2w_1||2w_4|)^{4\alpha^2}} \right]\,.
\eea
The second R\'enyi entropy  $\Delta S_A^{(2)}$ is obtained as usual from the normalized four-point correlator, naturally expressed in terms of the cross-ratios $(z, \bar{z})$ defined in eq.\eqref{defcrossratio}:
\begin{align}
\Delta S_A^{(2)} = -\ln\left[\frac{  1\,+\,|\kappa|^4 \,+\,{2|\kappa|^2}{|4\sqrt{z}|^{-4\alpha^2}}\, \left(|\sqrt{z}-1|^{8\alpha^2}\, +\,|\sqrt{z}+1|^{8\alpha^2} \right)}{(1+|\kappa|^2)^2}\right]\,.
\end{align}
The key point here is that $\sqrt{z}$ undergoes a sign flip (whilst $\sqrt{\bar z}$ does not) between the two regimes $t<l_{1,2}$ and $l_1<t<l_2$ as explained more generally in section \ref{sec:uniformmap}, which in turn leads to a discontinuity in the R\'enyi entropy as a function of time in the limit of zero width ($\epsilon\to 0$): 
\begin{align}\label{renyi2limit}
\Delta S_A^{(2;\,0)}\, = \,
\begin{cases}
0\,,\qquad & t<l_1\quad{\rm or}\quad t> l_2\,,\\\\
-\,\log\left(\frac{1+|\kappa|^4}{(1+|\kappa|^2)^2}\right)\,,  \qquad&l_1<t<l_2.
\end{cases} 
\end{align}
In this example, the quantum dimension of the operator ${\cal O}$ is given by $d_\mathcal{O}\, =\,\frac{(1+|\kappa|^4)}{(1+|\kappa|^2)^2}$. At order $\epsilon^2$ however, we encounter a surprising result, namely, that the correction {\em vanishes} at this order for $l_1 < t < l_2$:
\begin{align} \label{RE2_epsilon}
\Delta S_{A}^{(2)}\, =\,
\begin{cases}
\epsilon^2\,\alpha^2\,\frac{ |\kappa|^2}{(1+|\kappa|^2)^2 }\,\left[
{\cal S}_{l_1 l_2}(t)^2\,+\,{\cal S}_{l_1 l_2}(-t)^2 \right]\,,\qquad & t<l_{1}\quad {\rm or}\quad {t> l_2}\,,\\\\
-\, \ln \left[\frac{(1+|\kappa|^4)}{(1+|\kappa|^2)^2}\right] \,+\, {O}(\epsilon^4)\,,\qquad & l_1<t<l_2\,. 
\end{cases}
\end{align}
This does not agree with our universal arguments (which were verified for the minimal models above) that imply the same functional form for the correction at all times. Furthermore, the coefficient of the correction in this example, when $t<l_{1}$, depends on the operator ${\cal O}$ via $\kappa$, which is clearly non-universal. The reason for these discrepancies can be directly traced to the presence of the conserved 
$U(1)$ current in the free fermion theory. Using the OPE methods developed in section \ref{sec-OPE}, we can verify that there is in fact a  cancellation of the correction at $O(\epsilon^2)$ for $l_1<t<l_2$, due to a contribution from the $U(1)$ current which also affects the coefficient of the non-vanishing finite width correction at early times $t< l_1$.

To see the effect of the $U(1)$ current, we consider the OPE of the operators \eqref{def_expop} creating the localized quench on the uniformized $w$-plane. For the second R\'enyi entropy, there are four insertion points:
\be
\left(w_{p}^{(1)},\,\bar w_p^{(1)}\right)\,\equiv\,(w_p, \, \bar w_p)\,,\qquad
\left(w_{p}^{(2)},\,\bar w_p^{(2)}\right)\,\equiv\,(-w_p, \, -\bar w_p)\,,\qquad p=1,4\,.
\ee
As usual, these are the images of the two pairs of insertion points on the double-sheeted, branched thermal cylinder. For early times, $(t<l_1)$, the
 relevant OPE is determined by the limit $(w_4,\,\bar w_4)\to (w_1,\,\bar w_1)$:
\bea \label{OOOPE}
&&\mathcal{O}(w_1,\,\bar w_1) \,\mathcal{O}^\dagger (w_4,\,\bar w_4)\, \sim\, 
|w_1-w_4|^{-2 \alpha^2}\, \left[(1+|\kappa|^2)\right.\\\nonumber\\\nonumber
&&\left.\,-\, \alpha (1-|\kappa|^2)\left\{(w_4-w_1)\,J(w_1)\, +\,(\bar w_4-\bar w_1)\,\bar J(\bar w_1)\,+\,\tfrac{1}{2}(w_4-w_1)^2\,  \partial J(w_1)\,+\right.\right.\\\nonumber\\\nonumber
&&\left.\left.+\,\tfrac{1}{2}(\bar w_4-\bar w_1)^2 \, \bar{\partial}\bar J(\bar w_1)\right\}\,+\,\alpha^2 (1+|\kappa|^2)
\left\{(w_4-w_1)^2 \,T(w_1)\,+\,(\bar w_4-\bar w_1)^2 \,\overline T(\bar w_1)\right.\right.\nonumber\\\nonumber\\\nonumber
&&\left.\left. +\,|w_4-w_1|^2 J(w_1) \bar J(\bar w_1)\right\} ]\,+\,\ldots\right]
\eea
where we have defined the holomorphic and antiholomorphic $U(1)$ currents $(J,\,\bar J)$ and the stress tensor as,
\be
J\,=\,i\partial\varphi\,,\qquad \bar J\,=\,i\bar\partial\bar\varphi\,,\qquad
T\,=\,-\frac{1}{2}(\partial\varphi)^2\,,\qquad \overline T\,=\,-\frac{1}{2}(\bar\partial\bar\varphi)^2\,.
\ee
On the $w$-plane these must all have vanishing expectation values, by conformal invariance. However, it is clear that the OPE has  new contributions at order $\epsilon$ and $\epsilon^2$ from the currents $(J,\bar J)$. Therefore, accounting for the fact that all currents have vanishing one-point functions on the $w$-plane, the four-point function relevant for the second R\'enyi entropy in the early time regime $t<l_1$ is (expanding around $(z,\bar{z}) = (1,1)$),
\bea
&&\left\langle\prod_{j=1}^2 \mathcal{O}^\dagger  (w_4^{(j)},\,\bar w_4^{(j)}) \,\mathcal{O}(w_1^{(j)},\,\bar w_1^{(j)}) \right\rangle_w\, \simeq\, |w_1-w_4|^{-4\alpha^2}(-1)^{-2\alpha^2}\,(1+|\kappa|^2)^2  \times\\\nonumber
&& \left[ 1\,-\,\alpha^2\left(\frac{1-|\kappa|^2}{1+|\kappa|^2 }\right)^2 \left((w_4-w_1)^2 \,\langle J(w_1) J(-w_1) \rangle\, +\, (\bar w_4-\bar w_1)^2\, \langle \bar J(\bar w_1) \bar J(- \bar w_1) \rangle \right) \right]\,.
\eea
The correlator on the branched cylinder is then obtained by a conformal transformation on this result.  In the absence of the $U(1)$ current correlator $\sim \langle JJ\rangle$ in the above expression, the transformation would yield the  expected universal answer. Instead, the latter is now modified by the presence of the $U(1)$ current:
\bea \label{expop-OPE-before}
&&\left\langle\prod_{j=1}^2 \mathcal{O} (x_1^{(j)},\,\bar{x}_1^{(j)})\,\mathcal{O}^\dagger(x_4^{(j)},\,\bar{x}_4^{(j)}) \right\rangle_2\,\simeq\,\left( 2\epsilon^2 \right)^{-4\alpha^2}\,(1+|\kappa|^2)^2\,\times\\\nonumber\\\nonumber
&&\,\left[1\,-\,\tfrac{2}{3}\epsilon^2 \alpha^2\left(\{ w,x \}\,+\, \{ \bar w,\bar x \} \right) \right]_{x=0}\,
\left[ 
1\, +\, \frac{\alpha^2 \epsilon^2 }{4}\frac{(1-|\kappa|^2)^2}{(1+|\kappa|^2)^2} \left({\cal S}_{l_1l_2}(t)^2\,+ \,{\cal S}_{l_1l_2}(-t)^2\right) \right]\,.
\eea
The first factor in the second line, expressed in terms of the Schwarzian, arises from the conformal transformation from the $w$-plane to the branched surface, whilst the second factor originates directly from the current-current correlator (at order $\epsilon^2$). Recalling from eq.\eqref{schw} that the Schwarzian is also given by the universal function ${\cal S}_{l_1l_2}$, we obtain,
\bea
&&\left\langle\prod_{j=1}^2 \mathcal{O} (x_1^{(j)},\,\bar{x}_1^{(j)})\,\mathcal{O}^\dagger(x_4^{(j)},\,\bar{x}_4^{(j)}) \right\rangle_2\,\simeq\,\\\nonumber
&&\hspace{0.5in} \left( {2\epsilon^2} \right)^{-4\alpha^2}\,\left[ (1+|\kappa|^2)^2\left(1\,+\, \epsilon^2\alpha^2\tfrac{8\pi^2}{3\beta^2}\right) \,-\, \epsilon^2|\kappa|^2 \alpha^2  \,\left( {\cal S}_{l_1l_2}(t)^2\,+ \,{\cal S}_{l_1l_2}(-t)^2\right) \right]. \nonumber
\eea
Upon dividing by the square of the two-point correlator on the thermal cylinder, we obtain the second R\'enyi entropy. This explains the origin of the non-universal coefficient of the time dependent part of the R\'enyi entropy in \eqref{renyi2limit}, in the early time regime. For intermediate times, on the other hand, we need to consider the OPEs in the limit 
where $w_1\to -w_4 $ and $\bar w_1 \to \bar w_4$, which corresponds to an expansion about the branch point $(\sqrt z,\,\sqrt {\bar{z}}) \,=\, (-1,\,1)$ in the exact correlator. 
Noting our assumption that the weights satisfy $h>\frac{1}{4}$, 
and carefully repeating the steps of the OPE analysis above, we arrive  
at the result for  $l_1<t < l_2$\,:
\bea
&&\left\langle\prod_{j=1}^2 \mathcal{O} (x_1^{(j)},\,\bar{x}_1^{(j)})\,\mathcal{O}^\dagger(x_4^{(j)},\,\bar{x}_4^{(j)}) \right\rangle_2\,\simeq\, (1+|\kappa|^4) \times\\\nonumber
&& \times \,\left({2\epsilon^2} \right)^{-4\alpha^2}\, \left[1\,-\,\frac{2}{3}\epsilon^2 \alpha^2(\{ w,x \}+ \{ \bar w,\bar x \} ) \right]_{x=0}
\left[1\,+\,\epsilon^2 \frac{\alpha^2}{4} \left({\cal S}_{l_1l_2}(t)^2\,+\,{\cal S}_{l_1l_2}(-t)^2\right)\right]\,. \nonumber
\eea
The Schwarzian from the coordinate transformation now exactly cancels the contribution from the current-current correlator in this limit, so that the first finite-width correction to the R\'enyi entropy appears at order $\epsilon^4$.

\subsection{Large $c$ limit}
We now turn to discuss the large $c$ limit which is relevant from the point of view of CFTs with holographic gravity duals. There are two distinct types of situations in this context -- one where $c$ is taken large with conformal dimensions of the exciting operator $\Delta_{\cal O}$ and those of the primaries contributing to the Virasoro conformal block, $\Delta_b$ are kept fixed, i.e. $\frac{\Delta_b}{c}, \, \frac{\Delta_O}{c}<<1$ \cite{Caputa:2014vaa}. In this limit,  equations (\ref{vacexp}) and (\ref{vacexp1}) imply  that the vacuum block is unity,
\begin{align}
\tilde{F}_{\mathcal{O};{\,\rm vac}}\, =\, 1\,.
\end{align}
Therefore, our conformal block argument for the second R\'enyi entropy using  the contribution from the vacuum conformal block follows as before from equations (\ref{RE2confbG}), (\ref{G_confb}) and  (\ref{confb}). The second 
R\'enyi entropy at order $\epsilon^2$ is given precisely by the universal result in eq.(\ref{RE_confb0})  for times, $t<l_1$ and $t>l_2$. For intermediate  times when the excitation is within the entanglement interval, $l_1<t<l_2$, the cross-ratios are expanded around the limit $(u,\,\bar{u}) = (1,0)$ and the R\'enyi entropy is,
\bea
\Delta S_A^{(2)}&& \,=\, - \ln \left[u^{2h} (1-u)^{2h} (1-\bar{u})^{2h}\right]\\\nonumber\\\nonumber
&&\, \simeq\, 4h\, \ln \left[\frac{2\beta^2}{\epsilon\, \pi^2} \,{\cal S}_{l_1l_2}(t)\right] \,+\, \epsilon^2\,\frac{h}{2}
\left[{\cal S}_{l_1l_2}(t)^2\,+\,{\cal S}_{l_1l_2}(-t)^2\right]\,.
\eea
The term at order $\epsilon^2$ is precisely the universal correction we have argued in the general case, while the leading term is responsible for the late time logarithmic growth in R\'enyi entropies, argued on general grounds in \cite{Caputa:2014vaa} in this particular large-$c$ limit.

There is a second type of large-$c$ limit where operator dimensions scale with $c$, so that  $\frac{h}{c}$ is fixed. In this case  the R\'enyi and entanglement entropies are known \cite{Asplund:2014coa}. The entanglement entropy in particular is
\begin{align}
\Delta S_A^{(1)}\, =\, \frac{c}{6} \ln \left[ \frac{z^{\frac{1}{2}(1-\gamma)}\,\bar z^{\frac{1}{2}(1-\bar \gamma)} \,(1-z^\gamma)\, (1- \bar z^{ \bar \gamma}) }{\gamma \bar{\gamma}\,(1-z)\,(1-\bar{z})}  \right]\,,
\end{align}
where,
\begin{align}
\bar\gamma\,=\,\gamma\,=\, \sqrt{1-\frac{24 h}{c}}\,,
\end{align}
and $h\,=\,\Delta_{\cal O}/2$.
Using the cross-ratios defined in equation (\ref{defcrossratio}), expanding around the $\epsilon \to 0$ limit for $t<l_1$ or $t>l_2$, that is $(\sqrt {z},\sqrt{\bar{z}})\, =\, (1,1)$, we find that the change in the entanglement entropy in this regime is given by the universal result:
\begin{align}
\Delta S_A^{(1)} \,=\, \epsilon^2 \,\frac{{\Delta}_{\cal O} }{3 }\left[ {\cal S}_{l_1l_2}(t)^2\,+\,{\cal S}_{l_1l_2}(-t)^2
\right].
\end{align}
The large-$c$ limit with ${\Delta_{\cal O}}/c$ fixed is reproduced by the holographic gravity dual description 
\cite{Asplund:2014coa, Caputa:2014eta}. In both CFT and gravity pictures, the time dependence of the entanglement entry in the vicinity of the entry $(t\simeq l_1)$ and exit points $t\simeq l_2$ is nontrivial and also impacts the time dependence in the intermediate regime $l_1 < t < l_2$. 
We will discuss this in some more detail below.

\subsection{Check with holography} \label{holography-check}

The holographic dual to a thermal CFT state, excited by a finite width quench via some (heavy) local operator, is given by an AdS geometry with a point particle of mass $m$ falling from a radial distance $z \propto \epsilon$, near the boundary of AdS, towards a bulk black hole horizon \cite{Nozaki:2013wia, Caputa:2014eta}. 
The single interval entanglement entropy in the quenched state is computed  by the length of a geodesic with its end-points on the boundary in the backreacted geometry produced by the infalling massive particle. We will first summarise the results of \cite{Caputa:2014eta} wherein the holographic entanglement entropy at finite temperature was computed using the back-reacted metric of an infalling particle in the BTZ black hole background. We will also see  that the order $\epsilon^2$ correction to the holographic entanglement entropy, for times $t<l_1$ or $t>l_2$, is indeed the same as the universal  correction we have argued from within CFT. 

The mass of the infalling particle in the bulk is related to the conformal dimension of the local operator responsible for the field theory quench as, $m\,=\, \Delta_{\cal O}/R $ where $R$ is the AdS radius. The background geometry depends on the particle mass via a parameter $\widetilde m$, which is related to $m$ and $\Delta_{\cal O}$ as,
\begin{align}
\widetilde m\,\equiv\,8G_N R^2m\,,\qquad\qquad \widetilde m \,=\, \frac{12\Delta_{\cal O}R^2}{c}\,.
\end{align}
Here $G_N$ is Newton's constant in the gravity dual.
The backreacted geometry was obtained in \cite{Caputa:2014eta} by taking the solution for a point particle or defect at the origin of global AdS$_3$ spacetime and mapping it to BTZ coordinates to yield the backreacted solution for the infalling particle. The metric for a defect in global AdS$_3$ is,
\be
ds^2\, =\, -(r^2+R^2-\widetilde m)\,d\tilde{\tau}^2\, +\, \frac{R^2\, dr^2}{r^2+R^2-\widetilde m}\,+\,r^2\, d\phi^2\,,
\ee
where the angular and radial coordinate ranges are $0<\phi\leq 2\pi$ and $0<r< \infty$, respectively. The time dependent geometry produced by the falling particle in the BTZ coordinates results from the transformations,
\begin{align}
\sqrt{r^2+R^2}\,\sin \tilde{\tau} &\, =\, \frac{R}{\sqrt{M}z}\,\sqrt{1-Mz^2}\,\sinh\left(\sqrt{M}t\right)\,,\\\nonumber\\
\sqrt{r^2+R^2}\,\cos \tilde{\tau} & \,=\, \frac{R}{\sqrt{M}z}{ \left[\cosh(\lambda)\cosh\left(\sqrt{M}x\right)-\sqrt{1-Mz^2}\sinh(\lambda)\cosh\left(\sqrt{M}t\right) \right]}\,,\nonumber\\\nonumber\\
r \sin(\phi) & \,=\, \frac{R}{\sqrt{M}z}\sinh\left(\sqrt{M}x\right)\,,\nonumber\\\nonumber\\
r\cos(\phi)& \,=\, \frac{R}{\sqrt M z}{ \left[\cosh(\lambda)\sqrt{1-Mz^2}\,\cosh\left(\sqrt{M}t\right)\,-\,\sinh(\lambda)\cosh\left(\sqrt{M}x\right) \right]}\,.\nonumber
\end{align}
Here $(t,x)$  are boundary CFT coordinates and $z$, the radial coordinate (AdS boundary at $z=0$) in the ordinary BTZ black hole geometry which is obtained from the above transformations on global AdS$_3$ without a defect ($\widetilde m\,=\,0$):
\bea
ds^2\, =\, \frac{R^2}{z^2}\, \left(-(1-M z^2)\,dt^2\,+\,\frac{dz^2}{(1-Mz^2)}\,+\,dx^2 \right)\,.
\eea
The BTZ black hole mass is given by $M$ which is in turn related to the Hawking temperature as $\beta\,=\, 2\pi/\sqrt{M}$.
The parameter $\lambda$ generates a boost  which is a symmetry of the background in the absence of the defect. With $\widetilde m\neq 0$, the transformed geometry is time dependent and the boost parameter is linked to the width ($\epsilon$) of the excitation via
\begin{align}
\tanh \lambda \,=\, \sqrt{1-M\epsilon^2}\,.
\end{align}

The entanglement entropy of an interval can be obtained using the (covariant) prescription for  holographic entanglement entropy \cite{Ryu:2006bv, Hubeny:2007xt}. It is given by the length of a geodesic in the backreacted geometry, where the geodesic is anchored at the endpoints of the interval in the boundary CFT. The BTZ coordinates of these endpoints are $x_\infty^{(1)}$ and $x_\infty^{(2)}$ which correspond to the points $x=l_1$ and $x=l_2$, respectively in the CFT. At any given time $t$, the BTZ coordinates of these two points map to the global AdS coordinates $(\tilde \tau^{(i)},\,\phi^{(i)})$:
\bea
&& r_\infty^{(i)} \,=\, 
\frac{R \beta^2}{4\pi^2\epsilon\, z_\infty} \sqrt{\left(\tfrac{2\pi\epsilon}{\beta}\right)^2 \sinh^2  \tfrac{2\pi x_\infty^{(i)}}{\beta}\,+\,\left( \cosh\tfrac{2\pi t}{\beta} \,-\,\sqrt{1-\left(\tfrac{2\pi\epsilon}{\beta}\right)^2} \cosh \tfrac{2\pi x_\infty^{(i)}}{\beta}\right)^2}\,,\nonumber\\\nonumber\\\nonumber\\
&& \tan \left(\tilde{\tau}^{(i)}\right) \,=\, \frac{2\pi \epsilon}{\beta} \frac{\sinh \frac{2\pi t}{\beta}}{\cosh \frac{2\pi x_\infty^{(i)}}{\beta}\,-\, \sqrt{1-\left(\frac{2\pi\epsilon}{\beta}\right)^2}\cosh\frac{2\pi t}{\beta}}\,,\\\nonumber\\\nonumber\\
&& \tan \left(\phi^{(i)}\right)\, =\, \frac{2\pi \epsilon}{\beta} \frac{\sinh \frac{2\pi x_\infty^{(i)}}{\beta}}{\cosh  \frac{2\pi t}{\beta}\,-\, \sqrt{1-\left(\frac{2\pi\epsilon}{\beta}\right)^2}\cosh\frac{2\pi x_\infty^{(i)}}{\beta} }\,. \nonumber
\eea
The holographic entanglement entropy for the interval in the presence of the local quench is given by the formula (evaluated in global AdS),
\bea \label{EE_hol1}
\hat{S}_A \,=\, \frac{c}{6} \left( \ln \left(r_\infty^{(1)} \cdot r_\infty^{(2)}\right) \,+\,\ln \frac{2\cos\left(|\Delta \tilde{\tau}_\infty| \frac{\sqrt{R^2-\widetilde m}}{R}\right) - 2\cos\left(|\Delta \phi_\infty| \frac{\sqrt{R^2-\widetilde m}}{R}\right)}{R^2-\widetilde m}\right)\,.\qquad
\eea
Taking the $\epsilon \to 0$ limit of (\ref{EE_hol1}), and keeping terms upto $\epsilon^2$, we see that for $t<l_1$ and $t>l_2$, the time dependence of $\epsilon^2$ term in the holographic result matches with the universal result from CFT,
\begin{align}
\hat S_A \, =\, \frac{c}{6} \log \left[ \frac{\beta^2}{\pi^2 z_\infty^2} \,\sinh^2\tfrac{\pi}{\beta}(l_1-l_2) \right]\,+\,\epsilon^2\,\frac{\Delta_{\cal O}
}{3}\,\left[{\cal S}_{l_1l_2}(t)^2\,+\,{\cal S}_{l_1l_2}(-t)^2\right]\,,
\end{align}
\begin{figure}
\centering
\includegraphics[scale=1]{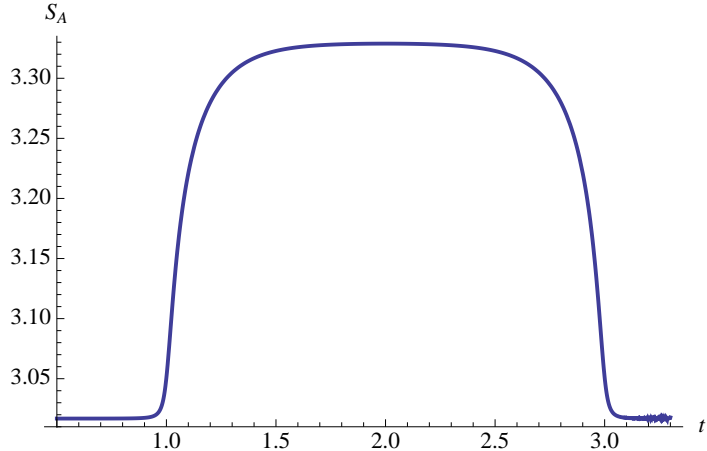}
\caption{\small{A plot of the holographic entanglement entropy with $l_1=1.0$, $l_2=3.0$, $\epsilon=0.03$ and $\beta=1$, for  CFTs in the large-$c$ limit with ${\Delta}_{\cal O}/c$ fixed. The plot is notable for the absence of the overshoots at $t=l_1$ and $t=l_2$ which characterised the minimal model example with fixed central charge. This is related to certain peculiarities of the large-$c$ limit relevant for holography.}}
\label{EEholo}
\end{figure}
where we have tacitly assumed that $(l_1 -t)\gg \epsilon$ and $(t-l_2)\gg \epsilon$, in the appropriate temporal ranges.
The behaviour in the vicinity of the endpoints of the interval and the intermediate time regime for $l_1< t < l_2$ is more nontrivial, and can be ascertained by carefully examining the holographic formula above. Specifically, in a narrow region near the endpoints of the interval for $l_1 <t$ and $l_2>t$, the entanglement entropy grows logarithmically as $\sim c\ln|l_{1,2}-t|/6$. The expected saturation of the holographic entanglement entropy correction $\Delta S^{(1)}_A$ for times $l_1< t < l_2$ does not actually occur at zero temperature, due to the fact that the quantum dimension of the quenching operator is effectively infinite in the appropriate large-$c$ limit. It is expected to become visible when nonperturbative corrections are included. What is important to note at this juncture is that the leading finite width correction to the holographic entanglement entropy for $t<l_1$ and $t>l_2$, precisely matches the universal CFT prediction.

\section{ Universal correction with chemical potential}

Two dimensional conformal field theories with ${\cal W}$-symmetries have been the subject of intense investigations following the proposal by \cite{Gaberdiel:2010pz} of holographic correspondence between ${\cal W}_N$ minimal model CFTs and higher spin theories of gravity in three dimensions. A particularly intriguing aspect of the correspondence is the possibility of analysing the respective CFTs in sectors with fixed higher spin charges by turning on conjugate chemical potentials. In the limit of large-$c$ such states are potentially dual to higher spin black hole solutions in AdS$_3$ \cite{Gutperle:2011kf} which exhibit rich and nontrivial thermodynamics 
\cite{deBoer:2013gz, David:2012iu}.

The first nontrivial correction to the single interval  R\'{e}nyi  entropy for CFTs with ${\cal W}$-symmetry in the presence of a chemical potential $\mu$ for spin-three charge, was calculated in \cite{Datta:2014ska}, and shown to be universal in \cite{Datta:2014uxa}. This correction, which appears at order $\mu^2$, also enabled a powerful check on the  proposal \cite{deBoer:2013vca,Ammon:2013hba} for holographic
entanglement entropy in higher spin theories given in terms of a Wilson line  in the Chern-Simons formulation of higher spin gravity in AdS$_3$.

In this section we will consider CFTs with a non-zero  spin-three chemical potential subjected to a finite width local quench. This situation is potentially interesting for various reasons. It is of general interest to understand the response to local quenches when conformal invariance is broken, even softly, since deformations of CFTs are often not analytically tractable. The chemical potential deformation appears to be analytically tractable within conformal perturbation theory because of the holomorphic nature of the perturbation, and therefore is worthy of study.  Finally, the effects of local quenches in the presence of higher spin chemical potential are accessible also within the dual holographic framework, which is the subject of ongoing work \cite{inprogress}.

We will show that the  leading $\mu$-dependent correction to the R\'{e}nyi entropies in the 
excited state appears at the order $\epsilon^2\mu^2$. The time dependence at order $\epsilon^2\mu^0$ is given by the universal results discussed in the previous sections. Furthermore, the step function at order $\epsilon^0$ is unaffected by the chemical potential at order $\mu^2$.

  The time dependence of the correction at order $\epsilon^2\mu^2$ is universal and is determined by the three-point function of the stress tensor 
with the higher spin currents on the $n$-sheeted cylinder. 
We determine this time dependence explicitly, 
 and perform a  cross check of this conclusion by    extracting the $\epsilon ^2 \mu^2$ 
correction to the second R\'{e}nyi entropy  for the free fermion theory.

\subsection{The general setup} \label{sec4-gen-arg}

Consider a conformal field theory which admits a higher spin (holomorphic) current $W(z)$ with spin $s$. We will restrict to spin $s=3$ at the end in order to be able to extract closed form expressions, but for now we assume $s$ to be any integer greater than two. 
A chemical potential for the corresponding charge can be introduced by deforming the action of the theory with the following term:
\begin{equation}\label{deform}
 \delta I\, =\, - \mu \int d^2 z \,\left( W(z)\, +\, \overline W(\bar z)\right)\,. 
\end{equation}
As emphasized in \cite{Datta:2014ska},  corrections to observables of the theory can be 
obtained as a perturbative expansion in the chemical potential $\mu$ for 
small enough $\mu$ relative to the inverse temperature e.g. for the spin three case where $\mu$ has dimensions of length, we require
$\mu/\beta \ll 1$. The resulting (holomorphic) conformal perturbation theory is analytically tractable, but requires a careful definition of the prescription for integrating holomorphic functions with singularities over the complex plane. Such  a prescription was made explicit in \cite{Datta:2014uxa,Datta:2014zpa}, yielding results  consistent with the canonical or Hamiltonian approach, and satisfying  various other consistency checks including holographic ones.
 
We wish to evaluate the $\mu$-dependent corrections to the 
R\'{e}nyi entropies in the presence of the local quench by the operator ${\cal O}$ with conformal dimension $\Delta_{\cal O}$.  For simplicity we focus attention only on  the holomorphic deformation  by $W(z)$ in eq.\eqref{deform}, to lowest nontrivial order. This will be sufficient at the order we work in, wherein a similar analysis can be performed for the anti-holomorphic deformation separately.  On the  replica geometry  relevant for the calculation R\'enyi entropies, the deformation must be introduced in each of the $n$ sheets so that we need to consider
\begin{equation}
\delta I\, =\, - \mu \sum_{i =1}^{n}  \int d^2 y  \,W(y^{(i)})\,, 
\end{equation}
where $W(y^{(i)} )$ refers to a current insertion in the 
$i$-th copy of the replica geometry and the integral is performed on that sheet.
From the definition of the correction to the R\'{e}nyi entropy in 
eq.\eqref{RE_n} we know that we first need to evaluate the 
correction to the $2n$-point function of the operator ${\cal O }$ in the 
deformed theory: 
\begin{eqnarray}\label{2npt-mu}
&&{\cal C}_n(\mu)\,\equiv\,\left\langle\prod_{j=1}^n \mathcal{O}^\dagger(x_4^{j},\,\bar x_4^{(j)}) \,\mathcal{O}(x_1^{(j)},\,\bar x_1^{(j)})\right\rangle_{n}^{(\mu)}  \\\nonumber\\\nonumber
&& \qquad\qquad\,=\,\frac{1}{Z^{(n)}(\mu)} \left\langle  \exp\left[  - \mu \sum_{i =1}^{n}  \int d^2 y  \,W(y^{(i)} )  \right]\,\,\prod_{j=1}^n \mathcal{O}^\dagger(x_4^{j},\,\bar x_4^{(j)}) \,\mathcal{O}(x_1^{(j)},\,\bar x_1^{(j)})\right 
\rangle_{n}\,. 
\end{eqnarray}
In the second line, $Z^{(n)}(\mu)$ denotes the partition function of the deformed theory on the replica geometry.
Our approach now is to perform conformal perturbation theory in $\mu$. To this end we will expand in powers of the chemical potential to second order in $\mu$. Simultaneously, we take $\epsilon$, the width of the quench to be small. Moreover, for a general (integer) spin $s>2$, the chemical potential $\mu$ has dimension $2-s$ and is irrelevant by power counting.   
Therefore, it is important to spell out the order of limits and the regime of validity of conformal perturbation theory.  Since we want to make use of CFT OPEs for the quenching operator ${\cal O}$, we require that the operator separation i.e. the quench width $\epsilon \gg \mu^{1/(s-2)}$. Similar reasoning applies to the length scale set by the inverse temperature which must be much larger than $\mu^{1/(s-2)}$. Finally, since we are also performing a small width expansion, we must have $\epsilon \ll \beta$. Therefore, we work in a parameter range given by,
\be\label{limitorder}
\beta\,\gg\,\epsilon  \,\gg\, \mu^{\frac{1}{s-2}}\,.
\ee
In this regime of approximation we apply the OPE \eqref{ope1p} at  leading nontrivial order in $\epsilon$.  we expect that the R\'enyi entropies are determined by the expectation value  of the stress tensor of the parent CFT as in eq.\eqref{general}, but with the chemical potential deformation switched on.

Expanding the correlator ${\cal C}_n(\mu)$ in powers of the chemical potential,
\be
{\cal C}_n(\mu)\,=\,{\cal C}_n^{(0)}\,-\,\mu\,{\cal C}_n^{(1)}\,+\,
\mu^2\,{\cal C}_n^{(2)}\,+\ldots\,,
\ee
the zeroth order term is the CFT result which we have already calculated whilst the first and second order corrections can be interpreted in terms of the one- and two-point functions of the current $W$ in the quenched state on the replica geometry, including the effect of the $\mu$-dependence of the deformed partition function. The corrections are computed by CFT correlators integrated over the cylinder. The first putative correction, linear in $\mu$, vanishes identically. In the early time regime $t<l_1$, using the OPE \eqref{2nptearly} we obtain, 
\bea
{\cal C}_n^{(1)}&&=\,\sum_{i =1}^{n}  \int d^2 y\,  \left\langle  \left[W(y^{(i)} )\,-\,\langle W(y^{(i)})\rangle_n\right]\,\prod_{j=1}^n
\mathcal{O}^\dagger(x_4^{j},\,\bar x_4^{(j)}) \,\mathcal{O}(x_1^{(j)},\,\bar x_1^{(j)})\right \rangle_{n}\\\nonumber\\\nonumber
&&\simeq -\, 4 \epsilon^2 \frac{\Delta_{\cal O}}{c}\,\frac{|w'_1\,w'_4|^{n\Delta_{\cal O}}}{ |w_4-w_1|^{2n\Delta_{\cal O}}}\,\times\\\nonumber\\\nonumber
&&\times\,\int d^2 y\,\sum_{i=1}^{n} (v^{\prime(i)})^ s\,\left\langle W(v^{(i)})\,
\sum_{j=1}^{n}  \left[ (w'_1{}^{(j)})^2\,   T(w_1^{(j)} )   \, +\, 
(\bar w'_1{}^{(j)})^2\, \overline T(\bar w_1^{(j)})\right] \right\rangle_w\,.
\eea
We have made use of the early time OPE and the uniformising map \eqref{uniform_map} from the branched cylinder to the complex plane. The points of insertion of the stress tensor are at $w^{(j)}\,=\, w(x^{(j)})$, while the insertions of the higher spin current are at $v^{(i)}\,=\, w(y^{(i)})$. As usual, primes refer to derivatives with respect to the appropriate arguments and $w_{1}, \, w_4$ are the images of the points $x_1^{(1)}$ and $x_4^{(1)}$, on the first sheet of the branched cylinder where the operators ${\cal O}$ and ${\cal O}^\dagger$ are inserted. Since $W(z)$ is a primary field and, along with the stress tensor, has vanishing expectation value on the uniformized plane, all the following correlators must vanish:
\be
\langle W(v)  \rangle\, =\, \langle W(v)\, T(w)  \rangle\,  =\,\langle \overline T(\bar w) \rangle\, =\, 
\langle W(v)\, \overline T(\bar w)  \rangle\, =\, 
0\,.
\ee
Therefore the putative correction at order $\mu$ vanishes and ${\cal C}_n^{(1)}=0$. We can use the above procedure to write down the formal expression for the 
first nontrivial correction which appears at order $\mu^2$:
 \bea
&&{\cal C}_n^{(2)} \simeq\,
-\, 2 \epsilon^2 \frac{\Delta_{\cal O}}{c}\,\frac{|w'_1\,w'_4|^{n\Delta_{\cal O}}}{ |w_4-w_1|^{2n\Delta_{\cal O}}} \,\sum_{i, j =1}^{n}\int d^2 y_1\int d^2 y_2\left(v^{\prime\,(i)}_1\, v^{\prime \,(j)}_2\right)^s\,
\times
\\\nonumber\\\nonumber
 & &  \,
\left[\sum_{k=1}^{n} \left\{ (w'_1{}^{(k)})^2\,   
\left\langle T(w_1^{(k)} )\, W(v^{(i)}_1)\,  W(v^{(j)}_2)  \right\rangle +
(\bar w'_1{}^{(k)})^2 \,\left 
\langle  \overline T(\bar w_1^{(k)} )\, W(v^{(i)}_1)\,  W(v^{(j)}_2)\right\rangle\right\} \right]\,. 
 \eea
As in the previous case, the correlators are evaluated on the uniformized plane and transformed to the $n$-sheeted cylinder.
Now, we may explicitly write out the $\epsilon$-dependence at quadratic order for all quantities in the above expression. We find,
\begin{eqnarray}
\label{2nptmu}
& &{\cal C}_n^{(2)}\,\simeq\,\,-\, 2 \epsilon^2 \frac{\Delta_{\cal O}}{c} 
\left(2 \epsilon\right)^{-2n \Delta_{\cal O}} \left[1\,-\,\tfrac{1}{3} \epsilon^2 n 
\Delta_{\cal O}\, \left(\{w,x\}|_{x=0} +\{\bar w,\bar{x}\}|_{\bar x =0}\right) 
\right]
\times\\\nonumber\\\nonumber&&
\sum_{i, j =1}^{n}\int d^2 y_1\int d^2 y_2\,\left(v^{\prime\,(i)}_1\, v^{\prime \,(j)}_2\right)^s
\,\left[  
\sum_{k=1}^{n}  (w'_1{}^{(k)})^2   
\left\langle T(w_1^{(k)} ) W(v^{(i)}_1)  W(v^{(j)}_2)  \right\rangle 
 \right]\,.
\end{eqnarray}
Here we have used the fact that  $\langle \overline T (\bar w) W(v_1) W(v_2) \rangle$ vanishes
on the complex plane since $W$'s are holomorphic currents, and that disconnected contributions are removed by normalising with the partition function of the deformed theory.
 The remaining correlators involving the higher spin currents  are nontrivial and can be obtained by application of the uniformization map from the multi-sheeted, branched cylinder to the complex plane. This procedure was outlined in 
\cite{Datta:2014uxa} and similar ideas were discussed in \cite{Long:2014oxa}. We explain the derivation of these correlators in some detail in appendix \ref{app:uniform}, the method being applicable for any holomorphic current with a given value of the spin. The basic idea is to obtain the correlation function between current insertions on different sheets of the $n$-sheeted Riemann surface by applying the uniformization map, and to subsequently  sum over images of the insertions on all sheets. 

Once the two-point function of the currents is known, the three-point function involving the stress tensor follows from the application of conformal Ward identities on the cylinder. Notice that the required correlation functions cannot have any branch cuts since each operator is accompanied by a sum over all sheets (see e.g. eq.\eqref{2nptmu}). The two-point function of a spin-$s$ current in the replica geometry is a polynomial in the conformal cross-ratios, as defined in eq.\eqref{generaleta}. Calculation of the correlators and corresponding integrals is technically complicated and can only be done on a case-by-case basis. We will focus attention on the spin-three case ($s=3$) below. Prior to this we will make make some general observations and derive some specific results, valid for general $s$.

Recall that the computation of the R\'enyi entropy requires one other ingredient, namely a normalisation by the correlator of the quenching operator on the unbranched thermal cylinder, so that 
\be
\Delta S_A^{(n)}\,=\,\frac{1}{1-n}\,\ln\left[\frac{{\cal C}_n(\mu)}{{\cal C}_1(\mu)^n}\right]\,.\label{renyidef}
\ee
From the formal expressions above, and the CFT OPEs in the limit \eqref{limitorder} for early times $t<l_1$, it is clear that the leading term  is the universal CFT result at order $\epsilon^2$, and the first correction of interest will appear at order $\mu^2\epsilon^2$:
\be
\Delta S_A^{(n)}\,=\,\epsilon^2\, \Delta S_A^{(n;\,2,\,0)}
\,+\,\mu^2\epsilon^2 \,\Delta S_A^{(n;\,2,\,2)}\,+\ldots
\ee
We use the superscript labels to identify  terms at a given order in the double expansion in $\mu$ and $\epsilon$.
The universal CFT result is given by
\be
 \Delta S_A^{(n;\,2,\,0)}\,=\,{\Delta_{\cal O}}\,\frac{1+n}{6n}\,\left[{\cal S}_{l_1 l_2}(t)^2\,+\,{\cal S}_{l_1 l_2}(-t)^2\right]\,.
\ee
We wish to calculate the  first non-trivial time dependent correction  arises at order $\mu^2\epsilon^2$.

To obtain the R\'{e}nyi entropy 
in the intermediate time domain $l_1 < t<l_2$, we need to use the OPE of insertions of the operator  ${\cal O}$ between 
neighbouring replica slices as discussed in section \ref{setup}. 
From (\ref{crosschan}) we see that the only difference in taking the OPE in this 
channel is the presence of the factor $F_{00}^{(n-1)}$ in the $2n$-point function, where
$F_{00}$ is the fusion matrix element corresponding to  the operator ${\cal O}$. 
Therefore for times $l_1 < t<l_2$, we get,
\begin{equation}
 \Delta S_A^{(n)} 
  \,=\,  - \log F_{00} \,+\, 
  \epsilon^2 \,\Delta S^{(n; \,2,\,0)}_A(t)\, 
  +\, \mu^2 \epsilon^2\, \Delta S^{(n;\, 2,\, 2)} _A( t)\,+ \cdots
\end{equation}

\subsection{Late time behaviour at ${O}(\epsilon^2\mu^2)$ for general $s$}
Determining the exact time dependence of the correction $\Delta S^{(n;\,2,\,2)}_A$ requires a technically involved, lengthy computation. However, we can easily obtain an exact result in the  infinite interval limit, at late times and for general values of the spin $s$. For this, we 
recall the general expression \eqref{general} which follows from the conformal OPE of the quenching operators, and which also determines the first finite width correction at order $\mu^2$. In particular, we may write
\bea
\Delta S^{(n;\,2,\,2)}_A &&=\,2\frac{\Delta_{\cal O}}{c}\,\frac{n}{n-1}\times\\\nonumber
&&\int d^2 y_1 d^2 y_2\left[\left\langle T(0) W(y_1) W(y_2)\right\rangle_n\,-\,\left\langle T(0) W(y_1) W(y_2\right)\rangle_\beta\right]\,+\,(t\to -t)\,.
\eea
The first term in the integrand is the correlator $\langle TWW\rangle $ on the branched cylinder, with an implicit sum  over all sheets. In the limit of large interval length, followed by a late time approximation, the correlator on the $n$-sheeted cover of the cylinder reduces to a thermal correlator at inverse temperature $n\beta$:
\be
\left\langle T(0) W(y_1) W(y_2)\right\rangle_n\left.\right|_{l_2\to\infty,\,t\gg l_1}\,\to\,\langle T(0) W(y_1) W(y_2)\rangle_{n\beta}\,.
\ee
This limit will allow us to calculate the correction to the saturation value of the R\'enyi entropies after the excitation enters the interval $A$. Up to a normalisation factor, the $WW$ correlator on the thermal cylinder is
\be\label{wwnorm}
\langle W(y_1)\,W(y_2)\rangle_\beta\,=\,\frac{\widetilde{\cal N}\,\pi^{2s}}{\beta^{2s}\,\left[{{\sinh}\frac{\pi}{\beta}(y_1-y_2)}\right]^{2s}}\,,
\ee
where $\widetilde {\cal N}$ is a normalisation factor which scales with the central charge  of the theory, e.g. in \cite{Datta:2014ska} where $\widetilde {\cal N}\,=\, 5c/6\pi^2$.
Next, we apply the conformal Ward identities on the cylinder:
\bea
\langle T(z) \,W(y_1)\, W(y_2)\rangle_\beta\,&&=\,\\\nonumber
\sum_{i=1,2}&&\left[\frac{\pi^2}{\beta^2}\frac{s}{\sinh^2\frac{\pi}{\beta}(z-y_i)} \,+\,\frac{\pi}{\beta}\,{\coth\tfrac{\pi}{\beta}(z-y_i)}\,\,\partial_{y_i}\right]
\langle W(y_1)\, W(y_2)\rangle_\beta\,.
\eea
To integrate the resulting holomorphic function over the complex $y_1$- and $y_2$-planes, we carefully follow the prescription laid out in the appendix of 
\cite{Datta:2014ska}. Note that the integrals are ambiguous and require a well-defined prescription for dealing with the contact term singularities (see also \cite{Datta:2014zpa}) .  The result of each integration is completely determined by the double pole and simple pole  singularities of the integrand. We find,
\bea
&&\int d^2 y_1\int d^2 y_2\,\langle T(z) \,W(y_1)\, W(y_2)\rangle_\beta\,=\,4\,\widetilde{\cal N}\, \frac{\pi^{2s}}{\beta^{2s-2}}\,(2s-1){\cal R}(s)\,,\\\nonumber\\\nonumber
&&{\cal R}(s)\,=\,-(-1)^s \,\frac{\Gamma(s)\,\Gamma\left(\tfrac{3}{2}\right)}{\Gamma\left(s+\tfrac{1}{2}\right)}\,.
\eea
The function ${\cal R}(s)$ determines the coefficient of the double pole singularity in the thermal correlator $\langle WW\rangle_\beta$.
Finally, the asymptotic value of the change in the R\'enyi entropies at order $\epsilon^2\mu^2$, for general $s$, is given by  
\be
\Delta S_A^{(n;\,2,\,2)}\left.\right|_{l_2\to \infty,\,t\gg l_1}\,=\,8\,\Delta_{\cal O}\left(\frac{\widetilde{\cal N}}{c}\right)\left(\frac{\pi^{2s}}{\beta^{2s-2}}\right)\frac{n(n^{2-2s}-1)}{{(n-1)}}\,(2s-1)\,{\cal R}(s)\,.\label{asymp-general}
\ee
The correction is independent of the central charge since the normalisation $\widetilde {\cal N}$ is proportional to $c$.
It is interesting to note that in the large $s$ limit, using Stirling's approximation for the gamma-functions, we find ${\cal R}(s)\sim (-1)^{s+1} \,\sqrt{\pi/s}$ and the entanglement entropy ($n\to1$) scales as $\sim s^{3/2}$.
 Below, we will see that the late time value of the correction for $s=3$ is reproduced by the asymptotics of the complete time dependent result, obtained after performing explicit integrations at finite time and finite interval length.
\subsection{Order $\epsilon^2\mu^2$ correction for spin three chemical potential}
Let us finally turn to the calculation of the integrals that determine the time dependent corrections in the presence of the spin-three deformation. The relevant set of correlators are evaluated in appendix \ref{appen-a}, by  transforming from the uniformised plane to the multi-sheeted cylinder. The two-point function of spin-three currents was already obtained in \cite{Datta:2014ska}:
\begin{eqnarray} \label{gen-cyl-corr}
&&\sum_{i, j =1}^{n}  (v^{(i)\prime}_1 v^{(i)\prime}_2)^ 3 
 \left\langle W(v^{(j)}_1)  W(v^{(j)}_2)\right\rangle\, =\\\nonumber
 &&\qquad\qquad\qquad\qquad \widetilde{\cal N}\left[n\, H^6(y_1-y_2)\,+\, \frac{\left(n^2-1\right)}{4n}\, \hat{I}_1\,+\,
 \frac{\left(n^2-1\right)\left(n^2-4\right)}{120n^3}\,I_2\right]\,.
 \eea
 The first of these terms is a disconnected contribution to the correlator and is simply a thermal correction (rather than the effect of entanglement). The second and third terms, $\hat I_1$ and $I_2$ control the higher spin corrections to the entanglement entropy and are defined in eq.\eqref{def-integrals}. The three-point function involving the stress tensor, which determines the finite width correction at order $\epsilon^2\mu^2$, is obtained by application of conformal Ward identities,
 \bea 
\sum_{i , j,k=1 }^{n} (w_1^{\prime\,(k)})^2 &&  \,(v^{\prime\,(i)}_1\, v^{\prime\,(i)}_2)^ 3\,   
\left\langle T(w_1^{(p)} )\, W(v^{(i)}_1) \, W(v^{(j)}_2)  \right\rangle\, = \, \\\nonumber\\\nonumber
&&{\widetilde {\cal N}}\,\left[3n \, I_1 \,+\, 
\frac{(n^2-1)}{4n}\, (2I_3+I_4)\, +\, \frac{\left(n^2-1\right)\left(n^2-4\right)}{120n^3} \,(I_5+2I_6) \right]\,.
\end{eqnarray}
The expressions for the holomorphic correlators  $I_1$, $I_3$, $I_4$ and $I_5$, containing the dependence on   
insertions of the higher spin current $W$, are listed in eq.\eqref{def-integrals}.  The first of these, namely $I_1$, gets cancelled off by an identical term from the normalisation ${\cal C}_1(\mu)^n$ in the definition of the R\'enyi entropies \eqref{renyidef}. Below, we use the symbol $\mathcal{I}_k$ to denote the integral over the  positions of the  
insertions of higher spin currents in a given 
holomorphic integrand $I_k$: 
\begin{align} \label{def-integral}
\mathcal{I}_k(t;\,l_1,\,l_2) \,=\, \int_{\mathbb R \times S^1_\beta } \int_{\mathbb R \times S^1_\beta } d^2y_1\, d^2 y_2 \, I_k(y_1,\,y_2;\,l_1,\,l_2)\,. 
\end{align} 
As pointed out earlier, these integrations of holomorphic functions over the complex plane/cylinder are performed using the prescriptions first put forward in \cite{Datta:2014ska} and 
developed further in \cite{Datta:2014zpa}. The result of the procedure  is listed  in appendix  \ref{fullint}, with the  time-dependent correction of interest taking the form, 
\be
 \Delta S^{(n; 2,2)}_A(t) \,=\,{\Delta_{\cal O}}\, \left(\frac{\widetilde{\cal N}}{c}\right) 
 \,\frac{(n+1)}{2n}  \left[ \left(2\,\mathcal{I}_3\,+\,{\cal I}_4\right)\, + \,\frac{
 \left(n^2-4\right)}{30n^2}\, \left(\mathcal{I}_5\,+\,2\,{\cal I}_6\right)\right]\,.\label{mu2full}
\ee
The actual functions, as listed in appendix \ref{fullint}, are involved and not particularly transparent. 
\begin{figure}
\centering
\includegraphics[scale=1.2]{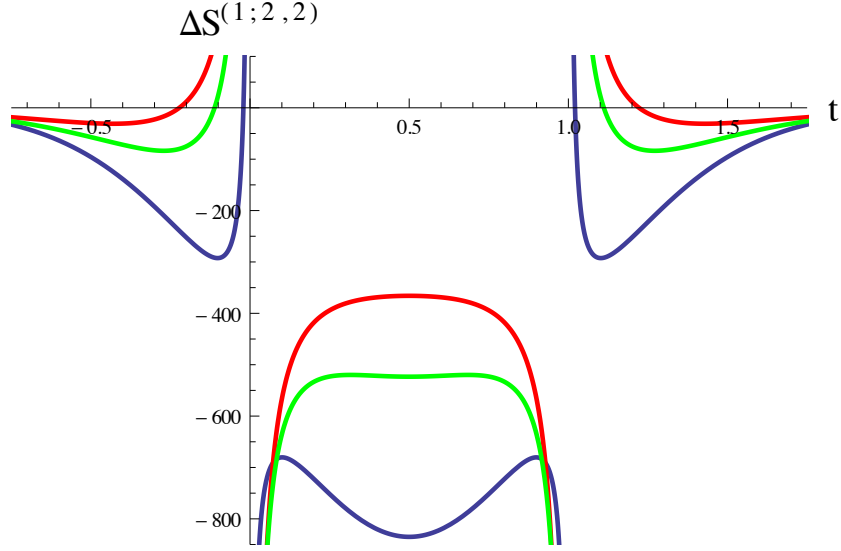}
\caption{Plot of $S_{(0, 1)}^{(1; 2, 2)}$, 
the $\mu^2 \epsilon^2$ correction to 
entanglement entropy as a function of time for the entanglement interval, $l_1=0$, $l_2=1$. 
The plots from bottom to top are for increasing values of $\beta = 1.8, \, 2.1, \, 2.4$, respectively.}
\label{Intro2}
\end{figure}
In order to convey the qualitative nature of the contribution, we have plotted the time evolution of the entanglement entropy (the limit $n\to1$)  in figure (\ref{Intro2}) for an interval of unit length,  at different values of the temperature. 
The divergences at the two  end-points of the interval are of the form $\sim 1/t$ and $ \sim 1/(t - 1)$
respectively. This behaviour is distinct from the order $\epsilon^2$ correction in the CFT which diverges at the end-points as $1/(t-l_{1,2})^2$.
 
 A relatively compact expression  for $\Delta S^{(n; \,2,\,2)}_A(t)$ emerges when the length
 of the interval is taken to be infinite i.e. in the limit $l_2\to\infty$.  This is because the integrals simplify considerably as shown in eq.\eqref{intlargel}. 
 The result for the semi-infinite interval, at all times $t$ is given by, 
\bea
&&\mu^2 \epsilon^2 \,\Delta S^{(n; \,2,\,2)}_A\left.\right|_{l_1=0,\,l_2\to\infty}\,=\,\mu^2\epsilon^2\,{\Delta_{\cal O}}\,\frac{\widetilde {\cal N}}{c}\,\frac{\pi^6}{\beta^4}\,\frac{(n +1)}{2n}\,\frac{1}{(1-e^{-2\pi t/\beta})^6}\times\nonumber\\\nonumber\\\nonumber
&&\left[-\tfrac{160}{3}\,-\,\tfrac{128}{3}\,\left(\tfrac{2\pi t}{\beta} \,-\,11\right)\,e^{-2\pi t/\beta}\,+\,64\,\left(\tfrac{16\pi^2 t^2}{\beta^2}\,-\,\tfrac{32\pi t}{\beta}\,-\,7\right)\,e^{-4\pi t/\beta}\,+ \right.\\\nonumber\\
&&\left.+\,\tfrac{128}{3}\,\left(\tfrac{36\pi^2 t^2}{\beta^2}\,+\,\tfrac{42\pi t}{\beta}\,-\,7\right)\,e^{-6\pi t/\beta}\,+\,\tfrac{32}{3}\,\left(\tfrac{32\pi t}{\beta}\,+\,31\right)\,e^{-8\pi t/\beta}\,+\right.\label{largel}
\\\nonumber\\\nonumber
&&\left.+\,\frac{\left(n^2-4\right)}{30n^2}\,\left\{320 \,+\,256\,\left(\tfrac{4\pi^2 t^2}{\beta^2}\,-\,\tfrac{14\pi t}{\beta}\,+\,1\right)\,e^{-2\pi t/\beta}\,
+\right.\right.\\\nonumber\\\nonumber
&&\left.\left.\,128\,\left(\tfrac{32\pi^2 t^2}{\beta^2}\,+\,\tfrac{16\pi t}{\beta}\,-\,11\right)\,e^{-4\pi t/\beta}\,+\,768\left(\tfrac{2\pi t}{\beta}\,+\,1\right)\,e^{-6\pi t/\beta}\,+\,64\,e^{-8\pi t/\beta}\right\}\right]\,.
\end{eqnarray}
The $n$-dependence of the terms above has been arranged so that one may immediately read off the values of the integrals $(2{\cal I}_2 + {\cal I}_3)$ and $({\cal I}_5 + 2{\cal I}_6)$, appearing in eq.\eqref{mu2full}.
The correction vanishes exponentially in the far past $(t<0)$, and for large positive $t$, approaches a constant negative 
value:
\be
\Delta S_A^{(n;\,2,\,2)}\left.\right|_{l_2\to \infty,\,t\to\infty}\,=\,-\,{\Delta_{\cal O}}\,\frac{\widetilde {\cal N}}{c}\,\frac{64\,(n^2+1)(n+1)}{3n^3}\frac{\pi^6}{\beta^4}\,,
\ee 
precisely matching the general prediction \eqref{asymp-general} when $s=3$.  This constitutes a check of our computation of the integrals leading to the full time dependent result. 
Although one may naively expect the expression in eq.(\ref{largel}) to have a pole at $t=0$ of order
6, careful analysis reveals that the pole at $t=0$ is in fact a simple pole. Another curious feature of the asymptotic value of the correction (\ref{largel})  for general $s$, is its alternating sign between even and odd values of $s$.
 It would be extremely interesting to reproduce these features within an appropriate holographic setup.

\subsection{Comparison with ${\cal W}_{1+\infty}$ or free fermion theory}
The free fermion theory is the simplest CFT admitting higher spin currents and ${\cal W}$-symmetry. The presence of the $U(1)$ current in the ${\cal W}_{1+\infty}$ algebra is a potential source for complication, but it is possible to isolate and extract the contributions of higher spin currents  to various physical observables (see \cite{Datta:2014ska, Datta:2014zpa} for related discussions).
Exact results for the chemical potential deformation of the correlator of equation 
\eqref{2nptmu} can be obtained for the $W_{1+\infty}$ theory, in bosonised language. 
We calculate the $O(\mu^2)$ correction to the correlation function of the quenching operator defined in equation \eqref{def_expop} where the (holomorphic) spin-three current takes the form,
\begin{align} \label{spin3}
W(x)\, =\,\frac{i}{\sqrt{6}} \left(\partial \varphi(x) \right)^3\,.
\end{align}
The normalisation constant is fixed by requiring the two-point function of the current to be normalised to unity.  In order to compare with the general results of section (\ref{sec4-gen-arg})  we will need to 
subtract out the contributions from the $U(1)$ current $J$ in the ${\cal W}_{1+\infty}$ algebra. 
Furthermore, we note that the spin-three current of this theory, defined in eq.\eqref{spin3},
is quasi-primary. Under a conformal transformation the current transforms as,
\begin{align} \label{Wtransf}
W(x)\, =\, w'(x)^3\, W(w)\, -\, \frac{1}{2} w'(x) \,\lbrace w(x),x \rbrace\, J(w)\,.
\end{align}
The $U(1)$ current $J$ is defined as
\begin{equation}
 J(x)\, =\, i \partial\varphi(x)\,, 
\end{equation}
in the bosonized description.
We will now evaluate the spin-3 correction to the second R\'{e}nyi  entropy 
for the excited state created by the operator 
given in (\ref{def_expop}). 
The $\epsilon^2 \mu^2$ correction  follows from the four-point function on the two-sheeted cylinder which we obtain as before by employing the uniformization map \eqref{uniform_map}: 
\bea\label{OOWWOO}
&&\sum_{i,j=1}^2  \left\langle \mathcal{O}(x_2^{(1)})\,\mathcal{O}(x_1^{(1)})\,W(y_1^{(i)})\,W(y_2^{(j)})\,\mathcal{O^\dagger}(x_4^{(2)})\,\mathcal{O^\dagger}(x_4^{(1)})\right\rangle_n \\\nonumber \\
&& =\, \left|w'(x_1^{(2)})\,w'(x_1^{(1)})\,w'(x_4^{(2)})\,w'(x_4^{(1)})\right| \,\sum_{i,j=1}^2\left[ w'(y_1^{(i)})^3 \,w'(y_2^{(j)})^3\,\,\times \right.\nonumber \\
&&\left.  \left\langle \mathcal{O}(w_1)\,\mathcal{O}(-w_1)\,W(v_1^{(i)})\,W(v_2^{(j)})\,\mathcal{O^\dagger}(w_4)\,\mathcal{O^\dagger}(-w_4)\right\rangle_w\,-\, \frac{1}{2} \left( w'(y_1^{(i)})^3\, w'(y_2^{(j)})\times\,\right.\right. \nonumber\\\nonumber \\
&&\left.  \left\{ w(y_2^{(j)}),\,y_2^{(j)} \right\}\,\left \langle \mathcal{O}(w_1)\,\mathcal{O}(-w_1)\,W(v_1^{(i)})\,J(v_2^{(j)})\,\mathcal{O^\dagger}(w_4)\mathcal{O^\dagger}(-w_4)\right\rangle_w\,  +\, y_1 \leftrightarrow y_2 \right) \nonumber \\\nonumber \\ 
&& +\, \frac{1}{4} \,w'(y_1^{(i)}) \,\left\{ w(y_1^{(i)}),\,y_1^{(i)} \right\}\, w'(y_2^{(j)})\, \left\{ w(y_2^{(j)}),\,y_2^{(j)} \right\}\,
\times\nonumber \\\nonumber \\ 
&& \left.\left\langle \mathcal{O}(w_1)\,\mathcal{O}(-w_1)\,J(v_1^{(i)})\,J(v_2^{(j)})\mathcal{O^\dagger}(w_4)\,\mathcal{O^\dagger}(-w_4)\right\rangle_w \right]\,.\nonumber
\eea
We have used the transformation property of  the spin-three current given in  eq.(\ref{Wtransf}). 
Our general argument for the time dependence of the
$\mu^2\epsilon^2$ correction was made for a theory which admits higher spin 
currents but without a 
 $U(1)$ symmetry. For a finite value of the width $\epsilon$,  the above correlator in \eqref{OOWWOO} can be determined by performing Wick contractions which lead to the result \eqref{Wperturbfinal}. 
Expanding \eqref{Wperturbfinal} for small $\epsilon$  and keeping only the connected contributions we find that the corresponding correction to the second R\'enyi entropy is,
\bea \label{2n-W-fermion}
\Delta S_A^{(2;\,2,\,2)}\,=\, 12 \mu^2 \epsilon^2 \alpha^2 \begin{cases}
\left(\frac{1}{8}\,{\cal I}_3\,+\,\frac{2|\kappa|^2}{8(1+|\kappa|^2)^2}\,{\cal I} _4\right)\qquad & t<l_1\quad {\rm or} \quad t>l_2\,,\\\\
\frac{1}{8}\frac{(1+|\kappa|^4)}{(1+|\kappa|^2)^2}\,{\cal I}_3 & l_1<t<l_2\,.
\end{cases}
\eea
This is clearly not in agreement with the general formula \eqref{mu2full}\footnote{Here the correction term does exhibit a jump, but that it is an effect due to the U(1) current.}. But there is a way to identify the $U(1)$ contributions and subtract them off from  the \eqref{2n-W-fermion} and show that the result reduces to 
eq.\eqref{mu2full} for $n=2$.  Note also that, precisely when $n=2$, one set of terms in \eqref{mu2full} vanishes.

In order to extract the contribution from the current $J$, we examine the  OPE (\ref{OOOPE}) which determines the four-point correlator relevant for the correction to the second R\'enyi entropy at order $\mu^2\epsilon^2$. As usual, this is achieved by transforming from the uniformized plane to the two-sheeted,  branched cylinder, keeping only connected contributions. For the early time regime $t< l_1$ we get,
\bea \label{opewwz1}
&& \frac{\mu^2}{2} \sum_{i,j=1}^2  (v_1^{\prime\,(i)})^3 \,(v_2^{\prime\,(j)})^3 \,\left \langle \mathcal{O}(w_1)\,\mathcal{O}(-w_1)\,W(v_1^{(i)})\,W(v_2^{(j)})\,\mathcal{O^\dagger}(w_4)\,\mathcal{O^\dagger}(-w_4)\right\rangle_w\\
&& =\,\frac{\mu^2}{2} \,\left|w_1-w_4\right|^{-4 \alpha^2}\,  (1+|\kappa|^2)^2 \,
\sum_{i,j=1}^2  (v_1^{\prime\,(i)})^3 \,(v_2^{\prime\,(j)})^3\,\times  
\nonumber\\\nonumber\\
&&\left(\,2 \epsilon^2 \alpha^2\, (w'_1)^2\,\left[\left\langle (\partial \varphi(w_1))^2\, W(v_1^{(i)})\,W(v_2^{(j)}) \right\rangle_w\,  +\, \left\langle (\partial \varphi(-w_1))^2\, W(v_1^{(i)})\,W(v_2^{(j)})\right\rangle_w\right] \right.\nonumber\\\nonumber\\
&&\left. -\, 4 \epsilon^2 \alpha^2\,  (w'_1)^2 \,\left(\frac{1-|\kappa|^2}{ 1+ |\kappa|^2}\right)^2\left\langle \partial \varphi(w_1) \partial \varphi(-w_1)\, W(v_1^{(i)})\,W(v_2^{(j)})\right\rangle_w\right)\,.   \nonumber 
\eea
Correlators of other operators which appear in the OPE above vanish.
This contains contributions from $\langle JJWW \rangle$  in addition to $\langle TWW \rangle$  which alone should lead to eq.\eqref{mu2full}. Hence, eq.\eqref{2n-W-fermion}  in its present form cannot be compared with the universal result of the previous section. 
From (\ref{opewwz1}) we see that the contribution from
the $U(1)$ current can be subtracted off by setting $|\kappa|=1$. Therefore, we should recover the universal result by setting $|\kappa|=1$ for $t< l_1$ in eq.\eqref{2n-W-fermion}. Indeed, this yields
\bea
  \Delta S_A^{(2;\,2,\,2)}\left.\right|_{|\kappa|=1}\,=\,   \frac{3}{4} \mu^2 \epsilon^2 \alpha^2 \,( 2\, {\cal I}_3\,  +\, {\cal I}_4 )\,, 
\eea
agreeing precisely with  eq.\eqref{mu2full}, for $n=2$. 
The analysis can be repeated for the time interval $l_1<t<l_2$ by using the OPEs appropriate for this regime. 
\section{Conclusions}
In this paper we have focussed our attention on a universal aspect of finite width local quenches generated by primary fields in CFTs in 1+1 dimensions and chiral deformations thereof. The effects described above apply to theories with both fixed and large central charges, and capture a particular growth and relaxational regime for entanglement entropies, immediately preceding and following the step function jump \cite{He:2014mwa} in a zero width local quench. The time dependence of  the finite width effects we have discussed strongly indicate that the step function jump (at zero width) is smoothed out such that it must be accompanied by an ``overshoot''. This ``overshoot'' effect can clearly be seen in the exact minimal model computations, illustrated in figure \ref{min_pos}. It is certainly interesting to ask whether the overshoot and its peak value encode interesting information regarding the CFT and/or the local operator generating the quench, just as the magnitude of the jump at zero width is given by the quantum dimension of the operator in question. A related question concerns the status of the jump for large-$c$ CFTs and their holographic duals wherein the quantum dimension diverges and is expected to be rendered finite by $1/c$ and/or non-perturbative effects. While the universal finite width effect we have discussed appears in the large-$c$ results for  RE/EE prior to the onset of the jump in this quantity, the overshoot and subsequent relaxation to  the value set by the quantum dimension are not visible in the strict large-$c$ limit. It would be interesting to understand the precise mechanism for its resolution.

A significant motivation for this work was to understand the effect of CFT deformations on the zero-width jump and on finite width effects. We focussed attention on chiral deformations, i.e. deformations by holomorphic currents because the resulting corrections are analytically tractable within the framework of conformal perturbation theory. We have seen that to lowest nontrivial order in the deformation parameter or chemical potential $\mu$, the zero width jump remains unaffected. This is in line with related observations in \cite{Chen:2015usa}. It would certainly be of interest to understand what happens to the size of the jump at higher orders in the deformation and perhaps even non-perturbatively, for appropriately chosen deformations. This brings us to the second crucial reason for discussing deformations of CFTs by holomorphic currents. CFTs endowed with such currents exhibit ${\cal W}$-symmetry and are dual to higher spin gravity on AdS$_3$. A major motivation of this work was to identify universal CFT predictions for local quenches in such situations, which can then be examined and studied using holographic dual descriptions. Much is now understood about holographic entanglement entropies in the higher spin setting \cite{deBoer:2013vca, Ammon:2013hba, Hegde:2015dqh, Chen:2016uvu}. The main outstanding question  in this context is whether a holographic description of the local quench can be made precise in the higher spin dual to a CFT with ${\cal W}$-symmetry, both with and without chemical potential deformations. We have been able to obtain the exact time dependence of RE/EE corrections at finite width in the case of the spin-three deformation and late time asymptotics for generic higher spin chemical potentials. It would be extremely interesting to verify these universal statements against corresponding holographic calculations.

\acknowledgments We would like to thank Shouvik Datta for   comments and suggestions 
on a previous version of this manuscript.  
SPK acknowledges financial support from STFC grant award ST/L000369/1.
\appendix
\section{Higher spin correlators on replica geometry} \label{app:uniform}
An $(h,0)$ operator on the complex $w$-plane  has the two-point correlator
\be
\langle W(w_1)W(w_2) \rangle\,=\,\frac{1}{(w_1-w_2)^{2h}}\,.
\ee
We avoid introducing unnecessary normalisation factors for the sake of clarity.
We consider the uniformizing map
\be
w\,=\,\left(\frac{z-a}{z-b}\right)^{1/n}\,,
\ee
from the $n$-sheeted Reimann surface, with branch points at $z=a$ and 
$z=b$, to the complex $w$-plane. Therefore
\be
\langle W(z_1)W(z_2) \rangle\,=\,w'(z_1)^{h}\,w'(z_2)^h\,\frac{1}{(w(z_1)-w(z_2))^{2h}}\,.
\ee
Now we use
\be
w'(z)\,=\,\frac{1}{n}\,w(z)\,\frac{(a-b)}{(z-a)(z-b)}\,,
\ee
to simplify the above expression so that
\bea
\langle W(z_1)W(z_2) \rangle\,=\,\frac{1}{n^{2h}}\,\frac{1}{(z_1-z_2)^{2h}}\,\frac{\left(\sqrt{x}-\frac{1}{\sqrt x}\right)^{2h}}{\left(x^{1/2n}-\frac{1}{x^{1/2n}}\right)^{2h}}\,,
\eea
where we have introduced the cross-ratio,
\be
x=\frac{(z_2-a)(z_1-b)}{(z_2-b)(z_1-a)}\,.
\ee
All non-trivial dependence on $n$ is therefore contained in the function of cross-ratios:
\bea
&&F_h(x)\,=\,\frac{1}{n^{2h}}\,\frac{\left(\sqrt{x}-\frac{1}{\sqrt x}\right)^{2h}}{\left(x^{1/2n}-\frac{1}{x^{1/2n}}\right)^{2h}}\,=\,\frac{1}{n^{2h}}\,\left(\sum_{k=-(n-1)/2}^{(n-1)/2}x^{k/n}\right)^{2h}\,.
\eea
This is a polynomial in the variable $x^{1/n}$ as is evident upon rewriting,
\be
F_h(x)\,=\,\frac{1}{n^{2h}}\,x^{h(1-n)/n}\,\left(\sum_{k=0}^{n-1}x^{k/n}\right)^{2h}\,.
\ee
We are only interested in correlators where the operators $W$ are introduced in each sheet and therefore a sum over images on different sheets is implied. Moving to a different sheet is achieved  by multiplying  $x$ by a phase factor, $x\to x e^{2\pi i m}$ for $m=0,1,\ldots (n-1)$. We therefore need to consider the sum
\be
\sum_{i,j=1}^n\langle W(z_1^{(i)})\,W(z_2^{(j)})\rangle\,=\,\frac{n}{(z_1-z_2)^{2h}}\sum_{m=0}^{n-1} F_h(x \,e^{2\pi i m})\,.
\ee
This eliminates all terms in the polynomial (in $x^{1/n}$) above that involve fractional powers of $x$; only integer powers of $x$ survive.

\subsection{Connected $\langle TT \rangle$ correlator  ($h=2$)}
Let us look at the example with $h=2$ corresponding to the stress tensor.  
The required coefficients of integer powers of $x$ in $F_h(x)$ are 
\bea
&& x^2\,:\qquad 0\\\nonumber
&&x\,:\qquad \frac{(n^2-1)}{6\,n^2}\\\nonumber
&&x^0\,:\qquad 1 \,-\,2\frac{(n^2-1)}{6\,n^2}\,,\\\nonumber
&& x^{-1}\,: \qquad \frac{(n^2-1)}{6\,n^2}\,.
\eea
The highest integer powers (for any choice of $n$) are $x$ and $1/x$.
Therefore the connected contribution to the stress-tensor two-point function is 
\bea
\sum_{i,j=1}^n\langle T(z_1^{(i)})\,T(z_2^{(j)})\rangle\,=\,\frac{1}{(z_1-z_2)^{4}}\left[n\,+\,\left(x+\frac{1}{x} -2\right)\frac{(n^2-1)}{6\,n}\right]\,.
\eea
\subsection{ Spin-3 correlator $\langle WW \rangle$ with $(h=3)$ }
Now, the highest possible integer power in $F_h(x)$ is $x^2$.
We find the following values for the coefficients of the integer powers of $x$:
\bea
&& x^2\,:\qquad \frac{(n^2-1)(n^2-4)}{120\,n^4}\\\nonumber
&&x\,:\qquad \frac{(n^2-1)}{4\,n^2} \,-4\,\frac{(n^2-1)(n^2-4)}{120\,n^4}\\\nonumber
&&x^0\,:\qquad 1\,-\,2\frac{(n^2-1)}{4\,n^2}\, +\, 6 \frac{(n^2-1)(n^2-4)}{120\,n^4}\,.
\eea
Since the $x \to 1/x$ symmetry is inbuilt, we find (after summing over all   copies)
\bea
\sum_{i,j=1}^n\langle W(z_1^{(i)})\,W(z_2^{(j)})\rangle&&=\,\frac{1}{(z_1-z_2)^{6}}\left[ n\,+\,\eta\frac{(n^2-1)}{4\,n}\,+\,\eta^2\frac{(n^2-1)(n^2-4)}{120\,n^3}\right]\,,\nonumber\\
 \eta\,&&=\,x\,+\,\frac{1}{x}\,-\,2\,.
\eea
\subsection{ Result for general $h$}
The  general result for the two point correlator of two holomorphic operators of weight $h$ on the $n$-sheeted surface takes the form (after summing over all copies):
\bea
\langle W_h(z_1)\,W_h(z_2)\rangle\,=\,\frac{1}{(z_1-z_2)^{2h}} \sum_{n=0}^{h-1}\,f_k\,\eta^k\,.\label{generaleta}
\eea
The coefficients $\{f_k\}$ can be determined on a case-by-case basis. A general form valid for all $h$ and $n$ can also be obtained with a little effort. For example, the coefficient of the highest power of $\eta$ is:
\be
f_{h-1}\,=\,\frac{1}{(2h-1)!\,n^{2h-3}}\,\prod_{k=1}^{h-1}(n^2-k^2)\,.
\ee

\section{Correlators for the correction at  $O(\mu^2\epsilon^2)$} \label{appen-a}

In this section, we derive the correlators used in section (\ref{sec4-gen-arg}). Let us consider the following correlator of two spin-$3$ currents on an $n$-sheeted branched cover of the plane with branch points at $z=z_2$ and $z=z_3$,
\begin{align} \label{WW-nplane}
\sum_{i,j=0}^{n-1} \langle W(z_5^{(i)}) W(z_6^{(j)}) \rangle_n = \frac{1}{(z_5-z_6)^6} \left(n+\eta \frac{\left(n^2-1\right)}{4n}+\eta^2\frac{\left(n^2-1\right)\left(n^2-4\right)}{120n^3} \right),
\end{align}
where $\eta$ and the cross-ratio $x$, are defined as,
\begin{align}
\eta \, =\, x + \frac{1}{x}-2\,,\qquad\qquad
x  = \frac{(z_2-z_5)(z_3-z_6)}{(z_2-z_6)(z_3-z_5)}\,.
\end{align}
To obtain the correlator of the stress tensor with two spin-$3$ currents on the $n$-sheeted plane, we apply the conformal Ward identity on equation (\ref{WW-nplane}),
\begin{align}\label{app-TWWn}
& \sum_{i,j, k=1}^{n} \langle T(z_1^{(k)}) W(z_5^{(i)}) W(z_6^{(j)})\rangle_n\\
= & (z_2-z_3)^{2h_\sigma}\bigg(\frac{h_\sigma}{(z_1-z_2)^2} +\frac{1}{(z_1-z_3)} \partial_{z_2}+\frac{h_\sigma}{(z_1-z_3)^2} +\frac{1}{(z_1-z_3)} \partial_{z_3}+\frac{3}{(z_1-z_5)^2} \nonumber\\
& +\frac{1}{(z_1-z_5)} \partial_{z_5}+\frac{3}{(z_1-z_6)^2} +\frac{1}{(z_1-z_6)} \partial_{z_6}\bigg)\frac{1}{(z_2-z_3)^{2h_\sigma}} \sum_{i,j=0}^{n-1} \langle W(z_5^{(i)}) W(z_6^{(j)}) \rangle_n \nonumber\\
= &  \bigg[ \frac{2n^2}{n^2-1} \frac{h_\sigma}{(z_5-z_6)^6}   \{ w(z_1),z_1 \} + \frac{3}{(z_1-z_5)^2(z_1-z_6)^2(z_5-z_6)^4} \nonumber\\ 
& + \frac{1}{(z_5-z_6)^6} \left(\frac{1}{(z_1-z_2)} \partial_{z_2} +\frac{1}{(z_1-z_3)} \partial_{z_3}+\frac{1}{(z_1-z_5)} \partial_{z_5} +\frac{1}{(z_1-z_6)} \partial_{z_6}\right) \bigg] f(\eta),\nonumber
\end{align}
Here $h_\sigma = \frac{c}{24}\left(n-\frac{1}{n} \right)$ is the conformal dimension of the twist field which produces the branch cut.  $\{w(z),z\}$ is the Schwarzian for the map from the $n$-sheeted cover of the plane to the uniformized $w$-plane:
\be
w(z)\,=\,\left(\frac{z-z_2}{z-z_3}\right)^{1/n}\,.
\ee
The first two terms are obtained by the action of Ward identity on $(z_2-z_3)^{-2h_\sigma}(z_5-z_6)^{-6}$. The function $f(\eta)$ is defined as,
\begin{align}\label{feta}
f(\eta)=n+\eta \frac{\left(n^2-1\right)}{4n}+\eta^2\frac{\left(n^2-1\right)\left(n^2-4\right)}{120n^3}.
\end{align}
In the following, we find it useful to rewrite our expressions in terms of certain functions defined on the (multi-sheeted) plane.  The final results for the integrands will be expressed in terms of the same functions transformed to the multi-sheeted cylinder, the two descriptions being related by the exponential map. To this end, we introduce the functions
$\tilde{G}$, $\tilde{K}$, $\tilde{H}$ and $\tilde{u}$,
\bea
&&\tilde{H}(z)\, =\, \frac{1}{z}\,,\qquad\tilde{G}(z)\, =\, \frac{(z_2-z_3)}{(z-z_2)(z-z_3)}\,, 
 \qquad \tilde{K}(z)\,  =\, \frac{1}{(z-z_1)^2}\,,\\\nonumber\\
&&  \tilde{u}(z)\,  =\, \frac{(z-z_2)(z_1-z_3)}{(z-z_3)(z_1-z_2)}\,, \nonumber
\eea
and rewrite $\eta$ as:
 \bea
 \eta = \frac{1}{\tilde{H}(z_5-z_6)^2}\,\tilde{G}(z_5)\,\tilde{G}(z_6)\,.
 \eea
The $TWW$ correlator on the $n$-sheeted plane can then be rewritten as,
\bea
&& \sum_{i,j,k=1}^{n} \langle T(z_1^{(k)})\, W(z_5^{(i)})\, W(z_6^{(j)})\rangle_n\, =\,  3 \left(n \, J_1\, +\, \frac{(n^2-1)}{4n}\, J_3\, +\, \frac{\left(n^2-1\right)\left(n^2-4\right)}{120n^3}\, J_5 \right) \nonumber\\\nonumber\\
&&+\, \frac{c}{12} \frac{n}{(z_5-z_6)^6} \,   \{ w(z_1),z_1 \}\,f(\eta)\,  +\, \frac{(n^2-1)}{4n} \,\left( - J_3 +  J_4 \right) \\\nonumber\\\nonumber
&& + \frac{\left(n^2-1\right)\left(n^2-4\right)}{60 n^3}\left( - J_5 +  J_6 \right), \nonumber
\eea
where we have defined the $\{J_i\}$:
\bea
&&J_1  =  \tilde{H}^4(z_5-z_6) \tilde{K}(z_5) \tilde{K}(z_6)\\\nonumber\\
&&J_3  =  \tilde{H}^2(z_5-z_6)\tilde{K}(z_5)\tilde{K}(z_6)\tilde{G}(z_5)\tilde{G}(z_6) \nonumber\\\nonumber\\
&&J_4  =  \tilde{H}^4(z_5-z_6)\tilde{K}(z_5)\tilde{K}(z_6)\left(-4+\tilde{u}(z_5)+\tilde{u}(z_6)+\frac{1}{\tilde{u}(z_5)}+\frac{1}{\tilde{u}(z_6)}\right)
\nonumber\\\nonumber\\
&&J_5  =  \tilde{K}(z_5)\tilde{K}(z_6)\tilde{G}(z_5)^2G(z_6)^2\nonumber\\\nonumber\\
&&J_6  \,=\,  \tilde{H}^2(z_5-z_6)\tilde{K}(z_5)\tilde{K}(z_6)\tilde{G}(z_5)\tilde{G}(z_6)\left(-4+\tilde{u}(z_5)+\tilde{u}(z_6)+\frac{1}{\tilde{u}(z_5)}+\frac{1}{\tilde{u}(z_6)}\right)\nonumber
\eea
We then use the map $z\,=\,\exp(2\pi x/\beta )$ from the (branched) cylinder to the (branched) $z$-plane in order to transform the correlator above to the cylinder,
\bea\label{app-TWW-sch1}
&&\langle T(x_1^{(k)}) W(y_1^{(i)}) W(y_2^{(j)})\rangle_{n-{\rm Cyl}} \,=\, ({z'_1}^{(k)})^2\, ({z'_5}^{(i)})^3\, ({z'_6}^{(j)})^3\,  \langle T(z_1^{(k)}) W(z_5^{(i)}) W(z_6^{(j)})\rangle_n \nonumber \\\nonumber\\
&& \qquad \qquad \qquad \qquad   \,+\, \frac{c}{12} \{z(x),x\}\left.\right|_{x_1^{(k)}}\, ({z'_5}^{(i)})^3\, ({z'_6}^{(j)})^3\,  \langle  W(z_5^{(i)}) W(z_6^{(j)})\rangle_n\,.
\eea
We finally find, on the branched cover of the cylinder
\bea\label{app-TWW-sch3}
&& \sum_{i,j,k=1}^{n}\langle T(x_1^{(k)})\, W(y_1^{(i)}) W(y_2^{(j)})\rangle_{n} \,=\,\\
&& 3 \left(n \, I_1 + \frac{(n^2-1)}{4n} I_3 + \frac{\left(n^2-1\right)\left(n^2-4\right)}{120n^3} I_5 \right)  \,+\, \frac{(n^2-1)}{4n}  \left(- I_3 +  I_4 \right)\nonumber\\
&& + \frac{\left(n^2-1\right)\left(n^2-4\right)}{60 n^3} \left(- I_5 + I_6 \right) \,+\, \frac{n \, c}{12} \left( z'(x)^2 \{w(z),z\}+\{z(x),x\} \right)|_{x_1} H^6(y_1-y_2) f(\eta). \nonumber 
\eea
The functions $\{I_j\}$, $H$, etc. are simply transformations of the $\{J_i\}$, $\tilde H$, etc. to the cylinder:
\bea \label{def-integrals} 
&& I_1  =  H^4(y_1-y_2) K(y_1) K(y_2)\\
&&\hat{I}_1  =  H^4(y_1-y_2) G(y_1) G(y_2)\nonumber\\
&&I_2  =  H^2(y_1-y_2) G^2(y_1) G^2(y_2)\nonumber\\
&&I_3  =  H^2(y_1-y_2)K(y_1)K(y_2)G(y_1)G(y_2) \nonumber\\
&&I_4  =  H^4(y_1-y_2)K(y_1)K(y_2)\left(-4+u(y_1)+u(y_2)+\frac{1}{u(y_1)}+\frac{1}{u(y_2)}\right)\nonumber\\
&&I_5  =  K(y_1)K(y_2)G(y_1)^2G(y_2)^2\nonumber\\
&&I_6  =  H^2(y_1-y_2)K(y_1)K(y_2)G(y_1)G(y_2)\left(-4+u(y_1)+u(y_2)+\frac{1}{u(y_1)}+\frac{1}{u(y_2)}\right),\nonumber
\eea
with $G$, $K$, $H$ and $u$ defined as ,
\begin{align} \label{def-fns-cyl}
& G(y) = \frac{\pi \sinh \left( \frac{\pi}{\beta}(x_2-x_3)\right)}{\beta \sinh \left( \frac{\pi}{\beta}(y-x_2)\right)\sinh \left( \frac{\pi}{\beta}(y-x_3)\right)}\,,\\
& K(y)  = \frac{\pi^2}{\beta^2 \sinh \left( \frac{\pi}{\beta}y\right)^2}\,,\nonumber\\
& H(y) = \frac{\pi}{\beta \sinh \left( \frac{\pi}{\beta}y\right)},\nonumber\\
& u(y)  = \frac{\sinh \left( \frac{\pi}{\beta}(y-x_2)\right)\sinh \left( \frac{\pi}{\beta}(x_3)\right)}{\sinh \left( \frac{\pi}{\beta}(y-x_3)\right)\sinh \left( \frac{\pi}{\beta}(x_2)\right)}\,, \nonumber
\end{align}
The last term on the second line of eq.\eqref{app-TWW-sch3} is a disconnected contribution and gets subtracted off. This can also be seen by starting from the $TWW$ correlator on the unbranched or uniformized complex plane and then using the uniformization map \eqref{uniform_map} from the $n$-sheeted cylinder to the (unbranched) complex plane ${\cal C}$, to deduce the correlator on the cylinder. We find, 
\bea\label{app-TWW-sch2}
&& \sum_{k, i , j=1 }^{n-1} (w^{(k)\prime}_1)^2  (v^{(i)\prime}_1 v^{(j)\prime}_2)^ 3   \langle T(w_1^{(p)} ) W(v^{(i)}_1)  W(v^{(j)}_2)  \rangle_{\cal C}\\
&&  = \sum_{k, i , j=1 }^{n}  \langle T(x_1^{(k)}) W(y_1^{(i)}) W(y_2^{(j)})\rangle_n  - \frac{n \, c}{12} \sum_{i , j=1 }^{n}  \{w(x),x\}|_{x_1^{(p)}}  (v^{(i)\prime}_1 v^{(j)\prime}_2)^ 3   \langle W(v^{(i)}_1)  W(v^{(j)}_2)  \rangle_{\cal C}.\nonumber
\eea
Substituting equation (\ref{app-TWW-sch3}) in the above equation and using the following tranformation property of the Schwarzian, 
\begin{align}
\{w(x),x\} = z'(x)^2 \{w(z),z\}+\{z(x),x\},
\end{align}
the Schwarzians from equations (\ref{app-TWW-sch1}) and (\ref{app-TWW-sch3}) cancel to yield,
\bea
&& \sum_{k, i , j=1 }^{n} (w^{(k)\prime}_1)^2  (v^{(i)\prime}_1 v^{(j)\prime}_2)^ 3   \langle T(w_1^{(p)} ) W(v^{(i)}_1)  W(v^{(j)}_2)  \rangle_{\cal C} =\\
&&\qquad\qquad\qquad 3 \left(n \, I_1 + \frac{(n^2-1)}{4n} I_3 + \frac{\left(n^2-1\right)\left(n^2-4\right)}{120n^3} I_5 \right)  + \frac{(n^2-1)}{4n}  \left(- I_3 +  I_4 \right)\nonumber\\
&&\qquad\qquad\qquad + \frac{\left(n^2-1\right)\left(n^2-4\right)}{60 n^3} \left(- I_5 + I_6 \right). \nonumber
\eea
\section{Integrals over the cylinder} \label{fullint}
The holomorphic correlators deduced above must now be integrated over the cylinder to obtain the perturbative correction to RE/EE. We employ the integration prescription described in \cite{Datta:2014ska}.
The following are the final results of the integrations, as defined in equation \eqref{def-integral}, over the integrands listed in equation \eqref{def-fns-cyl}. We  have set $x_2=-t$ and $x_3=L-t$ to simplify the expressions:

{\scriptsize
\begin{align}\label{i3}
\mathcal{I}_3 & = \frac{2 \pi ^6}{\beta ^6} \sinh^2 \left(\frac{L \pi }{\beta }\right) \left\lbrace \beta ^2 \left[-6+5 \coth^2\left(\frac{\pi  (L-t)}{\beta }\right)-6 \coth\left(\frac{\pi  (L-t)}{\beta }\right) \coth\left(\frac{\pi  t}{\beta }\right)  + 5 \coth^2\left(\frac{\pi  t}{\beta }\right)\right] \right.\\
& \times \text{cosech}^2\left(\frac{\pi  (L-t)}{\beta }\right) \text{cosech}^2\left(\frac{\pi  t}{\beta }\right) +4 \pi ^2 (L-t) t \, \text{cosech}^4\left(\frac{L \pi }{\beta }\right) \text{cosech}^2\left(\frac{\pi  (L-t)}{\beta }\right) \text{cosech}^2\left(\frac{\pi  t}{\beta }\right) \nonumber\\
& -4 \pi  \beta  \text{cosech}\left(\frac{L \pi }{\beta }\right) \text{cosech}\left(\frac{\pi  (L-t)}{\beta }\right) \text{cosech}\left(\frac{\pi  t}{\beta }\right) \left[(L-t) \text{cosech}^4\left(\frac{\pi  (L-t)}{\beta }\right) +t \, \text{cosech}^4\left(\frac{\pi  t}{\beta }\right)\right]\nonumber\\
& +\text{cosech}^2\left(\frac{L \pi }{\beta }\right) \left[\left[-4 \pi ^2 t^2+\beta ^2+2 \pi ^2 t^2 \coth^2\left(\frac{L \pi }{\beta }\right)-8 \pi  t \beta  \coth\left(\frac{\pi  t}{\beta }\right)+6 \pi ^2 t^2 \coth^2\left(\frac{\pi  t}{\beta }\right)\right.\right.\nonumber\\
&\left.\left.+4 \pi  t \coth\left(\frac{L \pi }{\beta }\right) \left(-\beta +\pi  t \coth\left(\frac{\pi  t}{\beta }\right)\right)\right] \text{cosech}^4\left(\frac{\pi  t}{\beta }\right)+\text{cosech}^4\left(\frac{\pi  (L-t)}{\beta }\right) \bigg[-4 L^2 \pi ^2+8 L \pi ^2 t \right.\nonumber\\
& -4 \pi ^2 t^2 +\beta ^2+2 \pi ^2 (L-t)^2 \coth^2\left(\frac{L \pi }{\beta }\right)-8 \pi  (L-t) \beta  \coth\left(\frac{\pi  (L-t)}{\beta }\right)+6 \pi ^2 (L-t)^2 \coth^2\left(\frac{\pi  (L-t)}{\beta }\right)\nonumber\\
&\left. \left.\left.+4 \pi  (L-t) \coth\left(\frac{L \pi }{\beta }\right) \text{cosech}\left(\frac{\pi  (L-t)}{\beta }\right) \left(\pi  (L-t) \cosh\left(\frac{\pi  (L-t)}{\beta }\right) -\beta  \sinh \left(\frac{\pi  (L-t)}{\beta } \right) \right) \right] \right] \right\rbrace \nonumber 
\end{align}
\begin{align}\label{i4}
\mathcal{I}_4 & = \frac{4\pi ^6}{3 \beta^4}  \left\lbrace \left(\coth\left(\frac{\pi}{\beta}  (L-t)\right)+\coth^2\left(\frac{\pi}{\beta}  t\right) \right) \left[9 \coth\left(\frac{\pi}{\beta}  (L-t)\right) \left(\coth\left(\frac{\pi}{\beta}  (L-t)\right)-\coth\left(\pi  \frac{t}{\beta}\right)\right) \right.\right.\\
&\left.  +9 \, \text{cosech}^2\left(\pi \frac{t}{\beta}\right)-4\right]+\text{cosech}\left(\frac{\pi}{\beta}  (L-t)\right) \text{cosech}\left(\pi  \frac{t}{\beta}\right)\sinh \left(\frac{L}{\beta} \pi \right) \left[ \left(-8 \frac{\pi}{\beta}  (L-t)+3 \coth\left(\frac{\pi}{\beta}  (L-t)\right) \right) \right.\nonumber\\
&\left. \times\text{cosech}^2\left(\frac{\pi }{\beta} (L-t)\right)+12 \frac{\pi}{\beta}  (L-t) \left(-2+\frac{\pi}{\beta}  (L-t) \coth\left(\frac{\pi }{\beta} (L-t)\right) \right)\text{cosech}^4\left(\frac{\pi}{\beta}  (L-t)\right)\right.\nonumber\\
& \left. \left. +\text{cosech}^2\left(\pi  \frac{t}{\beta}\right) \left[-8 \pi \frac{t}{\beta}+3 \coth\left(\pi  \frac{t}{\beta}\right)  +12 \pi  \frac{t}{\beta} \left(-2+\pi  \frac{t}{\beta}\coth\left(\pi  \frac{t}{\beta}\right) \right) \text{cosech}^2\left(\pi \frac{t}{\beta}\right)\right]\right] \right\rbrace\nonumber
\end{align}
\begin{align}\label{i5}
{\cal I}_5 &= \frac{4\pi ^6}{\beta ^6}  \left\lbrace-\beta  \coth^2\left(\frac{\pi  (L-t)}{\beta }\right)-\beta  \coth^2\left(\frac{\pi  t}{\beta }\right)-\text{cosech}^2\left(\frac{\pi  (L-t)}{\beta }\right) \left[\beta +2 \pi  (L-t) \left(-\coth\left(\frac{L \pi }{\beta }\right) \right. \right.  \right. \\
& \left. \left.  -\coth\left(\frac{\pi  (L-t)}{\beta }\right)\right)\right] -\beta \, \text{cosech}^2\left(\frac{\pi  t}{\beta }\right)+2 \pi  t \coth\left(\frac{L \pi }{\beta }\right) \text{cosech}^2\left(\frac{\pi  t}{\beta }\right) \nonumber\\ 
& \left. -2 \coth\left(\frac{\pi  t}{\beta }\right) \left[\beta  \coth\left(\frac{\pi  (L-t)}{\beta }\right)-\pi  t \, \text{cosech}^2\left(\frac{\pi  t}{\beta }\right)\right]\right\rbrace^2 \nonumber
\end{align}
\begin{align}\label{i6}
\mathcal I_6 & =\frac{\pi ^6}{64 \beta ^6} \text{cosech}^2\left(\frac{L \pi }{\beta }\right) \text{cosech}^6\left(\frac{\pi  (L-t)}{\beta }\right) \text{cosech}^6\left(\frac{\pi  t}{\beta }\right) \bigg\lbrace -40 \pi ^2 \left(L^2+10 L t-10 t^2\right)-65 \beta ^2 \\
&+\left(-20 \pi ^2 \left(29 L^2-30 L t+30 t^2\right)+134 \beta ^2\right) \cosh\left(\frac{2 L \pi }{\beta }\right)+40 \left(2 \pi ^2 \left(L^2-3 L t+3 t^2\right)-\beta ^2\right) \cosh\left(\frac{4 L \pi }{\beta }\right) \nonumber\\
& -10 \left(2 \pi ^2 \left(L^2-2 L t+2 t^2\right)+3 \beta ^2\right) \cosh\left(\frac{6 L \pi }{\beta }\right)+\beta ^2 \cosh\left(\frac{8 L \pi }{\beta }\right) -4 \left(L^2 \pi ^2+\beta ^2\right) \cosh\left(\frac{2 \pi  (L-4 t)}{\beta }\right) \nonumber\\
& +2 \left(2 L \pi ^2 (19 L-12 t)-5 \beta ^2\right) \cosh\left(\frac{2 \pi  (L-3 t)}{\beta }\right)-4 \left(20 \pi ^2 \left(4 L^2-3 L t+3 t^2\right)+9 \beta ^2\right) \cosh\left(\frac{2 \pi  (L-2 t)}{\beta }\right)\nonumber\\
& +8 \left(-L^2 \pi ^2+\beta ^2\right) \cosh\left(\frac{4 \pi  (L-2 t)}{\beta }\right)+5 \left(4 \pi ^2 \left(30 L^2-26 L t-t^2\right)+\beta ^2\right) \cosh\left(\frac{2 \pi  (L-t)}{\beta }\right)\nonumber\\
& +\left(-4 \pi ^2 \left(5 L^2+46 L t-45 t^2\right)+53 \beta ^2\right) \cosh\left(\frac{4 \pi  (L-t)}{\beta }\right)+\left(4 L \pi ^2 (5 L-4 t)+7 \beta ^2\right) \cosh\left(\frac{6 \pi  (L-t)}{\beta }\right)\nonumber\\
& +5 \left(4 \pi ^2 \left(3 L^2+28 L t-t^2\right)+\beta ^2\right) \cosh\left(\frac{2 \pi  t}{\beta }\right)+\left(-4 \pi ^2 \left(6 L^2+44 L t-45 t^2\right)+53 \beta ^2\right) \cosh\left(\frac{4 \pi  t}{\beta }\right)\nonumber\\
& +\left(4 L \pi ^2 (L+4 t)+7 \beta ^2\right) \cosh\left(\frac{6 \pi  t}{\beta }\right)+\left(4 \pi ^2 \left(71 L^2-90 L t+9 t^2\right)-62 \beta ^2\right) \cosh\left(\frac{2 \pi  (L+t)}{\beta }\right)\nonumber\\
& +\left(12 \pi ^2 (L-t)^2-5 \beta ^2\right) \cosh\left(\frac{4 \pi  (L+t)}{\beta }\right)+4 \left(\pi ^2 \left(-13 L^2+28 L t-5 t^2\right)+14 \beta ^2\right) \cosh\left(\frac{2 \pi  (2 L+t)}{\beta }\right)\nonumber\\
& +\left(4 \pi ^2 (L-t)^2+\beta ^2\right) \cosh\left(\frac{2 \pi  (3 L+t)}{\beta }\right)-2 \left(4 \pi ^2 \left(7 L^2-11 L t+9 t^2\right)+15 \beta ^2\right) \cosh\left(\frac{2 \pi  (L+2 t)}{\beta }\right)\nonumber\\
& +3 \beta ^2 \cosh\left(\frac{2 \pi  (L+3 t)}{\beta }\right) -4 \left(L^2 \pi ^2+\beta ^2\right) \cosh\left(\frac{2\pi (3L -4t)}{\beta }\right)+2 \left(2 L \pi ^2 (7 L+12 t)-5 \beta ^2\right) \nonumber\\
& \times\cosh\left(\frac{2\pi (2 L -3 t)}{\beta }\right)+3 \beta ^2 \cosh\left(\frac{2\pi (4L-3t)}{\beta }\right)-2 \left(4 \pi ^2 \left(5 L^2-7 L t+9 t^2\right)+15 \beta ^2\right) \cosh\left(\frac{2\pi (3L -2t)}{\beta }\right)\nonumber\\
& +\left(12 \pi ^2 t^2-5 \beta ^2\right) \cosh\left(\frac{2\pi (4L -2t)}{\beta }\right)+\left(4 \pi ^2 \left(-10 L^2+72 L t+9 t^2\right)-62 \beta ^2\right) \cosh\left(\frac{2\pi (2L -t)}{\beta }\right)\nonumber\\
& +4 \left(\pi ^2 \left(10 L^2-18 L t-5 t^2\right)+14 \beta ^2\right) \cosh\left(\frac{2\pi (3L -t)}{\beta }\right)+\left(4 \pi ^2 t^2+\beta ^2\right) \cosh\left(\frac{2\pi (4L -t)}{\beta }\right)\nonumber\\
& +360 L \pi  \beta  \sinh\left(\frac{2 L \pi }{\beta }\right)+16 L \pi  \beta  \sinh\left(\frac{4 L \pi }{\beta }\right)+56 L \pi  \beta  \sinh\left(\frac{6 L \pi }{\beta }\right)-8 L \pi  \beta  \sinh\left(\frac{2 \pi  (L-4 t)}{\beta }\right)\nonumber\\
& +8 \pi  (-2 L+t) \beta  \sinh\left(\frac{2 \pi  (L-3 t)}{\beta }\right)+176 \pi  (L-2 t) \beta  \sinh\left(\frac{2 \pi  (L-2 t)}{\beta }\right)+20 \pi  (-19 L+5 t) \beta  \sinh\left(\frac{2 \pi  (L-t)}{\beta }\right)\nonumber\\
& +8 \pi  (-3 L+31 t) \beta  \sinh\left(\frac{4 \pi  (L-t)}{\beta }\right)-4 \pi  (11 L+3 t) \beta  \sinh\left(\frac{6 \pi  (L-t)}{\beta }\right)-20 \pi  (14 L+5 t) \beta  \sinh\left(\frac{2 \pi  t}{\beta }\right)\nonumber\\
& +8 \pi  (28 L-31 t) \beta  \sinh\left(\frac{4 \pi  t}{\beta }\right)+4 \pi  (-14 L+3 t) \beta  \sinh\left(\frac{6 \pi  t}{\beta }\right)+8 \pi  (-19 L+21 t) \beta  \sinh\left(\frac{2 \pi  (L+t)}{\beta }\right)\nonumber\\
& +8 \pi  (L-t) \beta  \sinh\left(\frac{4 \pi  (L+t)}{\beta }\right)-8 \pi  (3 L+10 t) \beta  \sinh\left(\frac{2 \pi  (2 L+t)}{\beta }\right)+12 \pi  (-L+t) \beta  \sinh\left(\frac{2 \pi  (3 L+t)}{\beta }\right)\nonumber\\
& +16 \pi  (L+5 t) \beta  \sinh\left(\frac{2 \pi  (L+2 t)}{\beta }\right)+4 \pi  (L-t) \beta  \sinh\left(\frac{2 \pi  (L+3 t)}{\beta }\right) \sinh\left(\frac{2\pi (3L -4t)}{\beta }\right) \nonumber
\\
& +8 \pi  (L+t) \beta  \sinh\left(\frac{2\pi (2L -3t)}{\beta }\right)+4 \pi  t \beta  \sinh\left(\frac{2\pi (4L -3t)}{\beta }\right)+16 \pi  (6 L-5 t) \beta  \sinh\left(\frac{2\pi (3L -2t)}{\beta }\right)\nonumber\\
&+8 \pi  t \beta   \sinh\left(\frac{2\pi (4L -2t)}{\beta }\right)+8 \pi  (2 L-21 t) \beta  \sinh\left(\frac{2\pi (2L -t)}{\beta }\right)+8 \pi  (-13 L+10 t) \beta  \sinh\left(\frac{2\pi (3L -t)}{\beta }\right)\nonumber\\
& \left. -12 \pi  t \beta  \sinh\left(\frac{2\pi (4L-t)}{\beta }\right) \right\rbrace \nonumber
\end{align}}
The following are the values of the above integrals when the endpoints of the entangling interval are at $l_1=0$ and $l_2=\infty$, that is, we take the limit, $L \to \infty$:
\begin{align} \label{intlargel}
\mathcal{I}_3 & = -\frac{16 e^{4 \pi  \frac{t}{\beta}} \pi ^6 }{ \beta^4 \left(-1+e^{2 \pi  \frac{t}{\beta}}\right)^6}\left(-1+e^{4 \pi  \frac{t}{\beta}}-4 e^{2 \pi  \frac{t}{\beta}} \pi  \frac{t}{\beta}\right) \left(7+e^{4 \pi  \frac{t}{\beta}}+8 \pi  \frac{t}{\beta}+4 e^{2 \pi  \frac{t}{\beta}} (-2+\pi  \frac{t}{\beta})\right)\\
\mathcal{I}_4 & = -\frac{64 e^{4 \pi  \frac{t}{\beta}} \pi ^6}{3 \beta^4 \left(-1+e^{2 \pi  \frac{t}{\beta}}\right)^6} \big(5-e^{8 \pi  \frac{t}{\beta}}+4 \pi  \frac{t}{\beta}+e^{6 \pi  \frac{t}{\beta}} (10-4 \pi  \frac{t}{\beta}) \nonumber\\
& \quad + 12 e^{4 \pi  \frac{t}{\beta}} (-1+\pi  \frac{t}{\beta} (-3+2 \pi  \frac{t}{\beta}))+e^{2 \pi  \frac{t}{\beta}} (-2+12 \pi  \frac{t}{\beta} (3+2 \pi  \frac{t}{\beta}))\big) \nonumber\\
\mathcal{I}_5 & = \frac{4 \pi^6}{\beta^4}\left(1+\coth\left(\pi  \frac{t}{\beta}\right)\right)^2 \text{cosech}\left(\pi  \frac{t}{\beta}\right)^4 \left(-2 \pi  \frac{t}{\beta}+\sinh\left(2 \pi  \frac{t}{\beta}\right)\right)^2\nonumber\\
\mathcal{I}_6 & = \frac{128 \pi ^6}{ \beta^4 \left(-1+e^{2 \pi  \frac{t}{\beta}}\right)^6} e^{6 \pi  \frac{t}{\beta}} \big(3+e^{6 \pi  \frac{t}{\beta}}+4 \pi  \frac{t}{\beta}+e^{4 \pi  \frac{t}{\beta}} (1+4 \pi  \frac{t}{\beta} (-3+\pi  \frac{t}{\beta}))\nonumber \\
& \quad +e^{2 \pi  \frac{t}{\beta}} (-5+4 \pi  \frac{t}{\beta} (2+3 \pi  \frac{t}{\beta}))\big). \nonumber
\end{align}
At late times $t \to \infty$, each of these integrals becomes independent of time  and saturate to the values, 
\begin{align}\label{larget}
\mathcal{I}_3 = -16 \frac{\pi^6}{\beta^4}, \qquad
\mathcal{I}_4  = -\frac{64}{3}\frac{\pi^6}{\beta^4}\,,\qquad
\mathcal{I}_5 = 64 \frac{\pi^6}{\beta^4}\,, \qquad \mathcal{I}_6  = 128\frac{\pi^6}{\beta^4}.
\end{align}
As a simple cross check of result for the  integrals in (\ref{i4}) , (\ref{i5}) 
and (\ref{i6}) we have first taken the limit
large interval length limit, and the large time limit and finally performed the integrals 
which then reduce to simple integrations. 
We have confirmed the that result agrees with the equation (\ref{larget}).

\section{Exact result for free fermion example}

The exact result for eq.\eqref{OOWWOO}, ignoring the $WJ$ and $JJ$ correlators on the uniformised plane is given by,
\begin{eqnarray} \label{Wperturbfinal}
&& \frac{ \sum_{i,j=1}^2 \langle \mathcal{O}(x_1^{(1)})\mathcal{O}(x_1^{(0)})W(y_1^{(i)})W(y_2^{(j)})\mathcal{O^\dagger}(x_4^{(1)})\mathcal{O^\dagger}(x_4^{(0)})\rangle }{\langle \mathcal{O}(x^{(0)}_1) \mathcal{O^\dagger}(x^{(0)}_4) \rangle^2}\,=\,\\
&&  \frac{2}{ (1+|\kappa|^2)^2} \left\{ - 9 \alpha^4 H^2(y_1-y_2)\hat{K}^2(y_1)\hat{K}^2(y_2) \left[\left[(1+|\kappa|^4) \right.\right.\right.\nonumber \\ 
&&\left.\left.\left. + 2|\kappa|^2 \left(\frac{B}{A} \frac{(\hat{u}_1-\sqrt{z})^2(\hat{u}_2-\sqrt{z})^2}{\hat{u}_1 \hat u_2(\sqrt{z}-1)^4} + \frac{C}{A} \frac{(\hat{u}_1+\sqrt{z})^2(\hat{u}_2+\sqrt{z})^2}{\hat{u}_1 \hat u_2(\sqrt{z}+1)^4}\right)\right] 
\right.\right.\nonumber\\
&-& 18 \alpha^2 H^4(y_1-y_2) \hat{K}(y_1) \hat{K}(y_2) \bigg[(1+|\kappa|^4)\left(1+\frac{1}{8}H^{-2}(y_1-y_2)G(y_1)G(y_2)\right)\nonumber\\
&& +2|\kappa|^2 \frac{1}{2}\left(\frac{1}{\hat{u}_1}+\frac{1}{\hat{u}_2}\right)\left(\frac{B}{A} \frac{(\hat{u}_1-\sqrt{z})(\hat{u}_2-\sqrt{z})}{(\sqrt{z}-1)^2} + \frac{C}{A} \frac{(\hat{u}_1+\sqrt{z})(\hat{u}_2+\sqrt{z})}{(\sqrt{z}+1)^2}\right)\bigg]\nonumber\\
&&\left.- 6 \alpha^2 H^6(y_1-y_2) \left(1+\frac{3}{16} H^{-2}(y_1-y_2)G(y_1)G(y_2)\right)\left[(1+|\kappa|^4)+2|\kappa|^2\left(\frac{B}{A}+\frac{C}{A}\right)\right]\right\}, \nonumber
\end{eqnarray}
The $WJ$ and $JJ$ correlators  are present in the full result because the spin-$3$ current does not transform as a primary in the $W_{1+\infty}$ theory 
(eq.\eqref{Wtransf}).
Here $z$ is the cross-ratio defined in equation \eqref{defcrossratio}. The functions $\hat{K}$ and $\hat{u}$ are,
\begin{align}
& \hat{K}(y) = \frac{\pi \sinh \left( \frac{\pi}{\beta}(x_1-x_4)\right)}{\beta \sinh \left( \frac{\pi}{\beta}(y-x_1)\right)\sinh \left( \frac{\pi}{\beta}(y-x_4)\right)},\\
& \hat{u}(y)  = \frac{\sinh \left( \frac{\pi}{\beta}(y-x_2)\right)\sinh \left( \frac{\pi}{\beta}(x_1-x_3)\right)}{\sinh \left( \frac{\pi}{\beta}(y-x_3)\right)\sinh \left( \frac{\pi}{\beta}(x_1-x_2)\right)},\nonumber
\end{align}
and the ratios $\frac{B}{A}$ and $\frac{C}{A}$ are,
\begin{align}
\frac{B}{A} = \frac{|\sqrt{z}-1|^{8 \alpha^2}}{(2\sqrt{2}|z|)^{2\alpha^2}}, \quad \frac{C}{A} = \frac{|\sqrt{z}+1|^{8 \alpha^2}}{(2\sqrt{2}|z|)^{2\alpha^2}}.
\end{align}

\end{document}